\documentclass[aps,                
prc,                
showpacs,           
twocolumn,          
preprintnumbers,
floatfix,           
superscriptaddress 
]
{revtex4-1}

\usepackage{graphicx} 
\usepackage{epsfig} 
\usepackage{bm} 
\usepackage{amsmath}
\usepackage{slashed} 
\usepackage{hyperref}
\usepackage{physics} 
\usepackage[usenames,dvipsnames,svgnames,table]{xcolor}
\usepackage{soul} 
\setstcolor{red}   

\hypersetup{colorlinks,linkcolor=MidnightBlue,citecolor=MidnightBlue,urlcolor=MidnightBlue}

\begin{document}


\title{Exploring the partonic phase at finite chemical potential within an extended off-shell transport approach}


\author{Pierre Moreau}\email{moreau@fias.uni-frankfurt.de}
\affiliation{Institute for Theoretical Physics, Johann Wolfgang
Goethe-Universit\"{a}t, Frankfurt am Main, Germany}

\author{Olga Soloveva}
\affiliation{Institute for Theoretical Physics, Johann Wolfgang
Goethe-Universit\"{a}t, Frankfurt am Main, Germany}

\author{Lucia Oliva}
\affiliation{Institute for Theoretical Physics, Johann Wolfgang
Goethe-Universit\"{a}t, Frankfurt am Main, Germany}
\affiliation{GSI
Helmholtzzentrum f\"{u}r Schwerionenforschung GmbH,
 Darmstadt, Germany}

\author{Taesoo Song}
\affiliation{GSI
Helmholtzzentrum f\"{u}r Schwerionenforschung GmbH,
 Darmstadt, Germany}


\author{Wolfgang Cassing}
\affiliation{Institut f\"{u}r Theoretische Physik, Universit\"{a}t
Gie\ss en, Germany}

\author{Elena Bratkovskaya}
\affiliation{Institute for Theoretical Physics, Johann Wolfgang
Goethe-Universit\"{a}t, Frankfurt am Main, Germany}
\affiliation{GSI
Helmholtzzentrum f\"{u}r Schwerionenforschung GmbH,
 Darmstadt, Germany}


\begin{abstract}
We extend the parton-hadron-string dynamics (PHSD) transport
approach in the partonic sector by explicitly calculating the total and
differential partonic scattering cross sections as a function of temperature $T$ and baryon chemical potential $\mu_B$ on the basis of the effective propagators and couplings from the dynamical quasiparticle model (DQPM) that is matched to reproduce the equation of state of the partonic system above the deconfinement
temperature $T_c$ from lattice QCD. 
We calculate the collisional widths for the partonic degrees of freedom at finite $T$ and $\mu_B$ in the timelike sector and conclude that the quasiparticle limit holds sufficiently well.   Furthermore, the ratio of shear viscosity $\eta$ over entropy density $s$, i.e.,  $\eta/s$, is evaluated using the collisional widths and compared to lattice QCD calculations for $\mu_B$ = 0 as well. We find that the novel ratio $\eta/s$  does not differ very much from that calculated within the original DQPM on the basis of the Kubo formalism. Furthermore, there is only a very modest change of $\eta/s$ with the baryon chemical $\mu_B$ as a function of the scaled temperature $T/T_c(\mu_B)$.   This also holds for a variety of hadronic observables from central $A+A$ collisions in the energy range
5 GeV $\leq \sqrt{s_{NN}} \leq$ 200 GeV when implementing the differential cross sections into the PHSD approach.
We only observe small differences in the antibaryon sector (${\bar p}, {\bar \Lambda}+{\bar \Sigma}^0$) at $\sqrt{s_{NN}}$ = 17.3 GeV and 200 GeV with practically no sensitivity of rapidity and $p_T$ distributions to the $\mu_B$ dependence of the partonic cross sections. Small variations in the strangeness sector are obtained
in all collisional systems studied ($A+A$ and C + Au); however, it will be very hard to extract a robust signal experimentally. Since we find only small traces of a $\mu_B$ dependence in heavy-ion observables --- although the effective partonic masses and widths as well as their partonic cross sections clearly depend on $\mu_B$ --- this implies that one needs a sizable partonic density and large space-time QGP volume to explore the dynamics in the partonic phase. These conditions are only fulfilled at high bombarding energies where $\mu_B$ is, however, rather low. On the other hand, when decreasing the bombarding energy and thus increasing $\mu_B$, the hadronic phase becomes dominant and accordingly it will be difficult to extract signals from the partonic dynamics based on "bulk" observables.
\end{abstract}


\maketitle

\section{Introduction}
\label{Section1}

Non-equilibrium many-body theory or transport theory has become a
major topic of research in nuclear physics, cosmological particle
physics, and condensed matter physics. The multidisciplinary
aspect arises due to a common interest in understanding the various
relaxation phenomena of quantum dissipative systems. Important
questions in nuclear and particle physics at the highest energy
densities are the following: (i) how do non-equilibrium systems in extreme
environments  evolve and eventually thermalize, (ii) what are the
macroscopic transport coefficients of the matter in equilibrium,
and (iii) what is the nature of possible phase transitions? The dyna\-mics of heavy-ion collisions at
various bombarding energies provide the laboratory of choice for
research on nonequilibrium quantum many-body physics and
relativistic quantum-field theories, since the initial state of a
collision resembles an extreme nonequilibrium configuration while the
final state might even exhibit some degree of thermalization.

For many decades, the powerful method of the Schwinger-Keldysh
\cite{Schwinger:1960qe,Bakshi:1963bn,Keldysh:1964ud,Cr68} or closed
time path (CTP) real-time Green's functions --- being the essential
degrees of freedom --- has been shown to provide an appropriate basis
for  the formulation of the complex problems in the various  areas
of nonequilibrium quantum many-body physics. Within this framework,
one can derive suitable approximations --- depending on the
problem under consideration --- by preserving  overall consistency
relations \cite{Bonitz}. Originally, the resulting causal Dyson-Schwinger equation
of motion for the one-particle Green's functions (or two-point
functions), i.e., the Kadanoff-Baym (KB) equations \cite{KadanoffBaym}, have
served as the underlying scheme for deriving various transport
phenomena and generalized transport equations. For review articles
on the Kadanoff-Baym equations in the various areas of
nonequilibrium quantum physics, we refer the reader to Refs.
\cite{dubois1967lectures,Danielewicz:1982ca,Chou:1984es,Rammer:1986zz,Calzetta:1986cq,Haug}.

On the other hand, kinetic transport theory is a convenient method to
study many-body nonequilibrium systems. Kinetic equations, which do play the central role in
more or less all practical  simulations, can be derived from the KB
equations within suitable approximations. Hence, a major impetus in
the past has been to derive semiclassical Boltzmann-like transport
equations within the standard quasiparticle approximation.
Additionally, off-shell extensions by means of a gradient expansion
in the space-time inhomogeneities --- as already introduced by
Kadanoff and Baym \cite{KadanoffBaym} --- have been formulated for various
directions in physics, from a relativistic electron-photon plasma
\cite{BEZZERIDES197210} to the transport of nucleons at intermediate heavy-ion
reactions \cite{Botermans:1990qi} to the transport of
partons in high-energy heavy-ion reactions
\cite{Mrowczynski:1989bu,Makhlin:1998zi,Makhlin:1994ew,Geiger:1995ak,Geiger:1996ym,Brown:1998zx,Blaizot:1999xk,Cassing:1999wx,Cassing:1999mh}.
We recall that on the formal level of the KB equations the various
forms assumed for the self-energy have to fulfill consistency
relations in order to preserve symmetries of the fundamental
Lagrangian \cite{KadanoffBaym,Ivanov:1998nv,Knoll:2001jx}. This allows for a unified
treatment of stable and unstable (resonance) particles also out of equilibrium. 

The possibilities of solving in particular QCD in Minkowski space for out-of-equilibrium
configurations and nonvanishing quark (or baryon) densities will be
low in the next years, such that effective approaches are
necessary to model the dominant properties of QCD in equilibrium,
i.e., the thermodynamic quantities and transport coefficients.
To this aim, the dynamical quasiparticle model (DQPM) has been
introduced \cite{Cassing:2008nn}, which is based on partonic propagators with sizable
imaginary parts of the self-energies incorporated. Whereas the real part of
the self-energies can be attributed to a dynamically generated mass (squared), the imaginary parts
contain the information about the interaction rates in the system \cite{Weldon:1983jn,Lebedev:1989ev,Braaten:1990it,Braaten:1992gd,Pisarski:1993rf,Jeon:1994if,Wang:1995qf,Thoma:1993vs}. Furthermore,
the imaginary parts of the propagators define the spectral functions of the degrees of freedom
which might show narrow (or broad) quasiparticle peaks \cite{Liu:2017qah}. A further advantage of a propagator-based approach is that one can formulate a consistent thermodynamics \cite{Vanderheyden:1998ph} as well as a causal theory for nonequilibrium configurations on the basis of KB equations.

In order to explore the phase diagram of strongly interacting matter as a
function of temperature $T$ and baryon chemical potential $\mu_B$, different strategies are employed at present:
(i) Lattice calculations of quantum chromodynamics
(lQCD)~\cite{Bernard:2004je,Aoki:2006we,Bazavov:2011nk} show that the phase
transition from hadronic to partonic degrees of freedom (at
vanishing baryon chemical potential $\mu_B$ = 0) is a crossover. This
phase transition is expected to turn into a first-order transition
at a critical point $(T_r, \mu_r)$ in the phase diagram with
increasing baryon chemical potential $\mu_B$ \cite{Fischer:2012vc,Fischer:2018sdj,Senger:2011zza}. Furthermore, a nonvanishing magnetic field, as produced in heavy-ion collisions, can also influence the position of the critical point \cite{Ruggieri:2014bqa}. Since this latter cannot be determined theoretically in a reliable way by lQCD calculations, experimental information
from relativistic nucleus-nucleus collisions has to be obtained. In this respect, (ii) the beam
energy scan (BES) program --- performed at the  relativistic heavy-ion collider (RHIC) ---  aims to find the
critical point and the phase boundary by gradually decreasing the
collision energy~\cite{Mohanty:2011nm,Kumar:2011us} and thus increasing
the average baryon chemical potential.
Additionally, new facilities such as the Facility for Antiproton and Ion
Research (FAIR) and Nuclotron-based Ion Collider fAcility (NICA) are
under construction to explore in particular the intermediate energy
range of 4 GeV $\leq \sqrt{s_{NN}} \leq$ 20 GeV, where one might study also the competition between chiral
symmetry restoration and deconfinement \cite{Cassing:2015owa,Palmese:2016rtq}.

Accordingly, the partonic and hadronic
dynamics at finite or large baryon densities (or chemical potentials)
are of actual interest and are addressed also in various
hydrodynamical models \cite{Shen:2018pty,Li:2018fow,Ivanov:2005yw,Ivanov:2015vna},
hydrodynamical + hadron transport models \cite{Denicol:2018wdp,Petersen:2008dd,Karpenko:2018xam}, and more parametric approaches \cite{Li:2018ini}. However, as found in Ref. \cite{Denicol:2018wdp}, the inclusion of baryon diffusion leads only to a small effect on the ``bulk'' observables at BES RHIC energies.


About a decade ago, the parton-hadron-string dynamics (PHSD) transport approach was introduced, which
differs from the conventional Boltzmann-type models in the
aspect~\cite{Cassing:2009vt} that the degrees of freedom for the QGP
phase are off-shell massive strongly interacting quasiparticles
that generate their own mean-field potential. The masses of the
dynamical quarks and gluons in the QGP are distributed according to
spectral functions whose pole positions and widths, respectively,
are defined by the real and imaginary parts of their self-energies
\cite{Linnyk:2015rco}.  The partonic propagators and self-energies,
furthermore, are defined in the DQPM in which the strong coupling and the self-energies are fitted to lattice QCD results \cite{Cassing:2008nn}
assuming an ansatz for the mass and width
dependencies on temperature $T$ and quark chemical potential $\mu_q$
inspired by the hard-thermal-loop (HTL) approach.

In the past, the PHSD transport model, based on temperature-dependent DQPM masses,
widths, and cross sections \cite{Ozvenchuk:2012fn}, has successfully described numerous
experimental data in relativistic heavy-ion collisions from the
Alternating Gradient Synchrotron (AGS), Super-Proton Synchrotron (SPS), RHIC, and Large Hadron Collider (LHC)
energies~\cite{Cassing:2009vt,Linnyk:2012pu,Bratkovskaya:2011wp,Konchakovski:2014fya,Konchakovski:2011qa,Konchakovski:2012yg,Konchakovski:2014gda,Linnyk:2015rco}.

The {\it goals of this study} are to explore on a microscopic level the partonic phase
at finite baryonic chemical potential $\mu_B$ and different temperatures $T$
and to find traces of the $\mu_B$ dependence in observables.
Although the extension of the DQPM model to finite baryon
chemical potentials has been realized previously
\cite{Berrehrah:2015vhe,Berrehrah:2016vzw} and the
($T,\mu_B$) dependence of the transport coefficients (such as shear
and bulk viscosities or electric conductivity) for the equilibrated
QGP matter have been calculated \cite{Berrehrah:2015vhe,Berrehrah:2016vzw},
the properties of the non equilibrium QGP at finite $\mu_B$ --- as created in heavy-ion
collisions (HICs) ---
were not addressed by microscopic calculations within the PHSD so far in a consistent fashion.

Although the DQPM inherits the information on the total interaction rates
of the degrees of freedom in terms of widths, it lacks the individual total as well as differential cross sections for different reaction channels with partons
that are needed in the collision terms of a consistent relativistic transport approach. In PHSD, these cross sections have been parametrized so far to comply with the individual widths of quarks, antiquarks, and gluons as a function of energy density (cf. Ref. \cite{Ozvenchuk:2012fn}), which can be related to the
temperature $T$ by the lQCD equation of state (EoS). In this study, we will calculate these total and differential cross sections in leading order for the individual partonic channels on the basis of the DQPM propagators and couplings. This will allow us to additionally explore the energy and angular dependence of partonic cross sections on their
$T$ and $\mu_B$ dependence.

Moreover, using these cross sections, we calculate the interaction
rates of quarks and gluons in the timelike sector to study the validity of the quasiparticle approximation.
Furthermore,  we evaluate the equilibrium shear viscosity $\eta(T,\mu_B)$  within the Kubo formalism and the relaxation time approximation (RTA) and compare to results from lQCD at $\mu_B$ = 0 for the ratio $\eta/s$.
The calculated total and differential cross sections as well as parton masses  --- depending on ($T,\mu_B$) --- have been implemented
in the PHSD and thus, we introduce a further step towards a consistent relativistic transport approach in the partonic
sector.

In order to extract the Lagrange parameters $\mu_B$ and $T$ from the
PHSD in heavy-ion collisions, we developed a practical method (based
on the expansion of thermodynamic quantities in terms of the baryon number susceptibilities) which allows us to relate the
energy density and baryon densities --- calculated in each cell in space-time during
heavy-ion collisions --- to a state-of-the-art lattice QCD EoS (practically identical to the DQPM EoS at small $\mu_B$).

Finally, we will search for  traces of the $\mu_B$ dependence in the
 QGP dynamics in ``bulk'' observables from relativistic heavy-ion collisions such as rapidity
distributions and $p_T$ spectra using the extended PHSD approach as
a working tool.

This paper is organized as follows: In Sec. \ref{Section2}, we will provide a brief reminder of the DQPM and its ingredients as well as its results for the partonic equation of state. Section \ref{Section3} will be devoted to the calculation of the partonic differential cross sections as a function of $T$ and $\mu_B$, employing the effective propagators and couplings from the DQPM. In Sec. \ref{Section4}, we will use these cross sections to evaluate partonic scattering rates for fixed $T$ and $\mu_B$ as well as compute transport coefficients in Sec. \ref{Section5} like the shear viscosity $\eta$  in comparison to calculations from lQCD at $\mu_B = 0$.  Section \ref{Section6} is devoted to the extraction of the local $T$ and $\mu_B$ in the actual transport approach and characteristic results will be presented for central collisions of Pb + Pb at $\sqrt{s_{NN}}$ = 17.3 and Au + Au 200 GeV. In Sec. \ref{Section7}, we will compare the results of the novel transport approach PHSD5.0 to those of PHSD4.0 and experimental data for central heavy-ion collisions from AGS to RHIC energies. Furthermore, we explore the sensitivity of rapidity distributions and transverse momentum spectra to the partonic scattering in asymmetric C + Au collisions at the top SPS and RHIC energies.  A summary of our study will be presented in Sec. \ref{Section8}, while technical details in the calculation of the matrix elements and differential cross sections are shifted to the \hyperref[Appendix]{Appendixes}.


\section{Reminder of the DQPM and its ingredients}
\label{Section2}

Early concepts of the quark-gluon-plasma (QGP) were guided by the
idea of a weakly interacting system of massless partons which might
be described by perturbative QCD (pQCD). However, experimental
observations at RHIC indicated that the new medium created in
ultrarelativistic Au + Au collisions is interacting more strongly than
hadronic matter. It is presently widely accepted that this medium is
a strongly interacting system of partons  as extracted experimentally
from the strong radial expansion and the scaling of the elliptic
flow $v_2(p_T)$ of mesons and baryons with the number of constituent
quarks and antiquarks \cite{Adler:2003kt}. At vanishing  chemical potential $\mu_B$,
the QCD problem can be addressed at zero and finite temperature by
lattice QCD calculations on a (3 + 1)-dimensional torus with a suitable
discretization of the QCD action on the Euclidean lattice. These
calculations so far have provided valuable information on the QCD
equation of state, chiral symmetry restoration, and various
correlators that can be attributed/ or related to transport
coefficients. Because of the Fermion sign problem, lQCD calculations at
finite $\mu_B$ are presently not robust and one has to rely on
nonperturbative --- but effective --- models to obtain information in
the ($T$, $\mu_B$) plane or for systems out of equilibrium.

\subsection{Quasiparticle properties}

As mentioned above in the KB theory the
field quanta are described in terms of dressed propagators with
complex self-energies \cite{Cassing:2008nn}. Whereas the real part of
the self-energies can be related to mean-field potentials (of Lorentz
scalar, vector, or tensor type), the imaginary parts  provide
information about the lifetime and/or reaction rates of timelike
particles. The determination and extraction of complex self-energies for the
partonic degrees of freedom can be performed within the DQPM by fitting lattice QCD calculations in
thermal equilibrium.

The basic ideas of the DQPM are as follows:\\
(i) Introduce an ansatz (with a few parameters) for the
($T$ and $\mu_B$) dependence of masses and widths of the dynamical
quasiparticles (quarks, antiquarks, and gluons) to define the self-energies.\\
(ii) Define the form of propagators for strongly interacting massive partons.\\
(iii) Evaluate the QGP thermodynamics in equilibrium using the Kadanoff-Baym (KB) theory and
calculate (in the 2PI approximation) the entropy density $s$ and other
thermodynamic quantities such as the pressure $P$ and energy density. \\
(iv) Compare the DQPM results with the lQCD ones at zero and finite $\mu_B$ and $T$
and fix the initial parameters to obtain the best reproduction
of the lQCD thermodynamics. \\
This defines the properties of the quasiparticles, their propagators and couplings.

We recall the main ingredients of the DQPM:\\
1) The DQPM postulates retarded propagators of the
quark and gluon degrees-of-freedom (for the QGP in equilibrium) in the form

\begin{equation}
\label{propdqpm} G^{R} (\omega, {\bf p}) = \frac{1}{\omega^2 - {\bf
p}^2 - M^2 + 2 i \gamma \omega}
\end{equation}
~\\
using $\omega=p_0$ for energy.\\
2) The coupling (squared) $g^2$, which is the essential quantity in the DQPM defining
the strength of the interaction  and enters the definition
of the DQPM thermal masses and widths, is extracted from lQCD. In our previous studies
\cite{Berrehrah:2013mua,Berrehrah:2014ysa,Berrehrah:2015ywa},
we used an ansatz for the $(T,\mu_B)$ dependence of the coupling
$g^2 = \alpha_s/(4\pi)$ and extracted the two parameters --- entering the parametrization
of $g^2$ --- from a global fit to the lQCD thermodynamics. Furthermore, $g^2$ was also
compared to quenched QCD results on $\alpha_s(T)$ at $\mu_B=0$
for the pure glue case ($N_f$ = 0) from Ref. \cite{Kaczmarek:2004gv}.

Here we follow alternatively a procedure similar to Refs.
\cite{Berrehrah:2015vhe,Berrehrah:2016vzw} to determine the
effective  coupling  (squared) $g^2$ as a function of temperature
$T$; i.e., the coupling is defined at $\mu_B = 0$ by a
parametrization of the entropy density from  lattice QCD in the
following way:

\begin{equation}
g^2(s/s_{SB}) = d \left( (s/s_{SB})^e -1 \right)^f
\label{coupling_DQPM}
\end{equation}
~\\
with  the Stefan-Boltzmann entropy density $s_{SB}^{QCD} = 19/9\pi^2 T^3$ and the parameters $d=169.934$, $e=-0.178434$ and $f=1.14631$. In the following, we use a parametrization of the entropy density at $\mu_B = 0$ calculated by lQCD from Refs. \cite{Borsanyi:2012cr,Borsanyi:2013bia} to determine the DQPM coupling constant as a function of temperature.

The extension to finite $\mu_B$ can be worked out in different scenarios. First of all, an expansion of the grand-canonical potential, i.e., the negative pressure $P$, in terms of $\mu_B/T$ can be performed and the expansion coefficients can be calculated by lQCD \cite{Karsch:2013fga,Bazavov:2017dus}. This provides a solid framework for small and moderate $\mu_B$. Alternatively, Maxwell relations can be employed to extract the thermodynamic potential at finite $\mu_B$ starting from the information given by lQCD at $\mu_B$ =0 \cite{Steinert:2018bma}. Both methods give almost the same results up to $\mu_B \approx$ 450 MeV \cite{Steinert:2018bma}. For practical purposes, the explicit results can be fitted by a scaling ansatz \cite{Berrehrah:2016vzw} which
works up to $\mu_B \approx$ 450 MeV and suggests that the phase transition to the QGP is a crossover up to such baryon chemical potentials. We mention that the experimental studies of the STAR Collaboration within the BES program down to bombarding energies of $\sqrt{s_{NN}}$ = 7.7 GeV --- corresponding to  $\mu_B \approx$ 450 MeV --- did not indicate any critical point in the QCD phase diagram \cite{Adamczyk:2017iwn} so far.

To obtain the coupling constant at finite baryon chemical potential $\mu_B$, the scaling hypothesis assumes that $g^2$ is a function of the ratio of the effective temperature $T^* = \sqrt{T^2+\mu^2_q/\pi^2}$ and the $\mu_B$-dependent critical temperature $T_c(\mu_B)$ as \cite{Cassing:2007nb}

\begin{equation}
g^2(T/T_c,\mu_B) = g^2\left(\frac{T^*}{T_c(\mu_B)},\mu_B =0 \right)
\label{coupling}
\end{equation}
~\\
\noindent with $\mu_B=3\mu_q$ and $T_c(\mu_B) = T_c \sqrt{1-\alpha \mu_B^2}$, where $T_c$ is
the critical temperature at vanishing chemical potential ($\approx 0.158$ GeV) and $\alpha = 0.974\ \text{GeV}^{-2}$.\\

\begin{figure}[h!]
	\centering
	\includegraphics[width=\columnwidth]{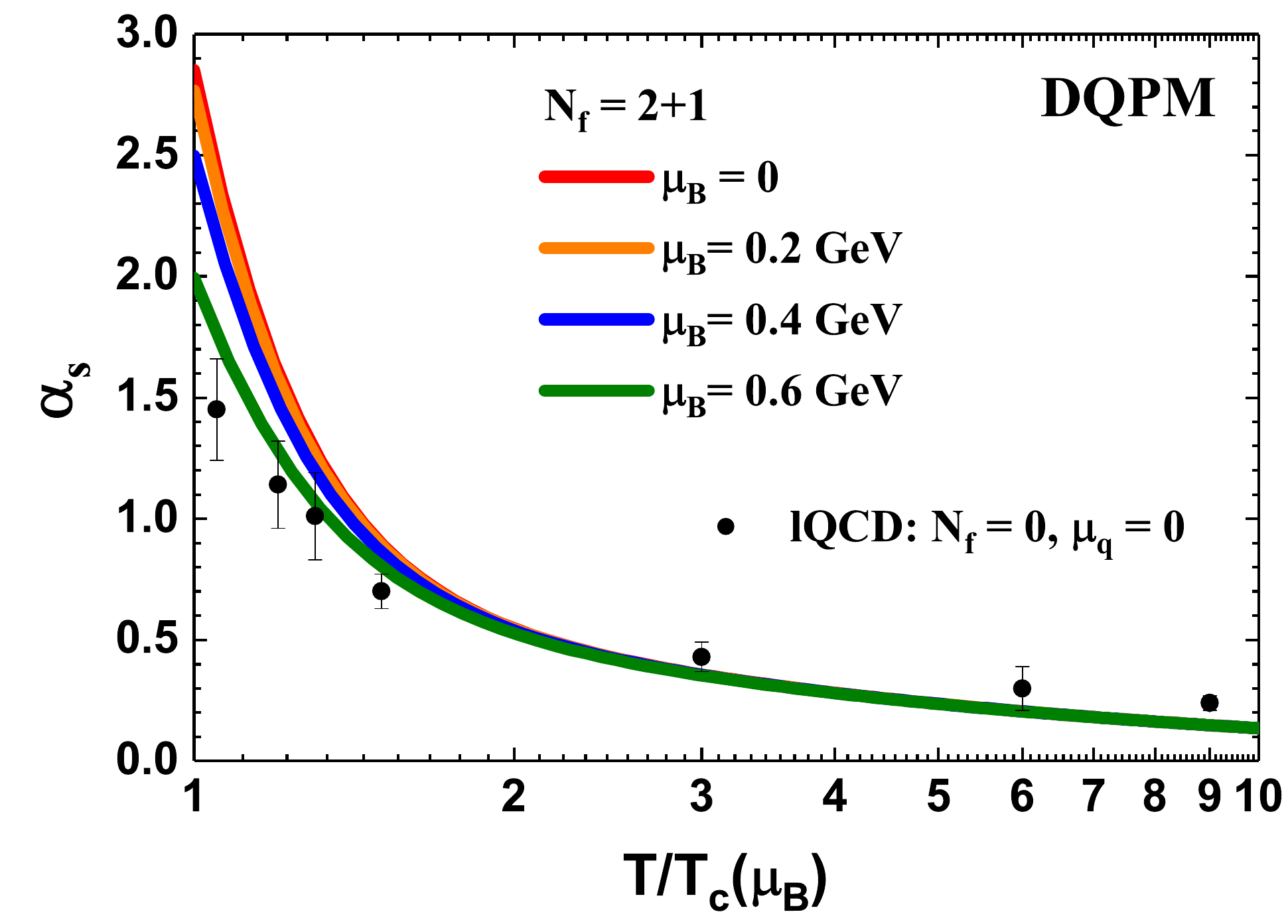}
	\caption{(Color online) The DQPM running coupling $\alpha_s = g^2(T,\mu_B)/(4\pi)$ (for $N_f$ = 2+1) as a function of the scaled temperature $T/T_c(\mu_B)$ for different values of the baryon chemical potential $\mu_B$. The lattice results for quenched QCD (for $N_f$ = 0) are taken from Ref. \cite{Kaczmarek:2004gv} and scaled by the critical temperature $T_c \approx$ 270 MeV.}
	\label{fig-DQPM-alphas}
\end{figure}

In Fig. \ref{fig-DQPM-alphas}, the DQPM running coupling $\alpha_s =
g^2(T,\mu_B)/(4\pi)$ is displayed as a function of the scaled
temperature $T/T_c(\mu_B)$ for different values of the baryon
chemical potential $\mu_B$. We find that with increasing $\mu_B$ the
effective coupling drops in the vicinity of the critical temperature
$T_c(\mu_B)$. This drop is rather moderate up to $\mu_B$ = 0.4 GeV
(adequate for central Au + Au collisions at 30 A GeV) but becomes
significant for $\mu_B$ = 0.6 GeV (roughly adequate for central
Au + Au collisions at 10 A GeV). A comparison to the lattice results
for quenched QCD from Ref. \cite{Kaczmarek:2004gv} --- scaled by $T_c
\approx$ 270 MeV --- shows that the DQPM coupling qualitatively
matches the lattice results but is slightly larger
for lower $\mu_B$. We note, however, that this comparison should be
taken only for orientation since the DQPM coupling corresponds to
unquenched QCD with three light flavors ($N_f$ = 2+1) whereas the
lattice results are for quenched QCD  ($N_f$ = 0).
Note that --- since the running
coupling (squared) $g^2 \sim (11 N_c-2 N_f)^{-1}$ ($N_c$ = 3) --- the coupling is larger for a
finite number of flavors $N_f$ compared to $N_f=0$.

With the coupling $g^2$ fixed from lQCD, one can now specify the dynamical quasiparticle mass (for gluons and quarks)
which is assumed to be given by the HTL  thermal mass in the asymptotic
high-momentum regime, i.e., for gluons by \cite{Bellac:2011kqa,Linnyk:2015rco}

\begin{equation}
M^2_{g}(T,\mu_B)=\frac{g^2(T,\mu_B)}{6}\left(\left(N_{c}+\frac{1}{2}N_{f}\right)T^2
+\frac{N_c}{2}\sum_{q}\frac{\mu^{2}_{q}}{\pi^2}\right)\ ,
\label{Mg9}
\end{equation}
~\\
and for quarks (antiquarks) by

\begin{equation}
M^2_{q(\bar q)}(T,\mu_B)=\frac{N^{2}_{c}-1}{8N_{c}}g^2(T,\mu_B)\left(T^2+\frac{\mu^{2}_{q}}{\pi^2}\right)\ ,
\label{Mq9}
\end{equation}
~\\
where $N_{c}=3$ stands for the number of colors while $N_{f}\ (=3)$
denotes the number of flavors. The dynamical masses (\ref{Mq9})
in the QGP are large compared to the bare masses of the light
($u,d$) quarks and adopted in the form (\ref{Mq9}) for the ($u,d$) quarks.
The strange quark has a larger bare mass, which also enters to some
extent the dynamical mass $M_s(T)$. This essentially suppresses the channel
$g \rightarrow s + {\bar s}$ relative to the channel $g \rightarrow u + {\bar u}$
or $d + {\bar d}$ and controls the strangeness ratio in the QGP. Empirically, we have used
$M_s(T,\mu_B)= M_u(T,\mu_B)+ \Delta M = M_d(T,\mu_B)+ \Delta M$, where $\Delta M$ =
30 MeV, which has been fixed once in comparison to experimental
data for the $K^+/\pi^+$ ratio in central Au + Au collisions at
$\sqrt{s_{NN}}$ = 200 GeV.  Furthermore, the effective
quarks, antiquarks, and gluons in the DQPM have finite widths, which are adopted in the form \cite{Linnyk:2015rco}

\begin{equation}
\label{widthg}
\gamma_{g}(T,\mu_B) = \frac{1}{3}N_{c}\frac{g^2(T,\mu_B)T}{8\pi}\ln\left(\frac{2c}{g^2(T,\mu_B)}+1\right),
\end{equation}
\begin{equation}
\label{widthq}
\gamma_{q(\bar
q)}(T,\mu_B)=\frac{1}{3}\frac{N^{2}_{c}-1}{2N_{c}}\frac{g^2(T,\mu_B)T}{8\pi}
\ln\left(\frac{2c}{g^2(T,\mu_B)}+1\right),
\end{equation}
~\\
where $c=14.4$  is related to a magnetic cutoff, which is an additional parameter
of the DQPM. Furthermore, we assume that the width
of the strange quark is the same as that for the light ($u,d$)
quarks.

\begin{figure}[h!]
\centering	
	\includegraphics[width=\columnwidth]{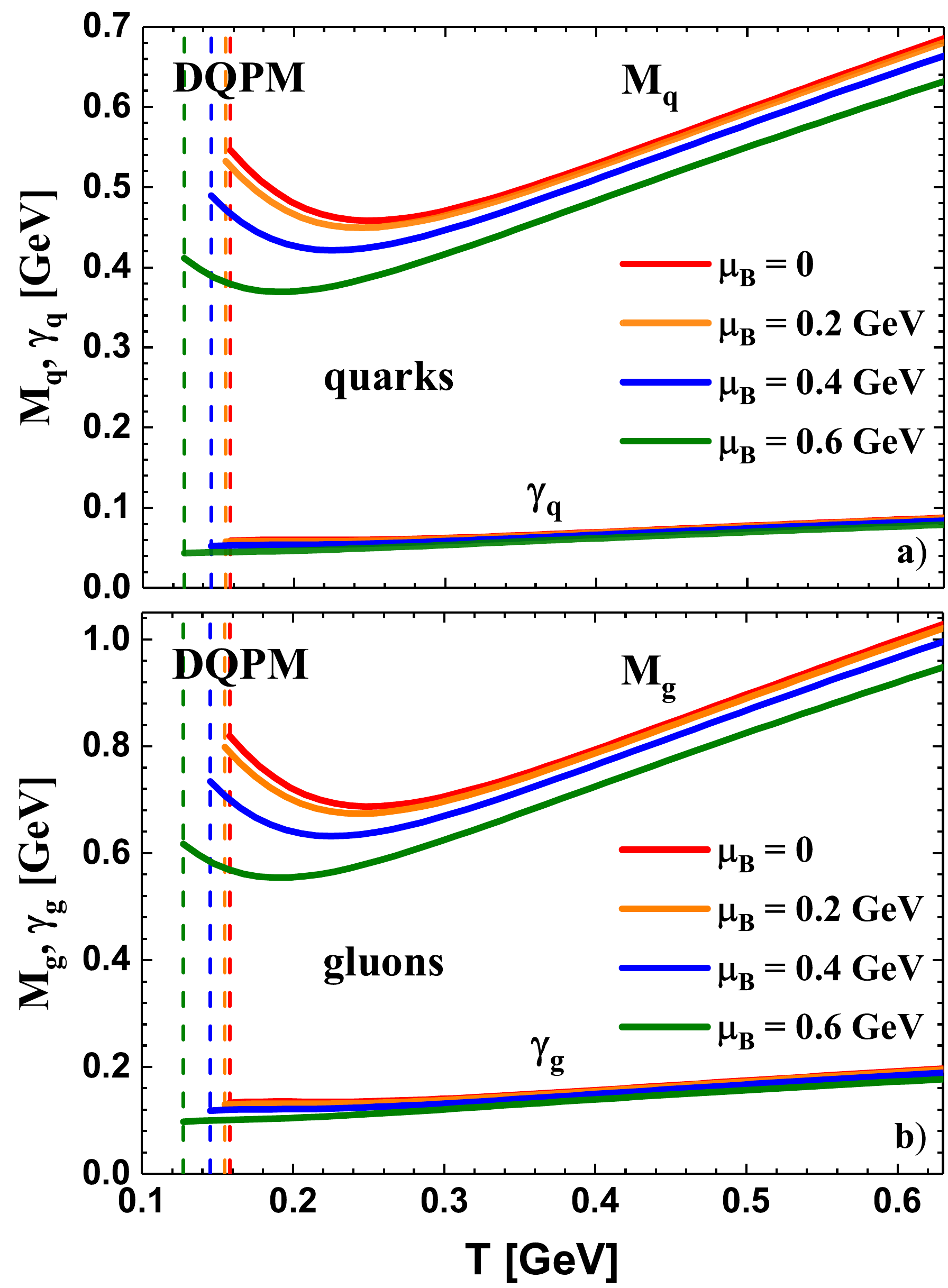}
	\caption{(Color online) The effective quark (a) and gluon (b) masses $M$ and widths $\gamma$ [from the parametrizations (\ref{widthg}) and (\ref{widthq})] as a function of the temperature $T$ for different $\mu_B$. The vertical dashed lines correspond to the DQPM $\mu_B$-dependent critical temperature $T_c(\mu_B)$.}
	\label{fig-DQPM-masses}
\end{figure}

The physical processes contributing to the width
$\gamma_g$ are both $gg \leftrightarrow gg$, $gq \leftrightarrow gq$
scattering as well as splitting and fusion reactions $gg
\leftrightarrow g$, $gg \leftrightarrow ggg$, $ggg \leftrightarrow
gggg$, or $g \leftrightarrow q \bar{q}$, etc. On the fermion side,
elastic fermion-fermion scattering $pp \leftrightarrow pp$, where
$p$ stands for a quark $q$ or antiquark $\bar{q}$, fermion-gluon
scattering $pg \leftrightarrow pg$, gluon bremsstrahlung $pp
\leftrightarrow pp+g$, or quark-antiquark fusion $q \bar{q}
\leftrightarrow g$, etc., emerge. Note, however, that the explicit
form of (\ref{widthg}) is derived for hard two-body scatterings
only. Furthermore, the widths $\gamma_{q(\bar
q)}(T)$ and $\gamma_{g}(T)$ provide only information on the total interaction rates and not on the individual differential cross sections. The computation of these cross sections will be carried out here in leading order on the basis of the propagators (\ref{propdqpm}) and coupling (\ref{coupling_DQPM}) and (\ref{coupling}) in Sec. \ref{Section3}, which in turn will allow us to recalculate the widths (\ref{widthg}) and (\ref{widthq}) and explore the validity of the quasiparticle limit in the timelike sector.

\subsection{Spectral functions}

In line with the propagator (\ref{propdqpm}), the parton spectral functions (or imaginary parts of the propagator $\rho = - 2 \Im G^R$) are no
longer $\delta$ functions in the invariant mass squared but given by
\begin{align}
\rho_{j}(\omega,{\bf p}) & = \frac{\gamma_{j}}{\tilde{E}_j}
\left(\frac{1}{(\omega-\tilde{E}_j)^2+\gamma^{2}_{j}}
-\frac{1}{(\omega+\tilde{E}_j)^2+\gamma^{2}_{j}}\right) \nonumber \\
& \equiv \frac{4\omega\gamma_j}{\left( \omega^2 - \mathbf{p}^2 - M^2_j \right)^2 + 4\gamma^2_j \omega^2}
\label{spectral_function}
\end{align}
~\\
separately for quarks, antiquarks, and gluons ($j = q,\bar q,g$).
Here, $\tilde{E}_{j}^2({\bf p})={\bf p}^2+M_{j}^{2}-\gamma_{j}^{2}$, where the
widths $\gamma_{j}$ and masses $M_{j}$ from the DQPM have been described above.
The spectral function (\ref{spectral_function}) is antisymmetric in $\omega$ and normalized as \cite{Pisarski:1989cs}

\begin{equation}
\int\limits_{-\infty}^{\infty}\frac{d\omega}{2\pi}\
 \omega \ \rho_{j}(\omega,{\bf p})=
\int\limits_{0}^{\infty}\frac{d\omega}{2\pi}\ 2\omega
\rho_{j}(\omega,{\bf p})=1\ ,
\end{equation} as mandatory for quantum field theory.

The actual quark mass $M_q$ and width $\gamma_q$ ---
employed as input in the PHSD calculations --- as well as the
gluon mass $M_g$ and width $\gamma_g$ are depicted in Fig. \ref{fig-DQPM-masses} as a function of $T/T_c$ and show an infrared enhancement close to $T_c$.
For $\mu_q=\mu_B/3 = 0$, the DQPM gives
\begin{equation} \label{qma}
M_q = \frac{2}{3} M_g, \hspace{1cm} \gamma_q = \frac{4}{9} \gamma_g
\ . \end{equation}

\subsection{Thermodynamics within the DQPM}

With the quasiparticle properties (or propagators) chosen as described above, one can
evaluate the entropy density $s(T,\mu_B)$, the pressure $P(T,\mu_B)$, and energy
density $\epsilon(T,\mu_B)$ in a straight forward manner by starting with
the entropy density and number density in the propagator representation from Baym \cite{Vanderheyden:1998ph,Blaizot:2000fc},

\begin{gather}
s^{dqp} = \label{sdqp} \\
\begin{align*}
& - \int \frac{d\omega}{2 \pi} \frac{d^3p}{(2 \pi)^3} \left[ d_g\ \frac{\partial n_B}{\partial T} \left( \Im(\ln -\Delta^{-1})+ \Im \Pi \Re \Delta \right) \right. \\
& + \sum_{q=u,d,s} d_q\ \frac{\partial n_F(\omega-\mu_q)}{\partial T} \left( \Im(\ln -S_q^{-1})+ \Im \Sigma_q \Re S_q \right) \\
&  + \sum_{\bar{q}={\bar{u}},{\bar{d}},{\bar{s}}} \left. d_{\bar{q}}\ \frac{\partial n_{F}(\omega+\mu_q)}{\partial T} \left( \Im(\ln -S_{\bar{q}}^{-1})+ \Im \Sigma_{\bar{q}} \Re S_{\bar{q}} \right) \right]
\end{align*}
\end{gather}

\begin{gather}
 n^{dqp} = - \int \frac{d\omega}{2 \pi} \frac{d^3p}{(2 \pi)^3} \label{nbdqp} \\
\begin{align*}
&  \left[ \sum_{q=u,d,s} d_q\ \frac{\partial n_F(\omega-\mu_q)}{\partial \mu_q} \left( \Im(\ln -S_q^{-1})+ \Im \Sigma_q \Re S_q \right) \right. \\
& \left. +  \sum_{\bar{q}={\bar{u}},{\bar{d}},{\bar{s}}} d_{\bar{q}}\  \frac{\partial n_{F}(\omega+\mu_q)}{\partial \mu_q} \left( \Im(\ln -S_{\bar{q}}^{-1})+ \Im \Sigma_{\bar{q}} \Re S_{\bar{q}} \right)  \right]
\end{align*}
\end{gather}
~\\
where $n_B(\omega) = [\exp(\omega/T)-1]^{-1}$ and
$n_F(\omega-\mu_q) = \{\exp[(\omega-\mu_q)/T]+1\}^{-1}$ denote the
Bose-Einstein and Fermi-Dirac distribution functions, respectively, while $\Delta
=(p^2-\Pi)^{-1}$, $S_q = (p^2-\Sigma_q)^{-1}$, and $S_{\bar q} =
(p^2-\Sigma_{\bar q})^{-1}$ stand for the full (scalar)
quasiparticle propagators of gluons $g$, quarks $q$, and antiquarks
${\bar q}$.  In Eqs. (\ref{sdqp}) and (\ref{nbdqp}), $\Pi$ and $\Sigma = \Sigma_q
\approx \Sigma_{\bar q}$ denote the (retarded) quasiparticle
self-energies. Furthermore, the number of transverse gluonic degrees of freedom is $d_g=2 \times (N_c^2-1)$
while for the fermion degrees of freedom we use $d_q= 2 \times N_c$ and $d_{\bar{q}}= 2 \times N_c$.

In principle, $\Pi$ as well as $\Delta$ are Lorentz
tensors and should be evaluated in a nonperturbative framework. The
DQPM treats these degrees of freedom as independent scalar fields (for each color and spin projection)
with scalar self-energies  which are assumed to be identical for
quarks and antiquarks. This is expected to hold well for the entropy and number density.
Note that one has to treat quarks and
antiquarks separately in Eqs. (\ref{sdqp}) and (\ref{nbdqp}) as their abundance differs
at finite quark chemical potential $\mu_q=\mu_B/3$. \\

With the choice (\ref{spectral_function}), the complex self-energies  $\Pi =
M_g^2-2i \omega \gamma_g$ and $\Sigma_{q} = M_{q}^2 - 2 i \omega \gamma_{q}$ are fully defined via (\ref{Mg9}),
(\ref{Mq9}), (\ref{widthg}), and (\ref{widthq}). Note that the retarded propagator
(\ref{propdqpm}) resembles the propagator of a damped
harmonic oscillator (with an additional ${\bf p}^2$)  and preserves
microcausality also for $\gamma > M$ \cite{Rauber:2014mca}, i.e., in
case of overdamped motion. Although the ansatz for the parton
propagators is not QCD it has been shown that a variety of QCD,
observables on the lattice are compatible with this choice \cite{Linnyk:2015rco}. \\

\begin{figure}[h!]
	\centering
	\includegraphics[width=\columnwidth]{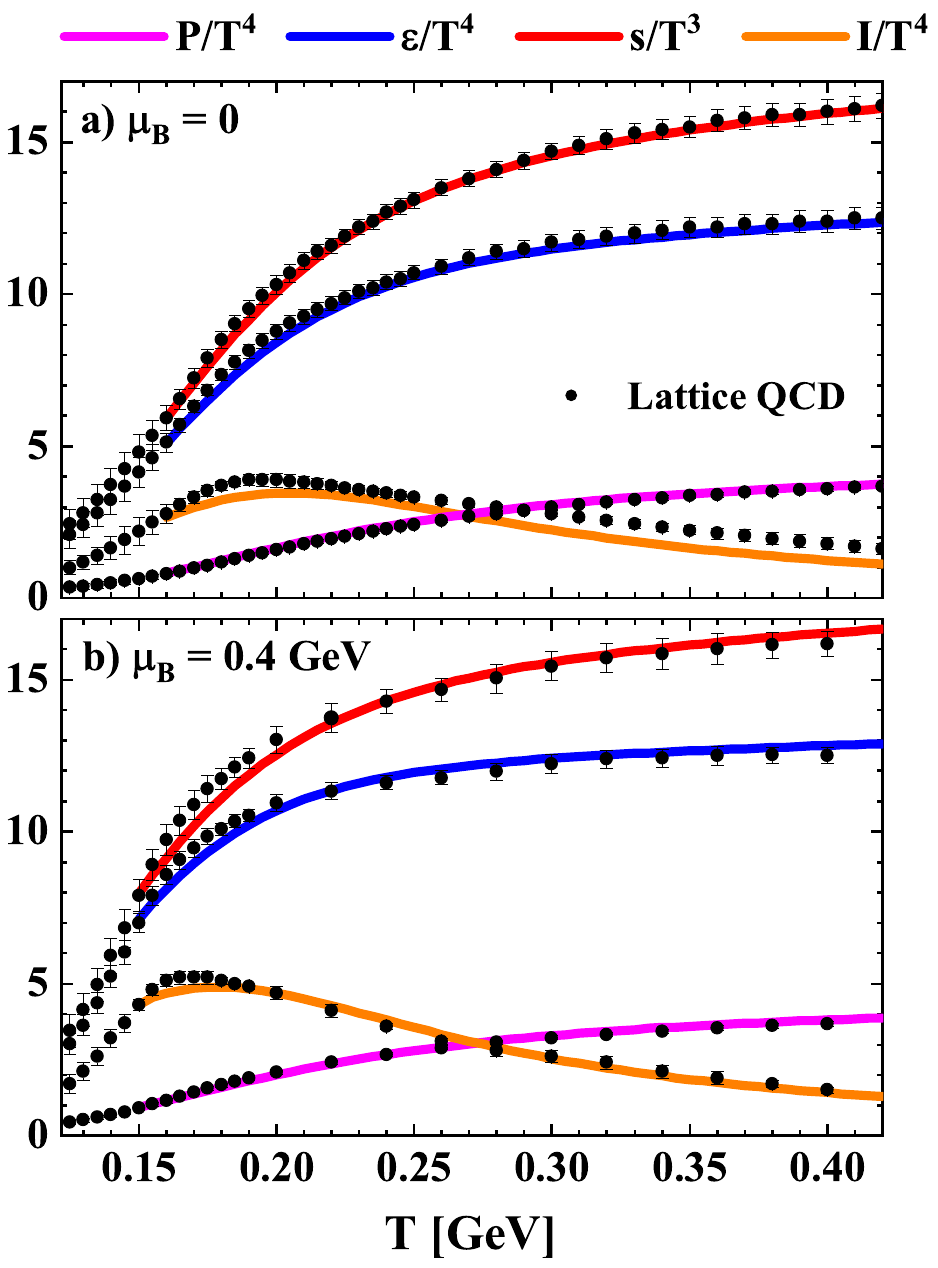}
	\caption{(Color online)
		The scaled pressure $P(T)/T^4$ (pink line), entropy density $s(T)/T^3$ (red line), scaled energy
		density $\epsilon(T)/T^4$ (blue line), and interaction measure (\ref{wint}) (orange line), from the DQPM in
        comparison to
		the lQCD results from Refs. \cite{Borsanyi:2012cr,Borsanyi:2013bia} (full dots) for $\mu_B$ = 0 (a) and $\mu_B$ = 400 MeV (b).}
	\label{fig-DQPM-thermo}
\end{figure}

In case the real and imaginary parts of the propagators
$\Delta$ and $S$ are fixed, the entropy density
(\ref{sdqp}) and number density (\ref{nbdqp}) can be evaluated numerically. As we deal with a
grand-canonical ensemble, the Maxwell relations give

\begin{equation}
 s =\frac{\partial P}{\partial T}\ \ ,\ \ n_B = \frac{\partial P}{\partial \mu_B} \ ,
\end{equation}
~\\
such that the pressure can be obtained by integration
of the entropy density $s$ over $T$ and of the baryon density $n_B$ over $\mu_B$ as

\begin{align}
P(T,\mu_B) =\ & P(T_0,0) + \int_{T_{0}}^{T} s(T',0)\ dT' \nonumber \\
& + \int_{0}^{\mu_B} n_B(T,\mu_B')\ d\mu_B' \ ,
\label{pressure}
\end{align}
~\\
where one identifies the full entropy density $s$ and baryon density $n_B$ with the quasiparticle entropy
density $s^{dqp}$ (\ref{sdqp}) and baryon density $n_B = n^{dqp}/3$ (\ref{nbdqp}). The starting point $T_{0}$ for the
integration in $T$ is chosen between 0.1 $< T < $ 0.15 GeV where
the entropy density is taken in accordance to the lattice QCD results from Ref. \cite{Borsanyi:2013bia}
in the hadronic sector.

The energy density $\epsilon$ then follows from the
thermodynamical relation
\begin{equation}
\label{eps} \epsilon = T s - P +\mu_B n_B
\end{equation}
~\\ and thus is also fixed by the entropy $s(T,\mu_B)$ and baryon density $n_B(T,\mu_B)$ as well as the
interaction measure

\begin{equation}
\label{wint} I: = \epsilon - 3P = Ts - 4P + \mu_B n_B
\end{equation}
~\\
that vanishes for massless and noninteracting degrees of freedom at $\mu_B = 0$. \\

A direct comparison of the resulting entropy density $s(T)$ (\ref{sdqp}), pressure $P(T)$ (\ref{pressure}),  energy
density $\epsilon(T)$ (\ref{eps}), and interaction measure (\ref{wint}) from the DQPM with lQCD results from the BMW
group \cite{Borsanyi:2012cr,Borsanyi:2013bia} at $\mu_B = 0$ (a) and $\mu_B$ = 400 MeV (b) is presented in Fig. \ref{fig-DQPM-thermo}. The dimensionless results $s/T^3$, $P/T^4$, and $\epsilon/T^4$ are shown to demonstrate the
scaling with temperature. The agreement is sufficiently good for the entropy and energy density as well as for the pressure. A satisfactory agreement also
holds for the dimensionless interaction measure, i.e., $(\epsilon - 3 P)/T^4$ (cf. orange line in Fig. \ref{fig-DQPM-thermo}).

\section{Differential cross sections for partonic interactions}
\label{Section3}

\subsection{Definitions}

\subsubsection{On-shell case}

The differential cross section for a $ 2 \rightarrow 2$ process of on-shell particles ($ 1 + 2 \rightarrow 3 + 4$) is given by:

\begin{align}
d\sigma^{\text{on}} = & \frac{d^3p_3}{(2\pi)^3 2E_3} \frac{d^3p_4}{(2\pi)^3 2E_4} \label{dsigma_on}   \\
& \times (2\pi)^4 \delta^{(4)}\left( p_1 + p_2 -p_3 -p_4 \right) \frac{|\bar{\mathcal{M}}|^2}{F} , \nonumber
\end{align}
~\\
where the flux is defined by $F = v_{\text{rel}}\ 2E_1\ 2E_2 $ with the definition $v_{\text{rel}} = |\vec{v}_1-\vec{v}_2|$, and the on-shell energies for the particle are defined as $E_j = \sqrt{\mathbf{p}_j^2+M_j^2}$. $|\bar{\mathcal{M}}|^2$ denotes the matrix element squared averaged over the color and spin of the incoming particles and summed over those of the final particles.  We want to evaluate the cross section in the rest frame of the heat bath where the Fermi-Dirac or Bose-Einstein functions describe the particle distributions. The only factor in Eq. (\ref{dsigma_on}) which is not Lorentz invariant is the flux factor $F$, while the other factors, the Lorentz invariant phase space (LIPS), the matrix element $|\bar{\mathcal{M}}|^2$, and the $\delta$ function for energy-momentum conservation, are invariant. This implies that the cross section can be calculated in any frame, but the flux factor has to be correctly taken into account according to the actual frame of interest.

The cross section is usually evaluated in the center-of-mass (c.m.) frame of the collision for simplicity. In this case, the momenta of the colliding particles obey $\mathbf{p}_1 + \mathbf{p}_2 = \mathbf{p}_3 + \mathbf{p}_4 = \mathbf{p} = \vec{0}$, and the notation $|\mathbf{p_1}| = |\mathbf{p_2}| = p_i$ and $|\mathbf{p_3}| = |\mathbf{p_4}| = p_f$ is used. The flux factor becomes $F^{\text{c.m.}} = 4 p_i \sqrt{s}$ and, after simplification, Eq. (\ref{dsigma_on}) reads

\begin{equation}
d\sigma^{\text{c.m.}}_{\text{on}} = \frac{p_f\ d \Omega}{16 \pi^2 \sqrt{s}} \frac{|\bar{\mathcal{M}}|^2}{F^{\text{c.m.}}} = \frac{d \Omega}{64 \pi^2 s} \frac{p_f}{p_i} |\bar{\mathcal{M}}|^2 ,
\label{dsigma_on_CM}
\end{equation}
~\\
where $s$ in the Mandelstam variable and $d\Omega$ is the differential solid angle corresponding to one of the final particle. The momenta of the initial ($i$) and final particles ($f$) in the c.m. frame is found to be:

\begin{equation}
p_{i,f} = \frac{\sqrt{\left(s-(M_{i,f}+M'_{i,f})^2\right)\left(s-(M_{i,f}-M'_{i,f})^2\right)}}{2\sqrt{s}} ,
\end{equation}
~\\
with $M_{i,f}$ and $M'_{i,f}$ being the masses of the colliding partons. The total cross section is obtained by performing the integral in Eq. (\ref{dsigma_on_CM}) over $d\Omega$ as

\begin{equation}
\sigma^{\text{c.m.}}_{\text{on}} = \frac{1}{32 \pi s} \frac{p_f}{p_i} \int_{-1}^{1} d \cos(\theta)  |\bar{\mathcal{M}}|^2 ,
\label{sigma_on_CM}
\end{equation}
~\\
where $\theta$ is the final polar angle of one of the final particle in the c.m. frame. In the c.m. frame, the collision is independent from the azimuthal angle $\phi$ and the corresponding integration gives a factor $2\pi$.

\subsubsection{Off-shell case}

In the off-shell case, the energy of the partons, as well their momenta, are independent degrees of freedom and a general definition of an ``off-shell cross section'' is not possible due to the lack of asymptotically stable states. However, transition matrix elements for different incoming and outgoing 4-momenta can be well defined also off shell. By transforming the Lorentz-invariant phase space in Eq. (\ref{dsigma_on}), one can include the off-shell effects for the scattering of timelike particles --- in the case of a well defined incoming flux $F = v_{\text{rel}}\ 2\omega_1\ 2\omega_2 $ --- by integrating over the energy of the final timelike particles as

\begin{align}
Fd\sigma^{\text{off}} =\ & \frac{d^4p_3}{(2\pi)^4} \ \frac{d^4p_4}{(2\pi)^4}\  {\tilde \rho}_3(\omega_3,\mathbf{p}_3)\ \theta(\omega_3)\ {\tilde \rho}_4(\omega_4,\mathbf{p}_4)\ \theta(\omega_4) \nonumber \\
& \times (2\pi)^4 \delta^{(4)}\left( p_1 + p_2 -p_3 -p_4 \right) |\bar{\mathcal{M}}|^2 \label{dsigma_off}
\end{align}
~\\
with the renormalized timelike spectral functions
\begin{equation}
\label{renorm}
\tilde{\rho}_j(\omega_j,\mathbf{p}_j) = \frac{\rho(\omega_j,\mathbf{p}_j)\ \theta(p_j^2)}{\int_{0}^\infty \frac{d\omega_j}{(2 \pi)} \ 2 \omega_j\ \rho(\omega_j,\mathbf{p}_j)\ \theta(p_j^2)} ,
\end{equation}
where the spectral function $\rho_i$ in (\ref{renorm}) --- corresponding to the parton type $i$ --- is taken from Eq. (\ref{spectral_function}). The final parton masses are defined as $m_i^2 = p^2_i = \omega_i^2 - \mathbf{p}^2_i$, where $p_i$ is the 4-momentum of particle $i$. One can verify that by replacing the spectral functions by their on-shell value:

\begin{align}
\lim\limits_{\gamma_j \rightarrow 0} &\ \rho_j(\omega,\mathbf{p}) = 2\pi\ \delta(\omega^2 - \mathbf{p}^2 - M_j^2) \\
& = \frac{\pi}{\omega}\ \left[ \delta(\omega - \sqrt{\mathbf{p}^2 + M_j^2}) + \delta(\omega + \sqrt{\mathbf{p}^2 + M_j^2})\right] \nonumber,
\end{align}
~\\
the off-shell cross section leads to the on-shell one as defined in the previous subsection from Eq. (\ref{dsigma_on}). \\

We follow the same strategy as in the previous subsection and evaluate the differential ``off-shell cross section'' for timelike quanta  in the center-of-mass system of the collision for convenience. By making use of the $\delta$ function in Eq. (\ref{dsigma_off}), one can integrate over $d^4p_4$ to obtain the total cross section in the c.m. frame by performing the integrations with the appropriate boundaries as:

\begin{align}
\label{sigma_off_CM}
F^{\text{c.m.}}\sigma^{\text{c.m.}}_{\text{off}} & = \frac{1}{(2\pi)^3} \int_0^{\sqrt{s}/2} p^2_f\ dp_f\ d\cos(\theta) \\
& \int_{p_f}^{\sqrt{s}-p_f} d\omega_3^{\text{c.m.}}    {\tilde \rho}_3(\omega_3,\mathbf{p}_3)\  {\tilde \rho}_4(\omega_4,\mathbf{p}_4)\  |\bar{\mathcal{M}}|^2  \nonumber
\end{align}
~\\ for $F^{c.m.}= 4 p^{\text{c.m.}}_i \sqrt{s}$.
Bear in mind that even if the calculation of the cross section is performed in the center-of-mass system, the energies and momenta entering the spectral functions (\ref{spectral_function}) should be expressed in the heat bath frame by applying the appropriate Lorentz transformations.

We mention that one can simplify the off-shell energy integration by an integration over the final masses of the partons in the non-relativistic limit. The off-shell cross section from Eq. (\ref{dsigma_off}) then becomes \cite{Berrehrah:2013mua}

\begin{equation}
\sigma^{\text{BW}}_{\text{off}} = \int_{0}^{\sqrt{s}} dm_3 \int_{0}^{\sqrt{s}-m_3} dm_4\ \rho^{\text{BW}}(m_3)\ \rho^{\text{BW}}(m_4) \int d\sigma^{\text{c.m.}}_{\text{on}} ,
\end{equation}
~\\
where the Breit-Wigner spectral function $\rho^{\text{BW}}$ in Ref. \cite{Berrehrah:2013mua} is obtained from Eq. (\ref{spectral_function}) in the limit $\omega \rightarrow m$:

\begin{equation}
\rho^{\text{BW}}_i(m) = \frac{2}{\pi} \frac{2m^2 \gamma_i}{(m^2-M_i^2)^2+(2m \gamma_i)^2}.
\label{Breit-Wigner}
\end{equation}
~\\
This distribution fulfills the normalization $\int_{0}^{\infty} dm\ \rho^{\text{BW}}(m) = 1$.

\subsection{Partonic scattering}

In the framework of the DQPM, quarks and gluons are massive with a
finite lifetime associated to their interaction width. In order to
calculate the matrix elements corresponding to a scattering of DQPM
partons, the scalar propagator (\ref{propdqpm}) has to be replaced by
the following propagators --- with full Lorentz
structure --- to describe a massive vector gluon and massive (spin-1/2) fermion
with a finite width \cite{Berrehrah:2013mua}:

\begin{equation}
\includegraphics[width=0.3\columnwidth]{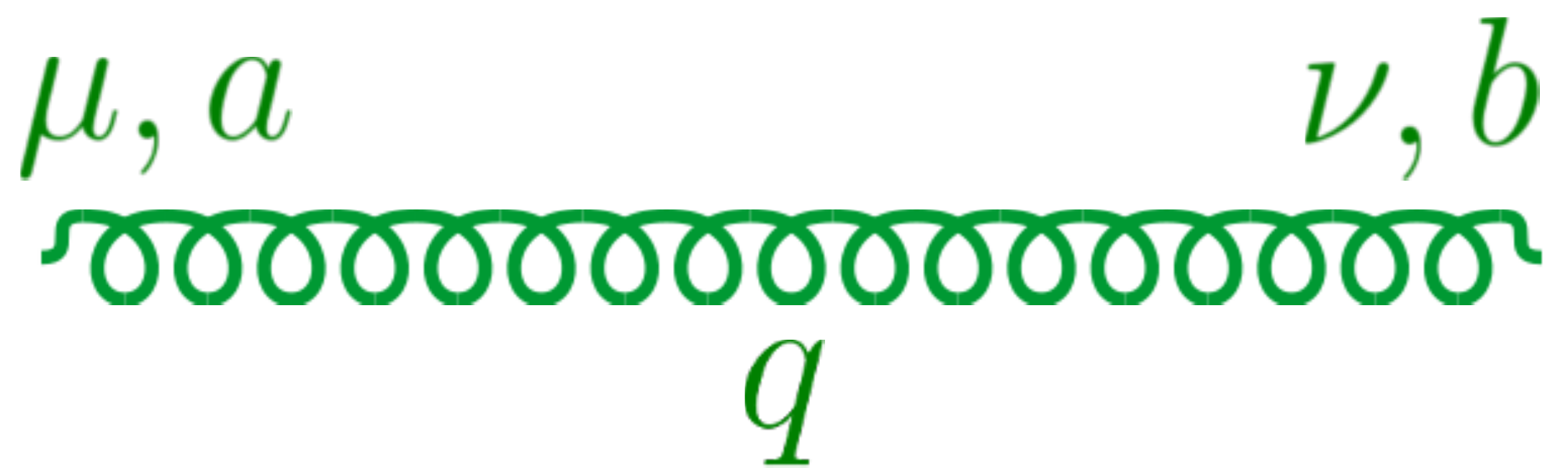} = -i \delta_{ab} \frac{g^{\mu \nu} - q^\mu q^\nu / M^2_g}{q^2 - M^2_g + 2i \gamma_g q_0} ,
\label{propg}
\end{equation}

\begin{equation}
\includegraphics[width=0.3\columnwidth]{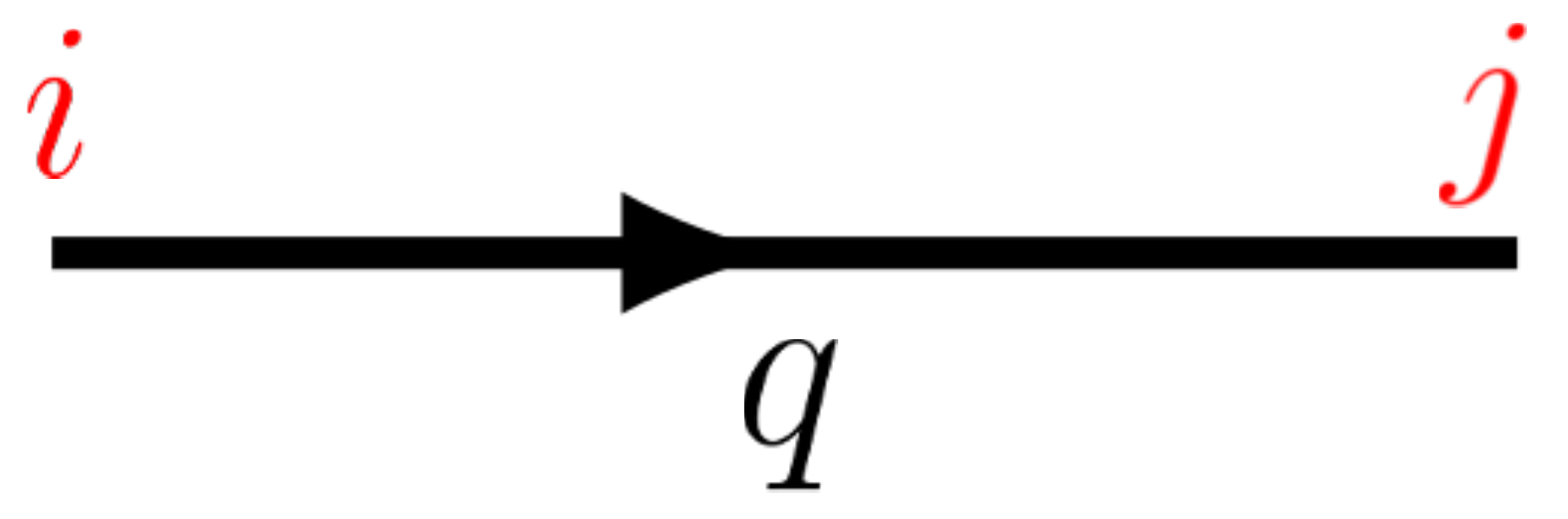} = i \delta_{ij} \frac{\slashed{q} + M_q}{q^2 - M^2_q + 2i \gamma_q q_0} ,
\label{propq}
\end{equation}
~\\
where $q$ is the 4-momentum of the exchanged particle. The $\delta$ functions ensure that the exchanged quark or gluon is connected with other parts of the diagram with the same color ($a,b$ for the gluon and $i,j$ for the quark). The invariant matrix element (squared) $|\bar{\mathcal{M}}|^2$, entering the differential cross section in Eqs. (\ref{dsigma_on})--(\ref{dsigma_off}), is calculated in leading order and is averaged over initial --- and summed over final --- spin and colors. In the following, we employ a degeneracy factor for spin and color of $d_q = 2 \times N_c = 6$ for quarks and $d_g = 2 \times (N_c^2-1) = 16$ for gluons in consistency with Eqs. (\ref{sdqp}) and (\ref{nbdqp}). Each matrix element can be decomposed into several channels known as $t-$, $u-$, and $s-$channels for quark-quark ($qq'$) and quark-gluon ($qg$) scatterings, as well as a four-point interaction for the case of gluon-gluon ($gg$) scattering. For details, we refer the reader to the Appendixes \ref{AppendixA}--\ref{AppendixC} and continue with the actual results.

\begin{widetext}

\begin{figure*}[h!]
	\centering
	\includegraphics[width=0.8\linewidth]{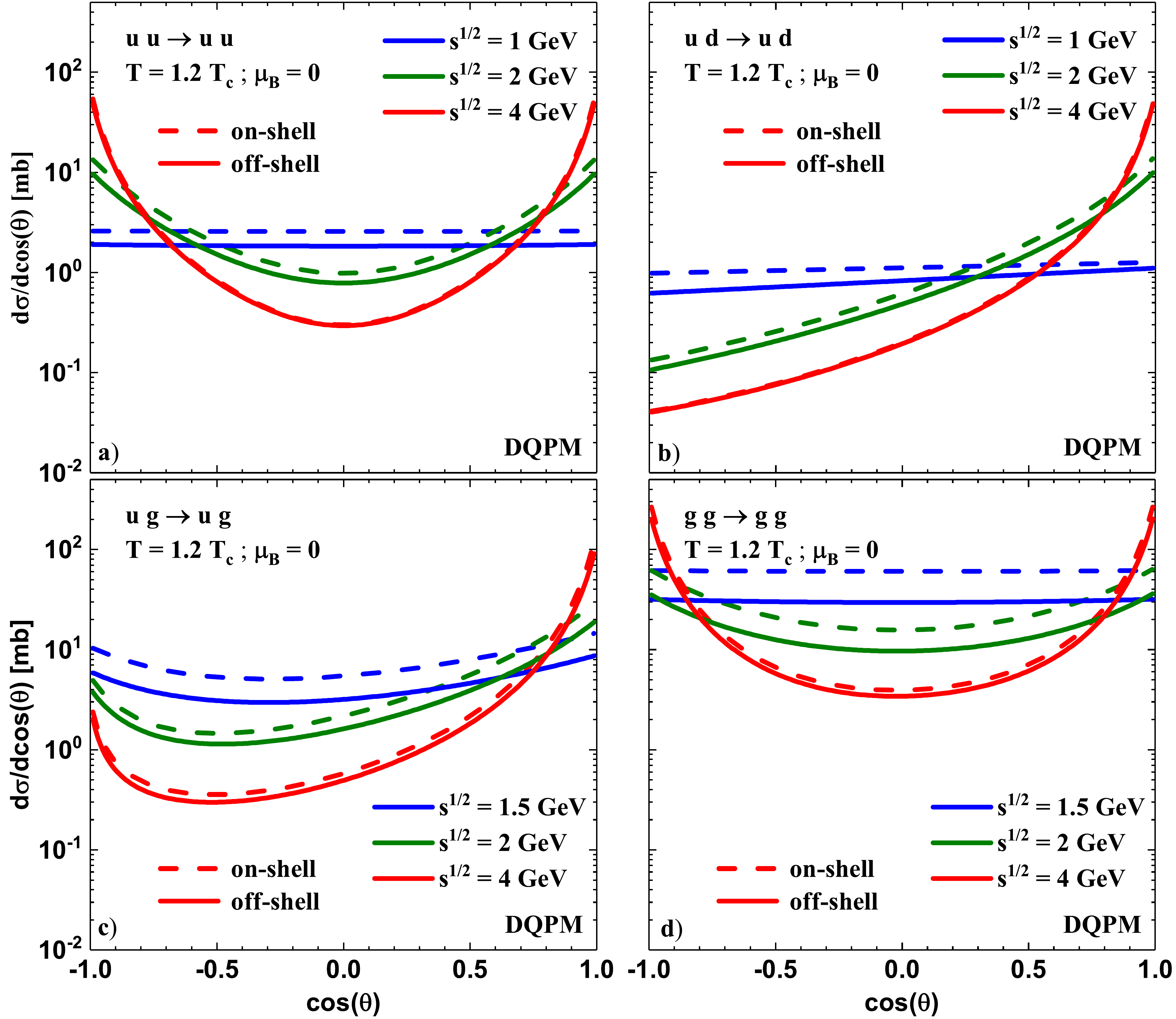}
	\caption{(Color online) Elastic differential cross sections between different partons for the on-shell case (dashed lines) from Eq. (\ref{dsigma_on_CM}) and the off-shell case (solid lines) evaluated in the center of mass of the collision system as a function of the angle $\cos(\theta)$ between the initial and final momenta of one of the partons for $T=1.2 T_c$ and $\mu_B$ = 0. The initial masses of the colliding partons are taken as the pole masses from Eqs. (\ref{Mg9}) and (\ref{Mq9}). The different lines correspond to different  collision energies $\sqrt{s}$ from 1 to 4 GeV (see legend).}
	\label{fig_dCS}
\end{figure*}

\begin{figure*}[h!]
	\centering
	\includegraphics[width=\linewidth]{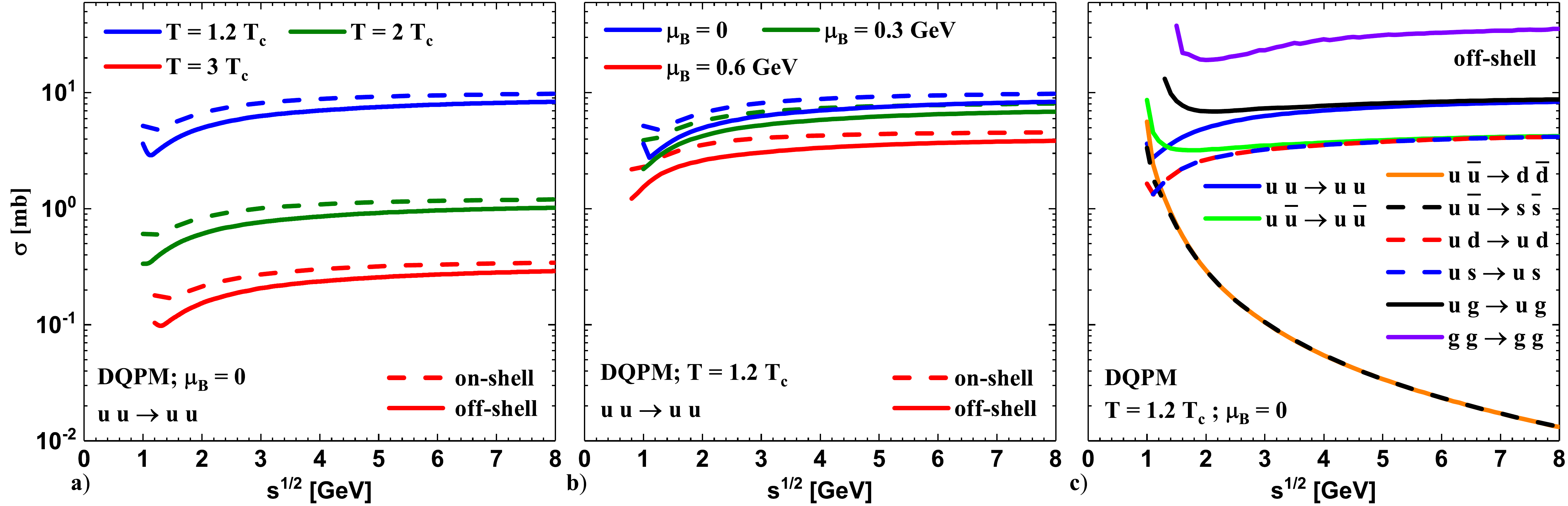}
	\caption{(Color online) Cross sections between different partons for the on-shell case (dashed lines) from Eq. (\ref{sigma_on_CM}) and the off-shell case (solid lines) from Eq. (\ref{sigma_off_CM}) evaluated in the center of mass of the collision system as a function of the collision energy $\sqrt{s}$ (see text for a detailed description). The initial masses of the colliding partons are taken as the pole masses from Eqs. (\ref{Mg9}) and (\ref{Mq9}).}
	\label{fig_CS}
\end{figure*}

\end{widetext}

In Fig. \ref{fig_dCS}, we show the differential cross sections between different partons for the on-shell case (dashed lines) from Eq. (\ref{dsigma_on_CM}) and the off-shell case (solid lines) evaluated in the center-of-mass of the collision system as a function of the collision energy $\sqrt{s}$. In these examples, the initial masses of the colliding partons are taken as the pole masses from Eqs. (\ref{Mg9}) and (\ref{Mq9}) to allow for a comparison between on-shell and off-shell scattering, i.e., for the same initial states. In all cases [$uu \rightarrow uu$ (a), $ud \rightarrow ud$ (b), $ug \rightarrow ug$ (c), $gg \rightarrow gg$ (d)], the cross sections are almost isotropic at the threshold energy $\sqrt{s} \approx$ 1 GeV and increase in anisotropy with increasing $\sqrt{s}$. Furthermore, the on-shell case (dashed lines) gives slightly larger cross sections than the off-shell case (solid lines), which will be discussed below.

Figure \ref{fig_CS} (a) displays the angle integrated cross section for $uu \rightarrow uu$ scattering as a function of $\sqrt{s}$ for temperatures of $T=1.2 T_c$ (blue), $T=2 T_c$ (green), and $T=3 T_c$ (red) for the on-shell (dashed lines) and off-shell (solid lines) cases at $\mu_B$ = 0. Again, the initial masses of the colliding partons are taken as the pole masses from Eqs. (\ref{Mg9}) and (\ref{Mq9}) to obtain the same initial flux $F$. At all temperatures $T$, the cross section does not change very much with collision energy $\sqrt{s}$ and the difference between the on-shell and off-shell case decreases with increasing $\sqrt{s}$.
The explicit dependence of the  $uu \rightarrow uu$ cross section on the chemical potential is shown in Fig. \ref{fig_CS} (b) for $\mu_B$ = 0 (blue),
$\mu_B$ = 0.3 GeV (green), and $\mu_B$ = 0.6 GeV (red) at $T=1.2 T_c$ for the on-shell (dashed lines) and off-shell (solid lines) cases. While the dependencies on $\sqrt{s}$ are similar, we find a decrease of the cross section with increasing chemical potential $\mu_B$ which can be traced back to a decreasing coupling with $\mu_B$ at
fixed temperature $T$ (see Fig. \ref{fig-DQPM-alphas}). Figure \ref{fig_CS} (c), furthermore, shows the dependence of all cross sections calculated on the collision energy $\sqrt{s}$ in the off-shell case for $T=1.2 T_c$ and $\mu_B$ = 0. While most of the channels do not change drastically with  $\sqrt{s}$ --- except for thresholds --- the flavor-changing processes $u {\bar u} \rightarrow d {\bar d}$ and  $u {\bar u} \rightarrow s {\bar s}$ drop quickly with increasing energy.

With all differential partonic cross sections fixed as a function of $T$ and $\mu_B$ (above the phase boundary), we can now continue with transport properties of the hot QGP as a function of  $T$ and $\mu_B$ employing the partonic energy-momentum distributions from the DQPM.

\section{Collisional widths of the hot and dense QGP}
\label{Section4}

\subsection{On-shell case}

In the on-shell case, all energies of the particles are taken to be $E^2 = \mathbf{p}^2 + M^2$, where $M$ is the pole mass. The on-shell interaction rate for the corresponding parton is given by \cite{Braaten:1991jj,Thoma:1993vs,Chakraborty:2010fr}

\begin{align}
\Gamma^{\text{on}}_i & (\mathbf{p}_i, T,\mu_q) = \frac{1}{2E_i} \sum_{j=q,\bar{q},g} \int \frac{d^3p_j}{(2\pi)^3 2E_j}\ d_j\ f_j(E_j,T,\mu_q)  \nonumber \\
& \ \ \ \ \ \ \ \ \ \ \ \ \times  \int \frac{d^3p_3}{(2\pi)^3 2E_3}  \int \frac{d^3p_4}{(2\pi)^3 2E_4} (1\pm f_3) (1\pm f_4) \nonumber \\
& \ \ \ \ \ \ \ \times |\bar{\mathcal{M}}|^2 (p_i,p_j,p_3,p_4)\ (2\pi)^4 \delta^{(4)}\left(p_i + p_j -p_3 -p_4 \right) \nonumber \\
= & \sum_{j=q,\bar{q},g} \int \frac{d^3p_j}{(2\pi)^3}\ d_j\ f_j\ v_{\text{rel}} \int d\sigma^{\text{on}}_{ij \rightarrow 34}\ (1\pm f_3) (1\pm f_4) ,
\label{Gamma_on}
\end{align}
~\\
where $d_j$ is the degeneracy factor for spin and color [for quarks $d_q = 2 \times N_c$ and for gluons $d_g =2 \times (N_c^2-1)$], and with the shorthand notation $f_j = f_j(E_j,T,\mu_q)$ for the distribution functions. In Eq. (\ref{Gamma_on}) and throughout this section, the notation $\sum_{j=q,\bar{q},g}$ includes the contribution from all possible partons, which in our case are the gluons and the (anti-)quarks of three different flavors ($u,d,s$). The Pauli-blocking ($-$) and Bose-enhancement (+) factors account for the available density of final states. Note that here all quantities have to be expressed in the rest frame of the heat bath, implying that the on-shell cross section $d\sigma^{\text{on}}$ from Eq. (\ref{dsigma_on_CM}) has to be modified according to the different fluxes:

\begin{equation}
F^{\text{HB}}\sigma^{\text{HB}} = \sigma^{\text{c.m.}} F^{\text{c.m.}} ,
\label{Sig_HB_CM}
\end{equation}
~\\
where the quantities denoted by HB are expressed in the rest frame of the heat bath and c.m. in the center-of-mass frame of the collision.

To evaluate the average width of the partons $i$, we finally have to average its interaction rate (\ref{Gamma_on}) over its momentum distribution,
\begin{align}
\Gamma^{\text{on}}_i(T,\mu_q) & = \frac{d_i}{n_i^{\text{on}}(T,\mu_q)} \int \frac{d^3p_i}{(2\pi)^3}\ f_i(E_i,T,\mu_q)  \nonumber \\ 
& \times \Gamma^{\text{on}}_i(\mathbf{p}_i,T,\mu_q)
\label{Gamma_on_avT}
\end{align}
~\\
with the on-shell density of partons $i$ at  $T$ and $\mu_q$ given by
\begin{equation}
n_i^{\text{on}}(T,\mu_q) = d_i \int \frac{d^3p_i}{(2\pi)^3}\  f_i(E_i,T,\mu_q) .
\label{n_on}
\end{equation}
~\\

\subsection{Off-shell case}

In order to obtain the width for the off-shell DQPM timelike partons, we have to calculate the interaction rate for the corresponding parton $i$ with momentum $\mathbf{p}_i$ due to collisions with  timelike particles $j$ leading to final timelike particles 3 and 4 by integrating additionally over all energies $\omega_j$ in the timelike sector:

\begin{align}
& \Gamma^{\text{off}}_i (\mathbf{p}_i, T,\mu_q) = \int_{0}^\infty \frac{d\omega_i}{(2 \pi)} \ \tilde{\rho}_i  \sum_{j=q,\bar{q},g} \int \frac{d^4p_j}{(2\pi)^4}\ \theta(\omega_j)\ d_j\ \tilde {\rho}_j\  f_j \nonumber \\
& \times \int \frac{d^4p_3}{(2\pi)^4} \ \theta(\omega_3) \ {\tilde \rho}_3 \int \frac{d^4p_4}{(2\pi)^4} \ \theta(\omega_4) \ \tilde{\rho}_4 (1\pm f_3) (1\pm f_4) \nonumber \\
& \ \ \ \times |\bar{\mathcal{M}}|^2 (p_i,p_j,p_3,p_4)\ (2\pi)^4 \delta^{(4)}\left(p_i + p_j -p_3 -p_4 \right) ,
\label{Gamma_off}
%
\end{align}
where the shorthand notation (\ref{renorm}) for the renormalized timelike spectral functions $\tilde{\rho}_j(\omega_j,\mathbf{p}_j)$ has been used and $f_j = f_j(\omega_j,T,\mu_q)$ for the distribution functions.
We mention that the limit (\ref{Gamma_off}) discards damping processes between the timelike and spacelike sector which are assumed to be subleading.
To evaluate the average timelike width of the partons $i$, we finally have to average its interaction rate as
\begin{align}
& \Gamma^{\text{off}}_i (T,\mu_q) =\frac{d_i}{n_i^{\text{off}}(T,\mu_q)} \int \frac{d^4p_i}{(2\pi)^4}\ \theta(\omega_i)\ \tilde {\rho}_i\  f_i(\omega_i,T,\mu_q) \nonumber  \\
& \times \sum_{j=q,\bar{q},g} \int \frac{d^4p_j}{(2\pi)^4}\ \theta(\omega_j)\ d_j\ \tilde {\rho}_j\  f_j \label{Gamma_off_avT} \\
 &  \times\int \frac{d^4p_3}{(2\pi)^4} \ \theta(\omega_3) \ {\tilde \rho}_3 \int \frac{d^4p_4}{(2\pi)^4} \ \theta(\omega_4) \ \tilde{\rho}_4 (1\pm f_3) (1\pm f_4) \nonumber \\
 & \ \ \  \times |\bar{\mathcal{M}}|^2 (p_i,p_j,p_3,p_4)\ (2\pi)^4 \delta^{(4)}\left(p_i + p_j -p_3 -p_4 \right) \nonumber ,
\end{align}
~\\
with the off-shell density of timelike partons $i$ given by
\begin{equation}
n_i^{\text{off}}(T,\mu_q) = d_i \int \frac{d^4p_i}{(2\pi)^4} \ \theta(\omega_i) \ 2\omega_i\ \tilde{\rho}_i\ f_i(T,\mu_q) .
\label{n_off}
\end{equation}

\begin{widetext}
	
	\begin{figure*}[h!]
		\centering
		\begin{tabular}{cc}
			\includegraphics[width=0.4\linewidth]{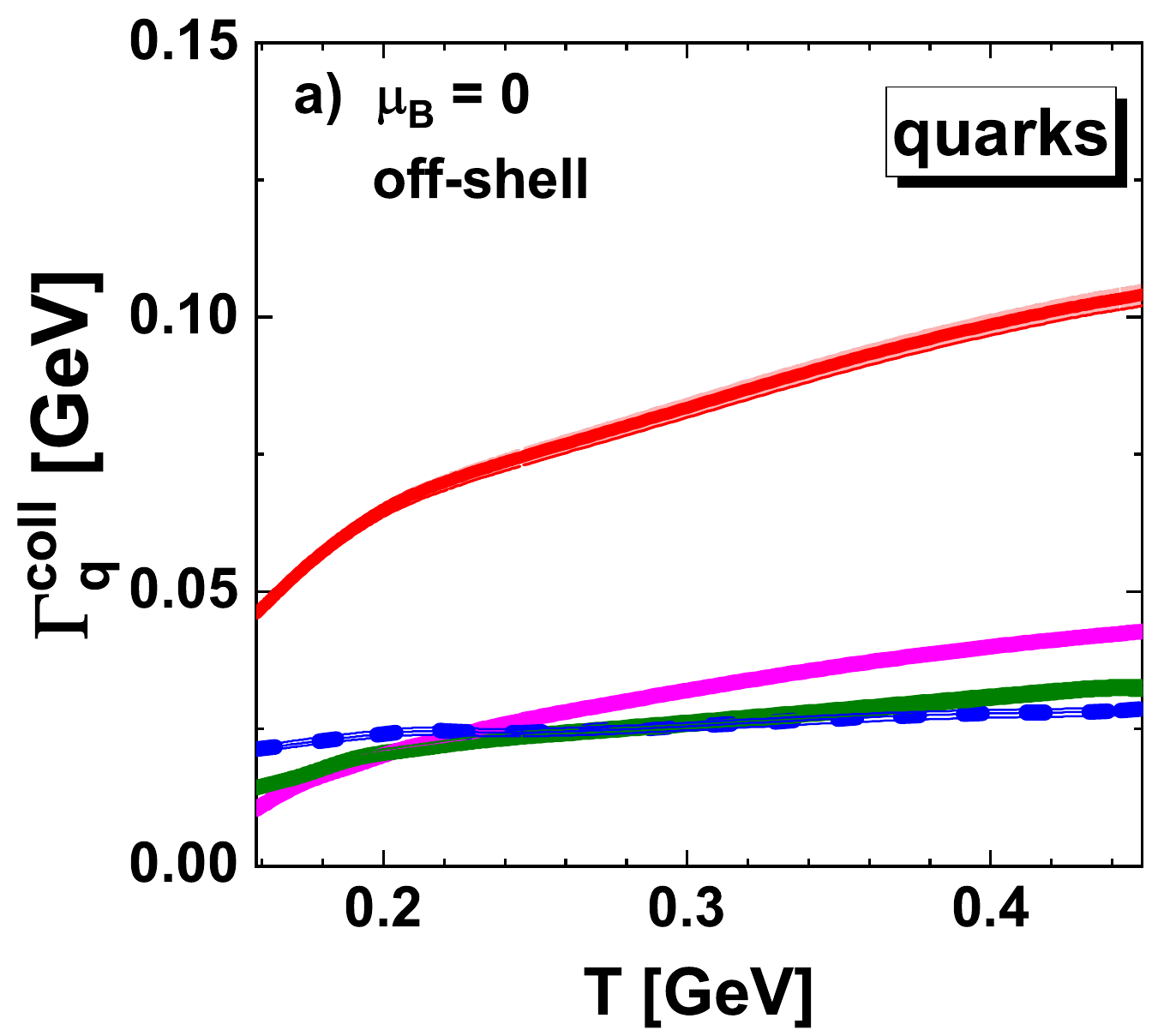}
			\includegraphics[width=0.4\linewidth]{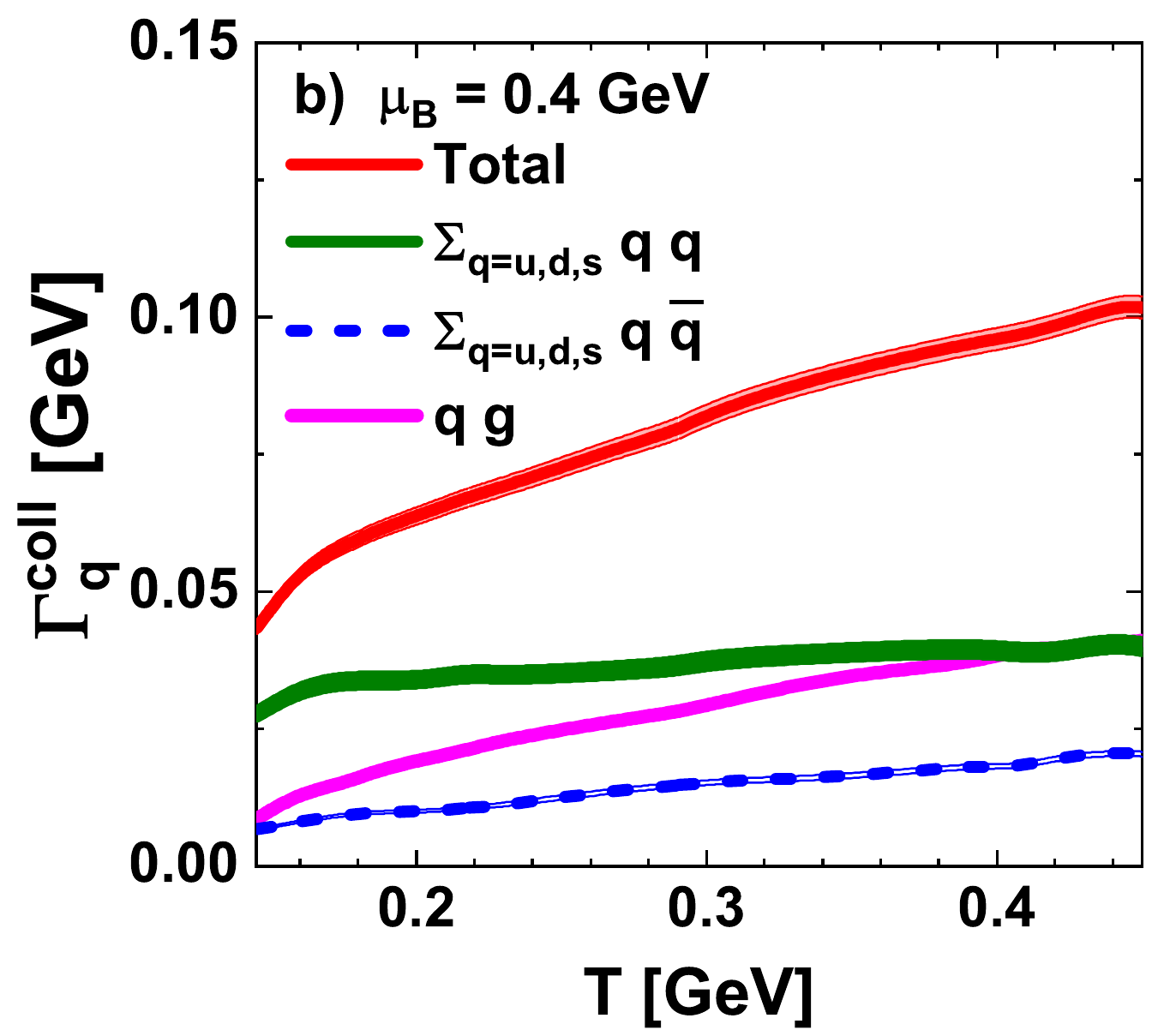}
		\end{tabular}
		\caption{(Color online) Off-shell collision rate from Eq. (\ref{Gamma_off_avT}) of a light quark $q$ as a function of the temperature $T$ for $\mu_B = 0$ (a) and $\mu_B = 0.4$ GeV (b) (blue lines).  The contributions from the scattering with light quarks (green), antiquarks (blue), and gluons (pink) are given by the lower hatched bands which arise from the finite statistics in the evaluation of the integrals by Monte Carlo.}
		\label{fig_gamma-q}
	\end{figure*}
	
	\begin{figure*}[h!]
		\centering
		\begin{tabular}{cc}
			\includegraphics[width=0.4\linewidth]{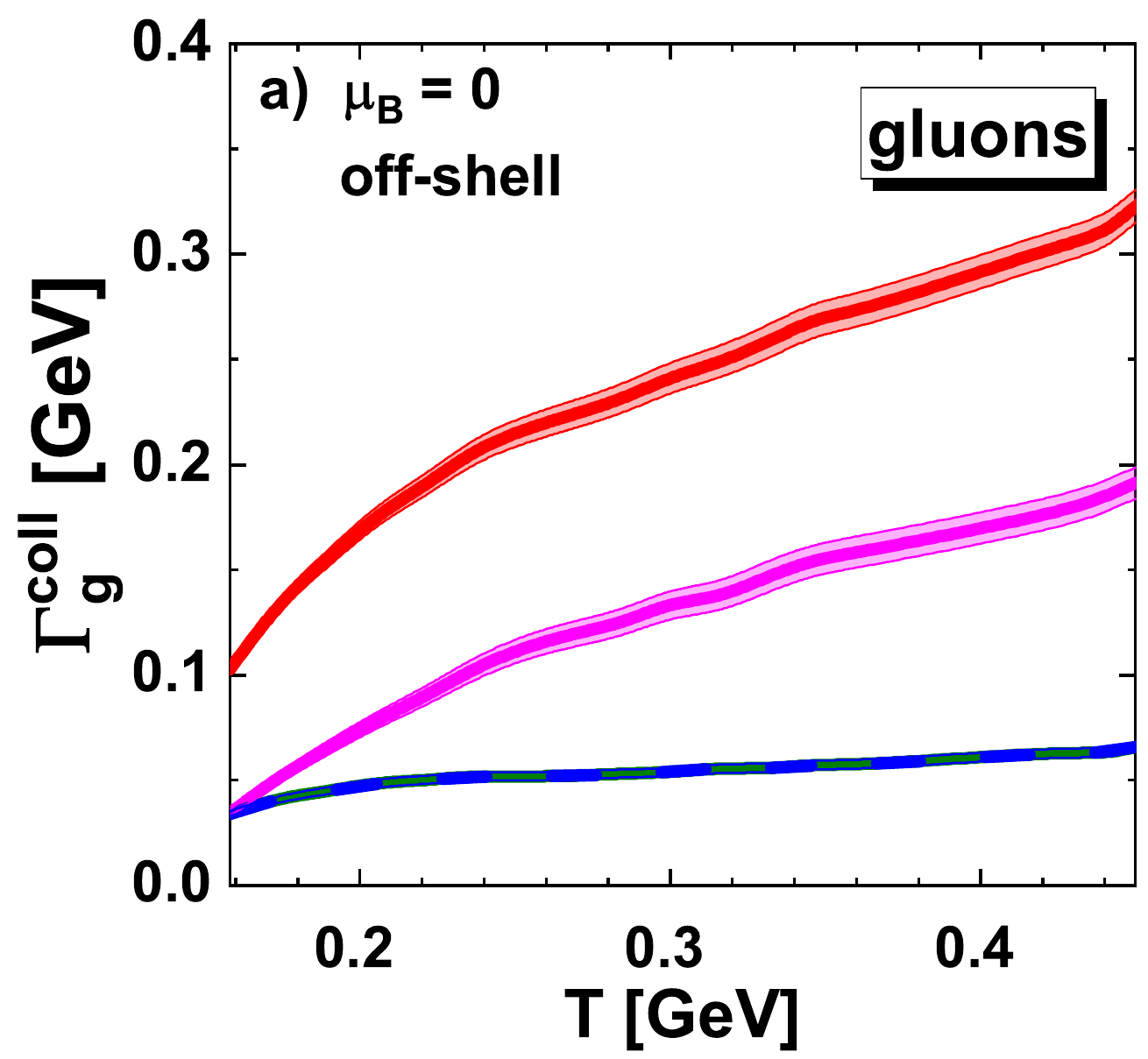}
			\includegraphics[width=0.4\linewidth]{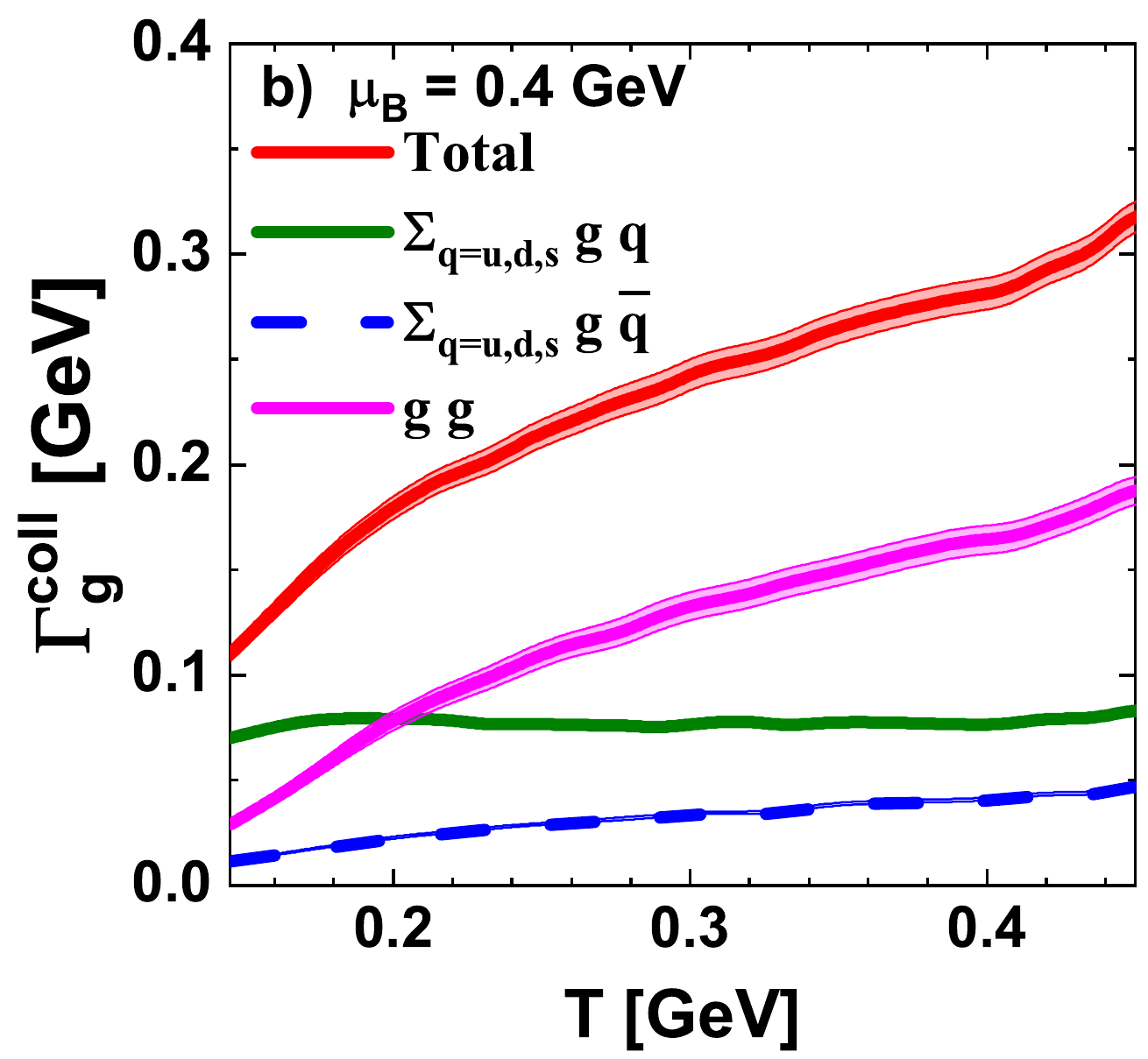}
		\end{tabular}
		\caption{(Color online) Off-shell collision rate from Eq. (\ref{Gamma_off_avT}) of a gluon $g$ as a function of the temperature $T$ for $\mu_B = 0$ (a) and $\mu_B = 0.4$ GeV (b).  The contributions from the scattering with light quarks (green), antiquarks (blue), and gluons (pink) are given by the lower hatched bands which arise from the finite statistics in the evaluation of the integrals by Monte Carlo.}
		\label{fig_gamma-g}
	\end{figure*}
	
	\begin{figure*}[h!]
		\centering
		\includegraphics[width=0.45\linewidth]{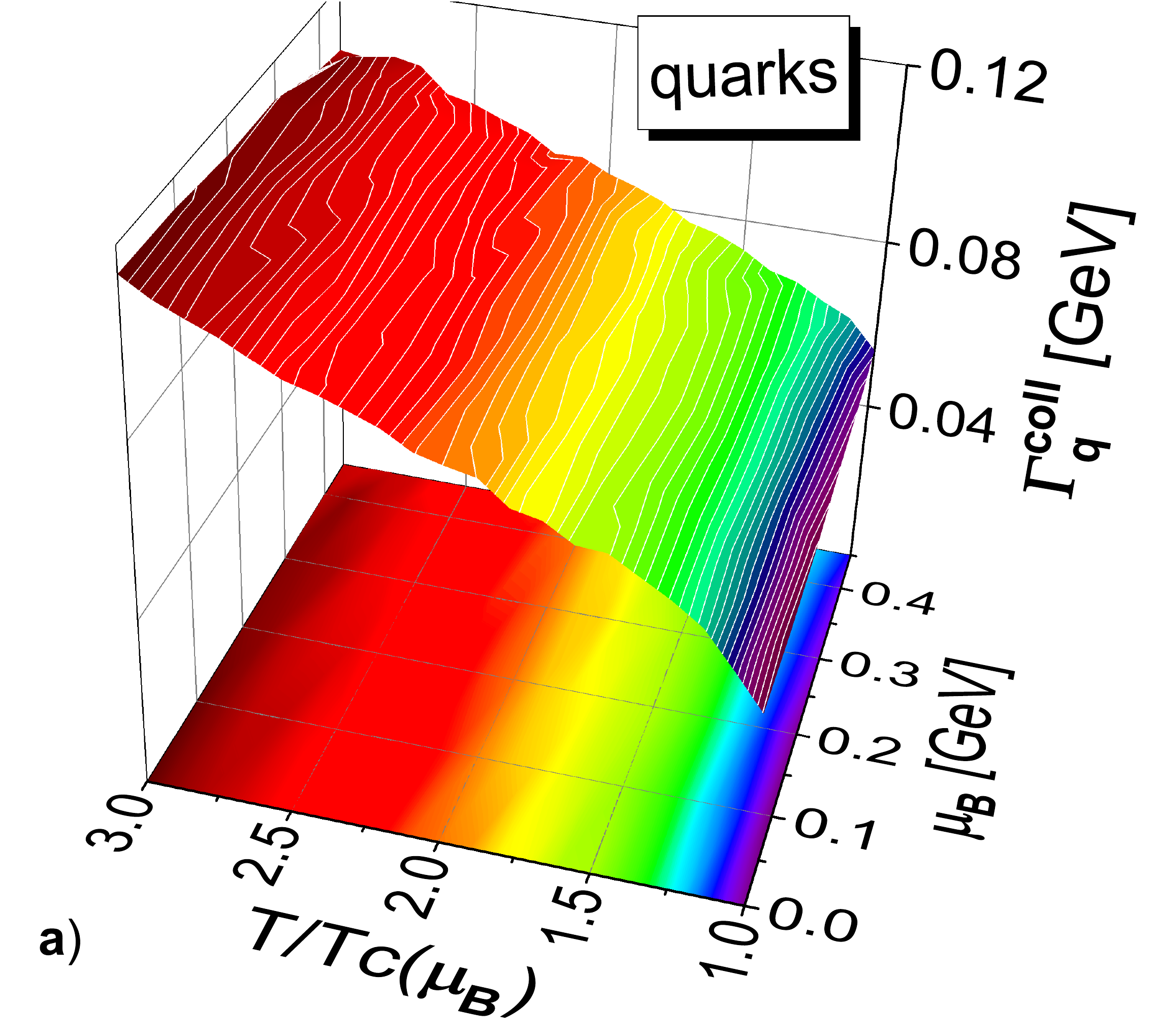}
		\includegraphics[width=0.45\linewidth]{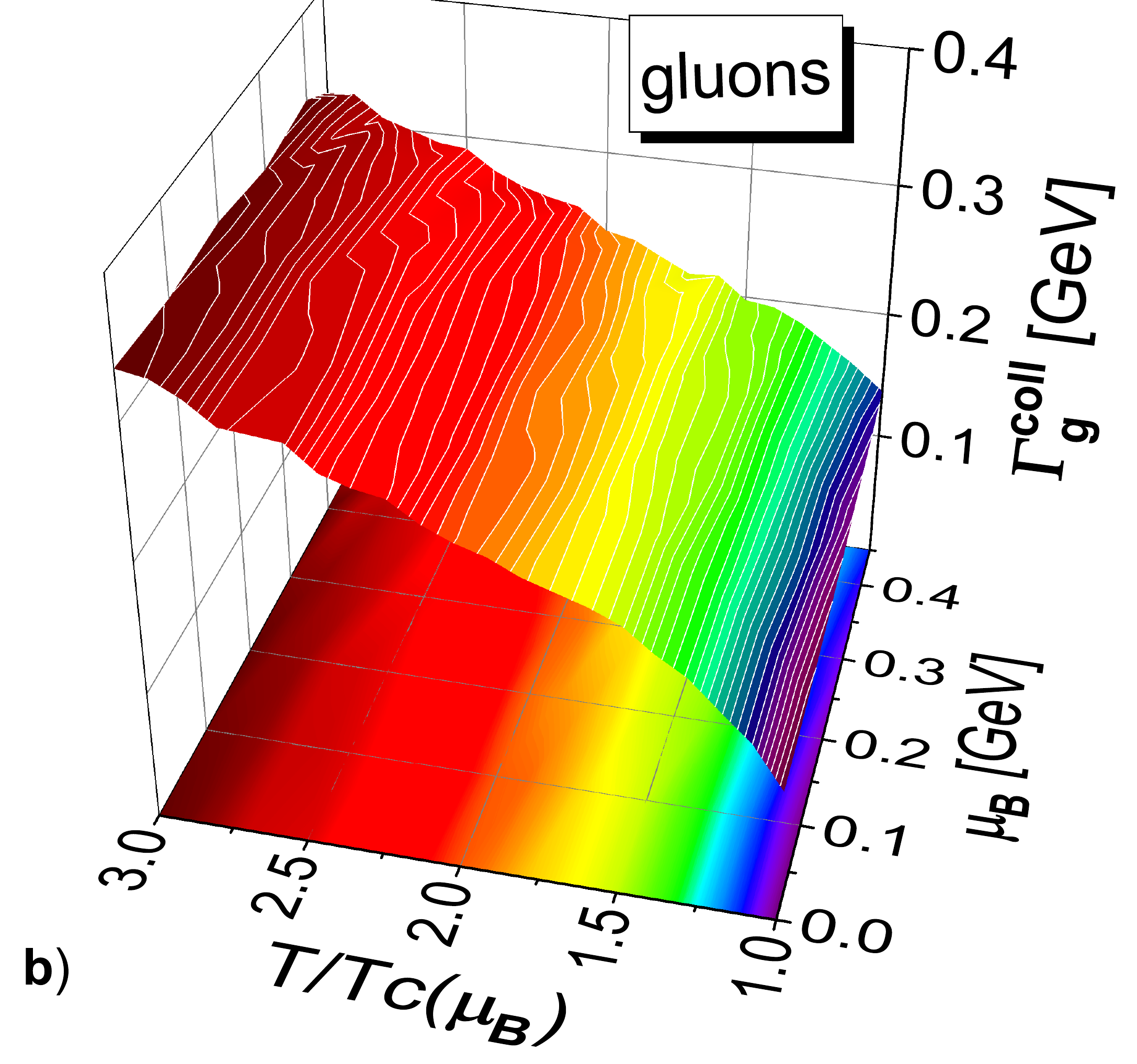}
		\caption{(Color online) Off-shell collision rate of a quark (a) and gluon (b) as a function of the scaled temperature $T/T_c(\mu_B)$ and the baryon chemical potential $\mu_B$ from Eq. (\ref{Gamma_off_avT}).}
		\label{fig_gamma-qg-3D}
	\end{figure*}
	
\end{widetext}

Figure \ref{fig_gamma-q} shows the ``off-shell interaction rate'' $\Gamma^{\text{coll}}_q$ of a light quark $q$ as a function of the  temperature $T$ for $\mu_B = 0$ (a) and $\mu_B = 0.4$ GeV (b).  The contributions from the scattering with light quarks (green), antiquarks (blue), and gluons (pink) are given by the lower hatched bands. At $\mu_B$ = 0, the total width $\Gamma^{\text{coll}}_q$ to a large extend stems from quark-gluon scattering and increases with temperature  while the contributions from scatterings with quarks and antiquarks are about equal and subdominant. At $\mu_B$ = 0.4 GeV, the quarks are more abundant than the antiquarks and the contributions from scatterings with quarks increase while that from collisions with antiquarks decrease relative to $\mu_B$ = 0. The contributions from collisions with gluons slightly decreases also with $\mu_B$, which can be attributed to a decrease of the cross sections with $\mu_B$ as noted before.  Figure \ref{fig_gamma-g} shows the off-shell interaction rate of a gluon $g$ as a function of the temperature $T$ for $\mu_B = 0$ (a) and $\mu_B = 0.4$ GeV (b) as in case of quark scattering in Fig.  \ref{fig_gamma-q}.  The contributions from the scattering with light quarks (green), antiquarks (blue), and gluons (pink) are given by the lower hatched bands. The discussion of the contributions to the total widths is very similar to the case of quark scattering and not repeated here.

In summarizing this section, we find that the collisional widths for timelike partons are sizable and increase with temperature (as in the DQPM) but still remain substantially smaller than the pole masses in Fig. \ref{fig-DQPM-masses}. Accordingly, a quasiparticle interpretation for timelike quanta should approximately hold.

Figure \ref{fig_gamma-qg-3D}, finally, gives an overview on the width $\Gamma_q$ (a) and width $\Gamma_g$ (b) as a function of the scaled temperature $T/T_c(\mu_B)$ and chemical potential $\mu_B$. While the dependencies on temperature are similar for fixed $\mu_B$, we see a general slight decrease of the total widths with $\mu_B$ for fixed $T/T_c(\mu_B)$ as discussed above. 

\section{Transport properties of the hot and dense QGP}
\label{Section5}

The starting point to evaluate viscosity coefficients of partonic matter is the Kubo formalism \cite{Kubo:1957mj,zubarev1996statistical,Aarts:2002cc,Iwasaki:2007iv,Lang:2012tt,Lang:2013lla,Haas:2013hpa,Christiansen:2014ypa}, which was used to calculate the viscosities for a previous version of the DQPM within the PHSD in a box with periodic boundary conditions (cf. Ref. \cite{Ozvenchuk:2012kh}). We focus here on the calculation of the shear viscosity based on Refs. \cite{Aarts:2002cc,Iwasaki:2007iv,Lang:2012tt,Lang:2013lla}, which reads

\begin{align}
\eta^{\text{Kubo}}(T,\mu_q) & = - \int \frac{d^4p}{(2\pi)^4}\ p_x^2 p_y^2 \sum_{i=q,\bar{q},g} d_i\ \frac{\partial f_i(\omega)}{\partial \omega}\ \rho_i(\omega,\mathbf{p})^2 \label{eta_Kubo} \\
=  \frac{1}{15T} \int & \frac{d^4p}{(2\pi)^4}\ \mathbf{p}^4 \sum_{i=q,\bar{q},g} d_i \left( (1 \pm f_i(\omega)) f_i(\omega) \right) \rho_i(\omega,\mathbf{p})^2 , \nonumber
\end{align}
~\\
where the notation $f_i(\omega) = f_i(\omega,T,\mu_q)$ is used for the distribution functions, and $\rho_i$ denotes the spectral functions from Eq. (\ref{spectral_function}). We note that the derivative of the distribution function accounts for the Pauli-blocking ($-$) and Bose-enhancement (+) factors. Following Ref. \cite{Lang:2012tt}, we can evaluate the integral over $\omega = p_0$ in Eq. (\ref{eta_Kubo}) by using the residue theorem. When keeping only the leading-order contribution in the width $\gamma(T,\mu_B)$ from the residue --- evaluated at the poles of the spectral function $\omega_i = \pm \tilde{E}(\mathbf{p}) \pm i \gamma$ --- we finally obtain
\begin{align}
& \eta^{\text{RTA}}(T,\mu_q)  = \frac{1}{15T} \int \frac{d^3p}{(2\pi)^3} \sum_{i=q,\bar{q},g} \label{eta_on} \\
\times  & \left( \frac{\mathbf{p}^4}{E_i^2 \ \Gamma_i(\mathbf{p}_i,T,\mu_q)}\ d_i \left( (1 \pm f_i(E_i)) f_i(E_i) \right) \right) + \order{\Gamma_i} , \nonumber
\end{align}
~\\
which corresponds to the expression derived in the relaxation-time approximation (RTA) \cite{Sasaki:2008fg,Sasaki:2008um,Bluhm:2009ef,Bluhm:2010qf,Albright:2015fpa} by identifying the interaction rate $\Gamma$ with $2\gamma$ as expected from transport theory in the quasiparticle limit \cite{Blaizot:1999xk}. This interaction rate $\Gamma_i(\mathbf{p}_i,T,\mu_q)$ (inverse relaxation time) is calculated microscopically by Eq. (\ref{Gamma_on}). We recall that the pole energy is $E_i^2 = p^2 + M_i^2$, where $M_i$ is the pole mass given in the DQPM by Eqs. (\ref{Mg9}) and (\ref{Mq9}). As in the previous section, we use here the notation $\sum_{j=q,\bar{q},g}$, which includes the contribution from all possible partons which in our case are the gluons and the (anti)quarks of three different flavors ($u,d,s$).

\begin{figure}
	\centering
	\includegraphics[width=0.95\linewidth]{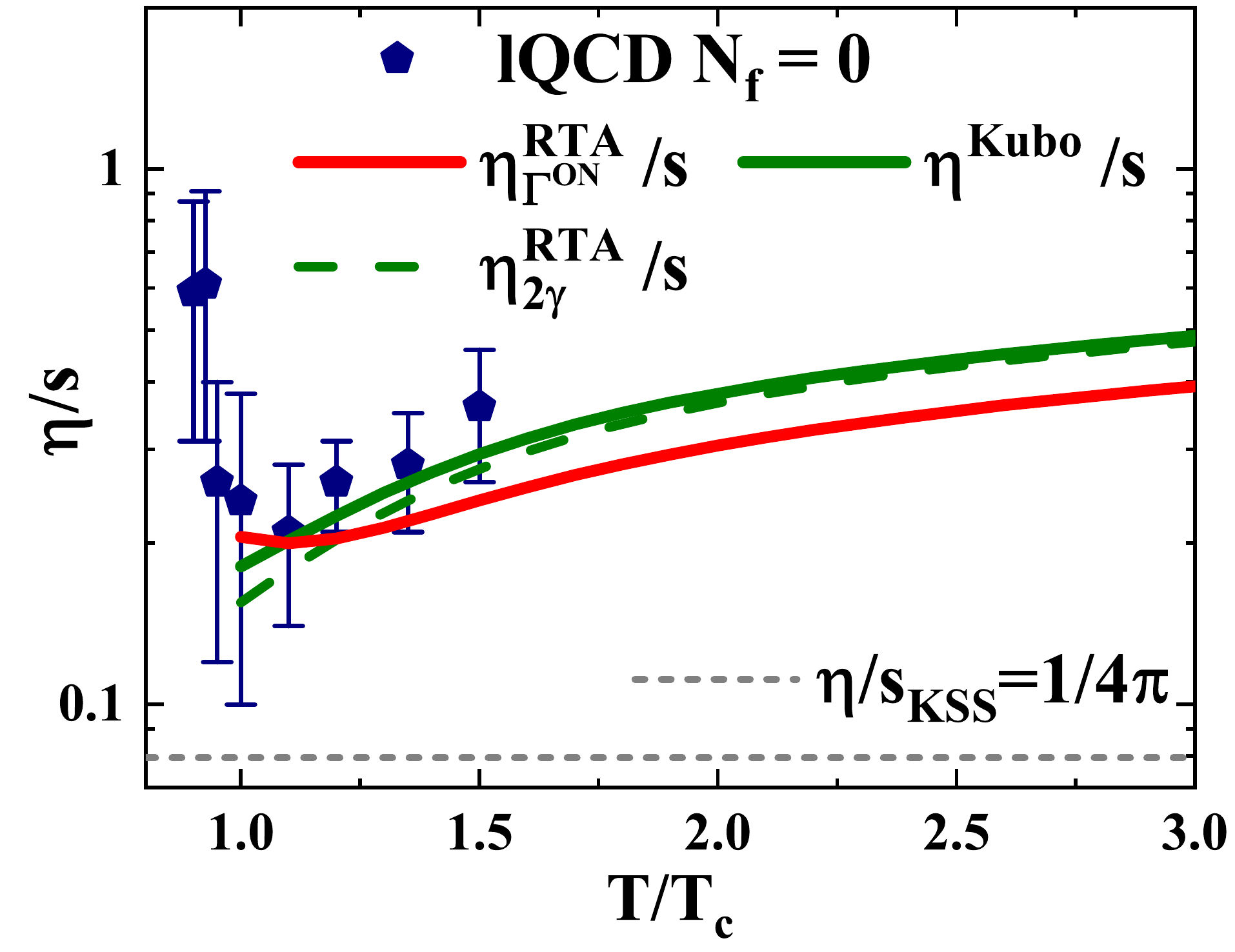}
    \caption{(Color online) The ratio of shear viscosity to entropy density as a function of the scaled temperature $T/T_c$ for $\mu_B = 0$
        from Eqs. (\ref{eta_Kubo}) and (\ref{eta_on}). The solid green line $(\eta^{\text{Kubo}}/s)$ shows the results from the original DQPM in the Kubo formalism while the dashed green line $(\eta^{\text{RTA}}_{2\gamma}/s)$ shows the same result in the quasiparticle approximation (\ref{eta_on}). The solid red line $(\eta^{\text{RTA}}_{\Gamma^{\text{on}}}/s)$ results from Eq. (\ref{eta_on}) using the interaction rate $\Gamma^{\text{on}}$ (\ref{Gamma_on}) calculated by the microscopic differential cross sections in the on-shell limit. The dashed gray line demonstrates the Kovtun-Son-Starinets bound \cite{Policastro:2001yc,Kovtun:2004de} $(\eta/s)_{\text{KSS}} = 1/(4\pi)$, and the symbols show lQCD data for pure SU(3) gauge theory taken from Ref. \cite{Astrakhantsev:2017nrs} (pentagons).}
	\label{fig_eta}
\end{figure}

The actual results are displayed in Fig. \ref{fig_eta} for the 
ratios of shear viscosity to entropy density $\eta/s$ as a function 
of the scaled temperature $T/T_c$ for $\mu_B$ = 0 in comparison to
those from lattice QCD \cite{Astrakhantsev:2017nrs}. 
The solid green line $(\eta^{\text{Kubo}}/s)$ shows the result from the original DQPM in the 
Kubo formalism  while the dashed green line $(\eta^{\text{RTA}}_{2\gamma}/s)$ shows the same result in 
the quasiparticle approximation (\ref{eta_on}) by replacing 
$\Gamma_i$ by $2\gamma_i$. The solid red line $(\eta^{\text{RTA}}_{\Gamma^{\text{on}}}/s)$ results from Eq.
(\ref{eta_on}) using the interaction rate $\Gamma^{\text{on}}$ 
(\ref{Gamma_on}) calculated by the microscopic differential cross 
sections in the on-shell limit. We find that --- apart from 
temperatures close to $T_c$ ---  the ratios $\eta/s$ do not differ 
very much and have a similar behavior as a function of temperature. 
The approximation (\ref{eta_on}) of the shear viscosity is found to 
be very close to the one from the Kubo formalism (\ref{eta_Kubo}), 
indicating that the quasiparticle limit ($\gamma \ll M$) holds in
the DQPM. 
We have also checked that the shear viscosity does not 
differ substantially if one uses the momentum-dependent interaction 
rate from Eq. (\ref{Gamma_on}) or the averaged one from Eq. 
(\ref{Gamma_on_avT}).

	\begin{figure*}
		\centering
		\includegraphics[width=0.49\linewidth]{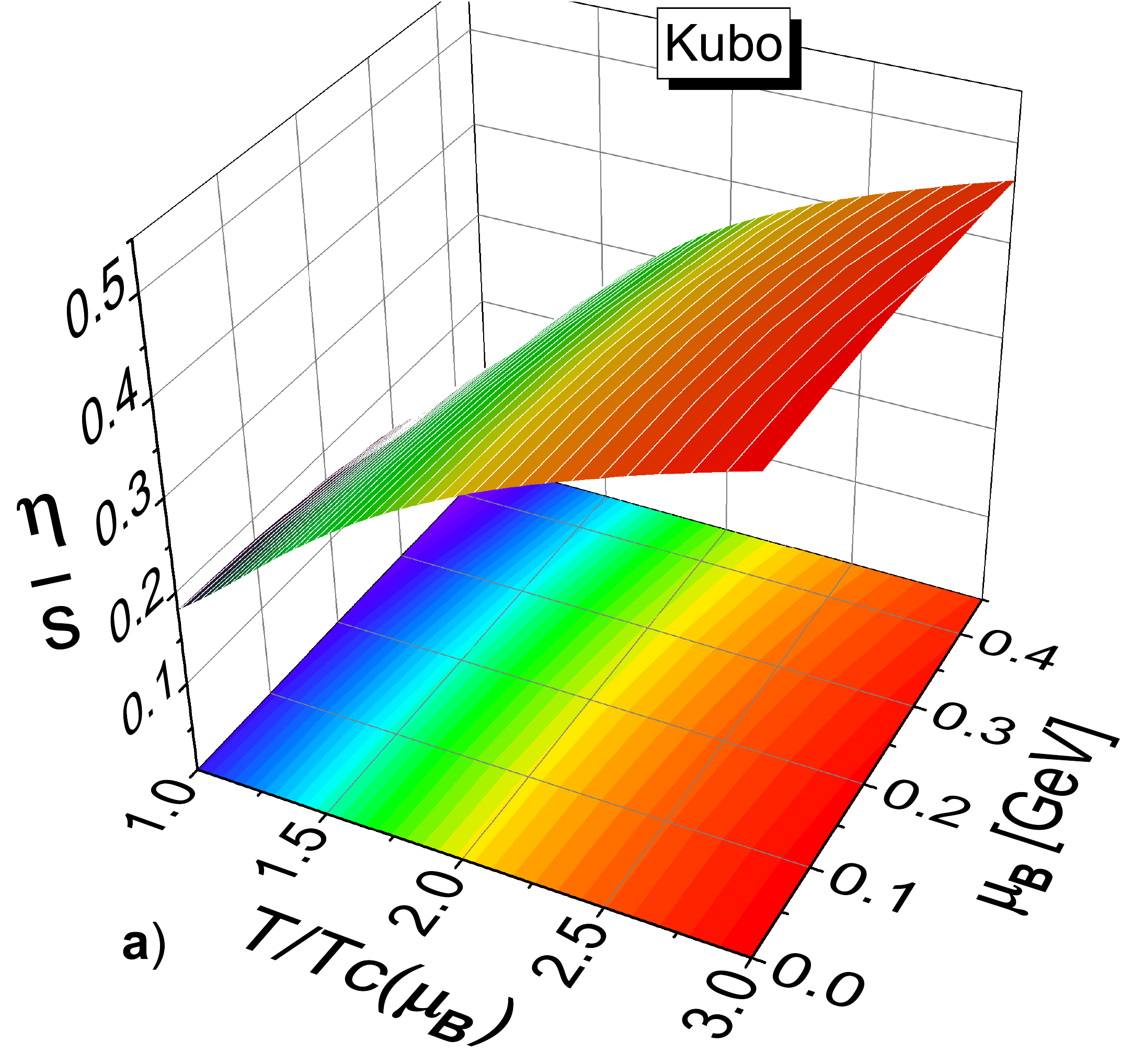}
		\includegraphics[width=0.49\linewidth]{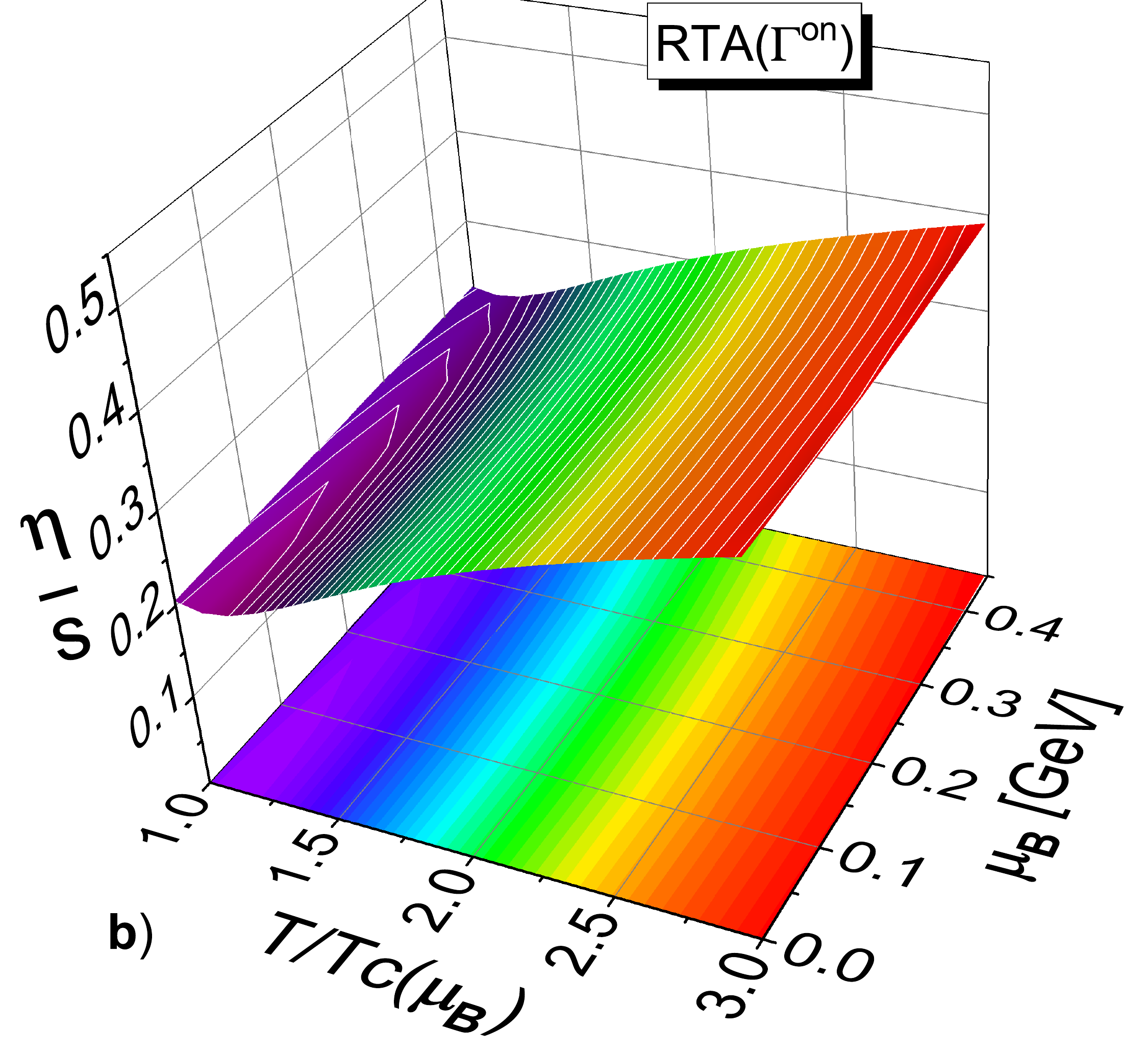} 
		\caption{(Color online) The ratio of shear viscosity to entropy density $\eta/s$ as a function of the scaled 
			temperature $T/T_c(\mu_B)$ and baryon chemical potential $\mu_B$ calculated within the Kubo formalism (a) from Eq. (\ref{eta_Kubo}) and in the relaxation time approximation (RTA) (b) from Eq. (\ref{eta_on}) using the on-shell interaction rate $\Gamma^{\text{on}}$ from Eq. (\ref{Gamma_on}).}
		\label{fig_eta_3d}
	\end{figure*}

An overview for the ratio of shear viscosity to entropy density $\eta/s$  as a function of the scaled temperature $T/T_c(\mu_B)$ and $\mu_B$ is given in Fig. \ref{fig_eta_3d} in case of the Kubo formalism (a) (\ref{eta_Kubo}) and the on-shell limit (\ref{eta_on}) (b). There is no strong variation with $\mu_B$ for fixed $T/T_c(\mu_B)$; however, the ratio increases slightly with $\mu_B$ in the on-shell limit while it slightly drops with $\mu_B$ in the Kubo formalism for the DQPM. Accordingly, there is some model uncertainty when extracting the shear viscosity in the different approximations.

In summarizing this section, we find that the results for the ratio of shear viscosity over entropy density from the original DQPM and those from the microscopic calculations are  similar and within error bars compatible with present results from lattice QCD. However, having the differential cross sections for each partonic channel at hand one might find substantial differences for nonequilibrium configurations as encountered in relativistic heavy-ion collisions where a QGP is formed initially out of equilibrium.

\section{Extraction of $T$ and $\mu_B$ from PHSD in heavy-ion collisions}
\label{Section6}

Since PHSD is a microscopic off-shell transport approach, it does not incorporate thermodynamic Lagrange parameters such as $T$ and $\mu_B$  that characterize the system in equilibrium.
In order to extract the required information (the temperature $T$ and baryon chemical potential $\mu_B$) ---  defining the parton properties and differential scattering processes in the PHSD space-time grid --- we use a parametrization of the lQCD equation of state from Ref. \cite{Gunther:2017sxn} where the pressure (negative thermodynamic potential) is expanded as:

\begin{align}
\frac{P}{T^4} =  c_0(T) + c_2(T) \left( \frac{\mu_B}{T} \right)^2 + c_4(T) \left( \frac{\mu_B}{T} \right)^4 + \order{\mu_B^6}. \label{lQCD_EOS}
\end{align}
~\\
This equation of state matches the conditions of a heavy-ion collision where strangeness neutrality $\langle n_S \rangle = 0$ is realized on average and where the relation between electric charge and baryon number $\langle n_Q \rangle = 0.4\ \langle n_B \rangle $ is fixed by the content of the initial nuclei. We mention that the inclusion of the sixth-order coefficient $c_6$ induces wiggles in the EoS due to oscillating contributions in Eq. (\ref{lQCD_EOS}) (see also Ref. \cite{Bazavov:2017dus}) but does not lead to considerable changes for $\mu_B/T<3$ and is discarded here. Note that the parametrization of the coefficients $c_i(T)$ in Eq. (\ref{lQCD_EOS}) is also in agreement with lQCD data below $T_c$ which allows for an evaluation of $T$ and $\mu_B$ also in the hot hadronic phase. We point out that these results have to be taken as estimates in the regions of large chemical potentials for $\mu_B/T > 3$. 

\begin{figure*}[t]
	\centering
	\includegraphics[width=\columnwidth]{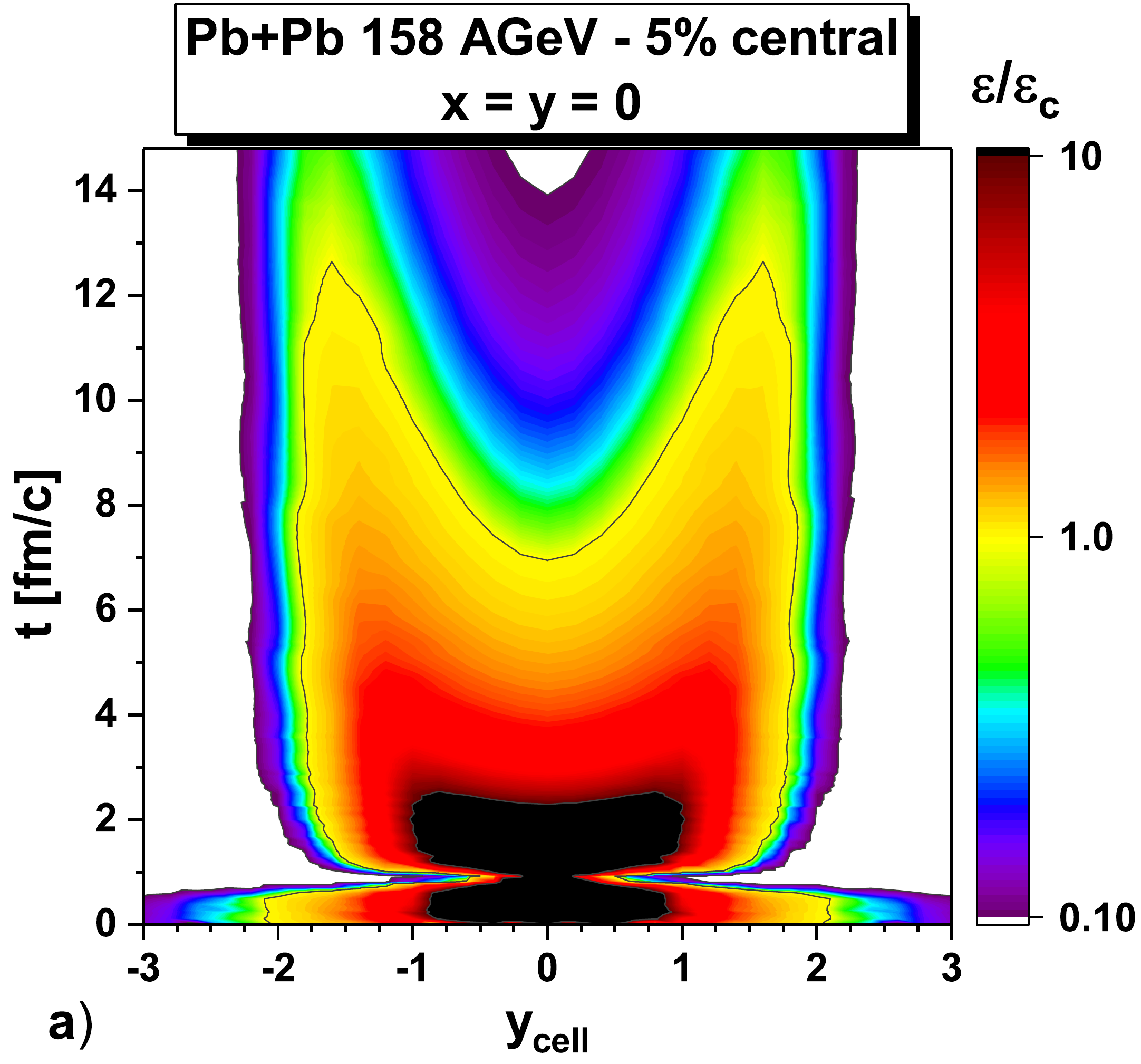}
	\includegraphics[width=\columnwidth]{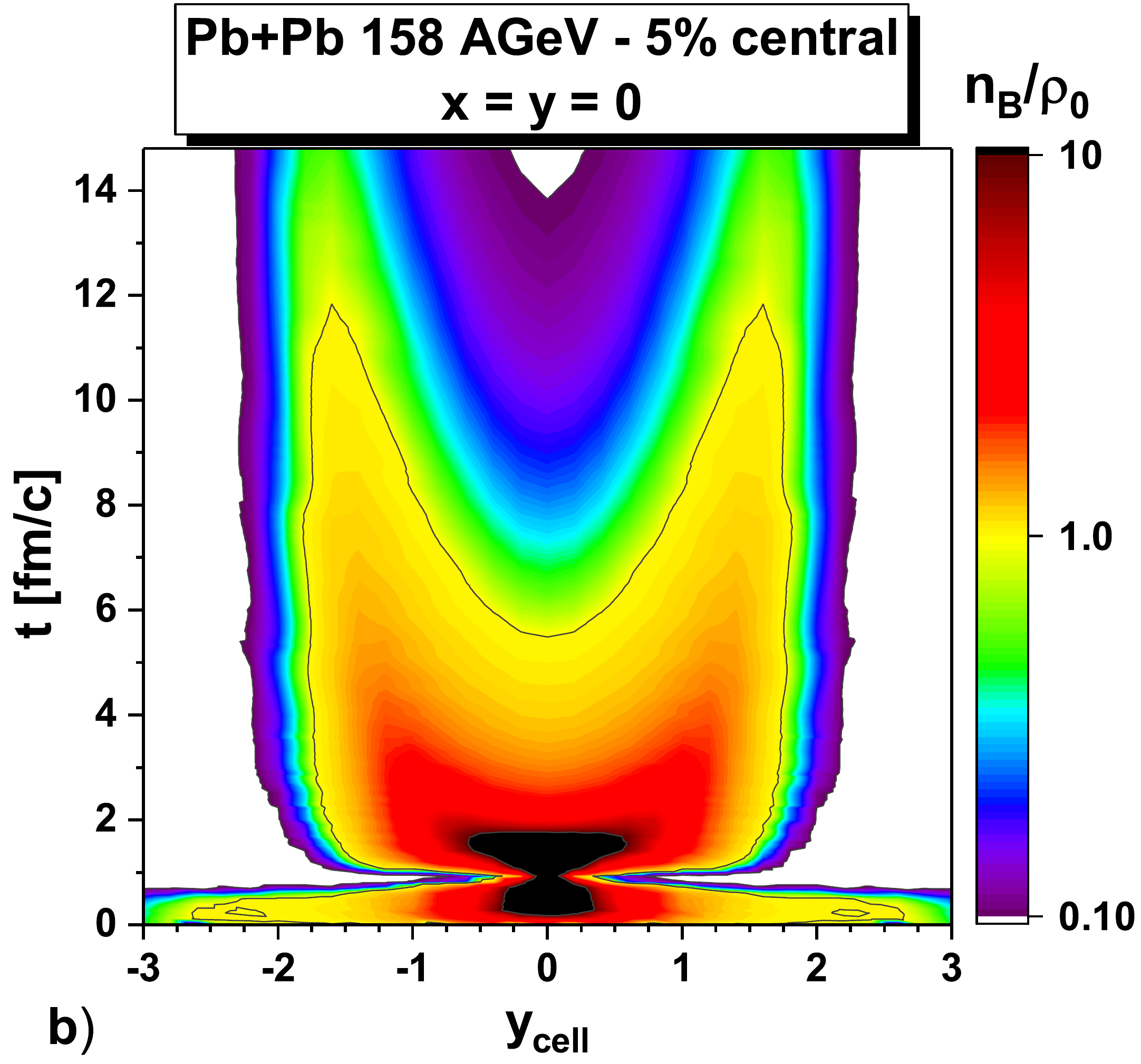}
	\caption{(Color online) The energy density $\epsilon$ (left) and baryon density $n_B$ (right) as a function of the cell rapidity $y_{\text{cell}}$ at different times $t$ for 5\% central Pb+Pb collisions at 158 A GeV extracted from the central cell $x = y= 0$ of the PHSD. The energy density and baryon density are divided by the critical energy density $\epsilon_c = 0.5$ GeV/fm$^3$ and the saturation density $\rho_0$ = 0.17 fm$^{-3}$, respectively. The initial time $t=0$ is systematically taken to be the time of the first nucleon-nucleon collision.}
	\label{fig-nB_158AGeV}
\end{figure*}

In each space-time cell of the PHSD grid, the thermodynamic quantities are calculated by the method developed in Ref. \cite{Xu:2017pna}, i.e., by diagonalization of the energy momentum tensor $T^{\mu \nu}$ as

\begin{equation}
T^{\mu \nu}\ (x_\nu)_i = \lambda_i\ (x^\mu)_i = \lambda_i\ g^{\mu \nu}\ (x_\nu)_i,
\label{TmunuL}
\end{equation}
~\\
with $i =0,1,2,3$, where $\lambda_i $ are the eigenvalues of $T^{\mu \nu}$ and $(x_\nu)_i$ are the corresponding eigenvectors. For $i = 0$, the local energy density $\epsilon$ is identified with the eigenvalue of $T^{\mu \nu}$ (Landau matching) and the corresponding timelike eigenvector is defined as the 4-velocity $u_\nu$:

\begin{equation}
T^{\mu \nu}u_\nu = \epsilon u^\mu = (\epsilon g^{\mu \nu})u_\nu
\label{Landau}
\end{equation}
~\\
using the normalization condition $u^\mu u_\mu =1$. The three other solutions are ($-P_i$), the pressure components expressed in the local rest frame of the cell. The energy-momentum tensor $T^{\mu \nu}$ is calculated in PHSD as

\begin{equation}
T^{\mu \nu} = \sum_{i} \frac{p_i^\mu p_i^\nu}{E_i},
\end{equation}
~\\
where the sum $i$ runs over all the particles in the considered cell.

To study the matter at finite baryon density, we calculate additionally the net-baryon current

\begin{equation}
J^\mu_B = \sum_{i} \frac{p_i^\mu}{E_i} \frac{(q_i - \bar{q}_i)}{3} ,
\end{equation}
~\\
with $q_i$ ($\bar{q}_i$) being the number of light quarks (antiquarks) within particle $i$. To obtain the local net-baryon density, we apply the Lorentz transformation defined via

\begin{equation}
n_B = \gamma_E \left( J^0_B - \vec{\beta_E} \cdot \vec{J_B} \right) = \frac{J^0_B}{\gamma_E},
\end{equation}
~\\
where $\vec{\beta_E} = \vec{J_B}/J^0_B$ is known as the Eckart velocity and $\gamma_E$ is the associated Lorentz factor. We mention that by using the Eckart velocity $\vec{\beta_E}$ to transform the net-baryon current, the spatial part of the latter automatically vanishes, whereas this is not guaranteed when employing the energy flow $u^\mu$ from Eq. (\ref{Landau}).

We illustrate in Fig. \ref{fig-nB_158AGeV} our extraction method and show the time and rapidity dependence of the energy density $\epsilon$ and baryon density $n_B$ extracted from the PHSD in the central cell $x = y = 0$ of 5\% central Pb + Pb collisions at 158 A\ GeV. We mention that in the PHSD the parallel ensemble method is used, implying that the densities are averaged over a large number of parallel events ($\sim$ 250 for AGS energies, $\sim$ 150 for top SPS and $\sim$ 30 for top RHIC). With this procedure, even if each event contains only a few particles per cell, the overall profile is relatively smooth in space-time. Furthermore, in the evaluation of the energy
density $\epsilon$ and baryon density $n_B$, leading quarks or diquarks are also included since they carry most of the baryon number in the string fragmentation picture. The cell rapi\-dity $y_{\text{cell}}$ is evaluated from the cell 4-velocity $u^\mu = \gamma(1,\vec{\beta})$ in Eq. (\ref{Landau}) as $y_{\text{cell}} = 1/2 \log [(1+\beta_z)/(1-\beta_z)]$. For illustration, we have scaled the energy and baryon densities by the critical energy density $\epsilon_c = 0.5$ GeV/fm$^3$ used in the PHSD and the saturation density $\rho_0 = 0.17$ fm$^{-3}$, respectively. One can see that at the collision time ($t = 0$) a high amount of energy is deposited into the midrapidity region whereas at higher rapidities the initial nuclei are still intact. Indeed, we observe that around the initial rapidity of the nuclei ($y_N \approx 2.9$) the baryon density is still close to the saturation density $\rho_0$. After the passing time $t = R/\gamma_N \approx 0.75$ fm/c, all the initial nucleons in the central cell $(x=y=0)$ have interacted and maximal values for the energy density and baryon density are reached which are of the order of $\epsilon \approx 5$ GeV/fm$^3$ and $n_B \approx 2$ fm$^{-3}$, respectively. We can identify approximatively constant values of $\epsilon$ and $n_B$ as a function of the cell rapidity $y_{\text{cell}}$ along parabolas of proper time $\tau \sim t/\cosh(y_{\text{cell}})$. After $\tau \approx$ 6 fm/c, the energy density reaches the critical value $\epsilon_c$, implying that quarks and gluons hadronize into hadrons, whereas the baryon density appears to be slightly lower than the saturation density $\rho_0$.

In the beginning of heavy-ion collisions, the created medium
is highly anisotropic due to the longitudinal expansion.
In order to correct for the anisotropy,
we apply the {\it shape generalized equation of state} developed in Ref. \cite{Ryblewski:2012rr} in order to extract values for the temperature $T$ and baryon chemical potential $\mu_B$. In this framework, the energy density $\epsilon^{\text{anis}}$ and pressure components of an anisotropic medium are evaluated by the following expressions:

\begin{align}
\epsilon^{\text{anis}} & = \epsilon^{\text{EoS}}\ \ r(x) \label{eanis}, \\
P_\perp & = P^{\text{EoS}} \ \left[ r(x) + 3xr'(x) \right] \label{Ptrans}, \\
P_{\parallel} & = P^{\text{EoS}} \ \left[ r(x) - 6xr'(x) \right] \label{Plong} ,
\end{align}
~\\
where $P_\perp$ and $P_\parallel$ are, respectively, the transverse and longitudinal pressures, and $\epsilon^{\text{EoS}}$ and $P^{\text{EoS}}$ are the equilibrium energy density and pressure from which a temperature $T$ and chemical potential $\mu_B$ can be extracted. The anisotropy parameter $x$ can be approximated as a function of the pressure components as $P_\parallel/P_\perp = x^{-3/4}$, and the function $r(x)$ reads

\begin{equation}
r(x) =
\left\{
\begin{aligned}
\frac{x^{-1/3}}{2}\left[1+\frac{x\ \text{arctanh}
	\sqrt{1-x}}{\sqrt{1-x}}\right] \text{for } x \leq 1  \\
\frac{x^{-1/3}}{2}\left[1+\frac{x \arctan
	\sqrt{x-1}}{\sqrt{x-1}}\right] \text{for } x \geq 1 \ .
\end{aligned}
\right.
\label{r(x)}
\end{equation}
~\\
In a PHSD simulation, we calculate in each of the cells the energy density $\epsilon^\text{PHSD}$, the baryon density $n_B^{\text{PHSD}}$, as well as the pressure components $P^\text{PHSD}_\perp$ and $P^\text{PHSD}_\parallel$, from which one can evaluate the function $r(x)$ in Eq. (\ref{r(x)}). In order to find the temperature $T$ and baryon chemical potential $\mu_B$ according to the EoS --- constructed at the beginning of this section --- we have to solve the following system of equations:

\begin{equation}
\label{iso}
\left\{
\begin{aligned}
\epsilon^{\text{EoS}}(T,\mu_B) & = \epsilon^\text{PHSD}/r(x)  \\
n_B^{\text{EoS}}(T,\mu_B) &  = n_B^\text{PHSD}
\end{aligned}
\right. ,
\end{equation}
~\\
where the left-hand sides represent the EoS which depends on the unknowns $T$ and $\mu_B$, whereas on the right-hand sides of these equations we have the energy density and baryon density evaluated in PHSD. In Eq. (\ref{iso}), the energy density from PHSD $\epsilon^\text{PHSD}$ is divided by the function $r(x)$ from Eq. (\ref{r(x)}) to account for the anisotropy of the considered cell according to the {\it shape generalized equation of state} in Eq. (\ref{eanis}). We solve this system by using the Newton-Raphson method \cite{newton,raphson}.


\begin{figure}
	\centering
	\includegraphics[width=0.99\columnwidth]{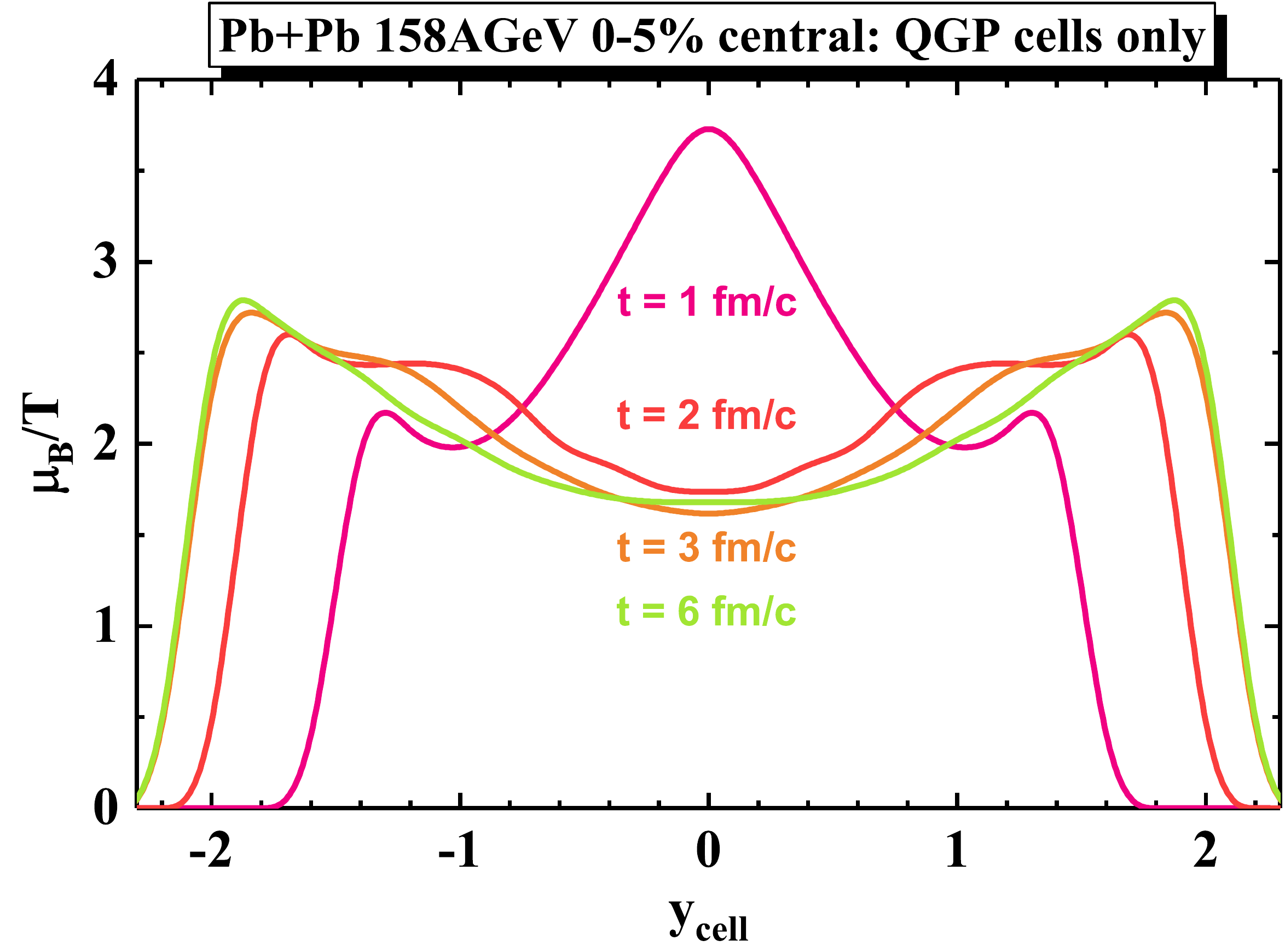}
	\caption{(Color online) The ratio $\mu_B/T$ as a function of the cell rapidity $y_{\text{cell}}$ at different times $t$ for 5\% central Pb + Pb collisions at 158 A GeV from PHSD.}
	\label{fig-muBoT_158AGeV}
\end{figure}

\begin{figure}
	\centering
	\includegraphics[width=0.99\columnwidth]{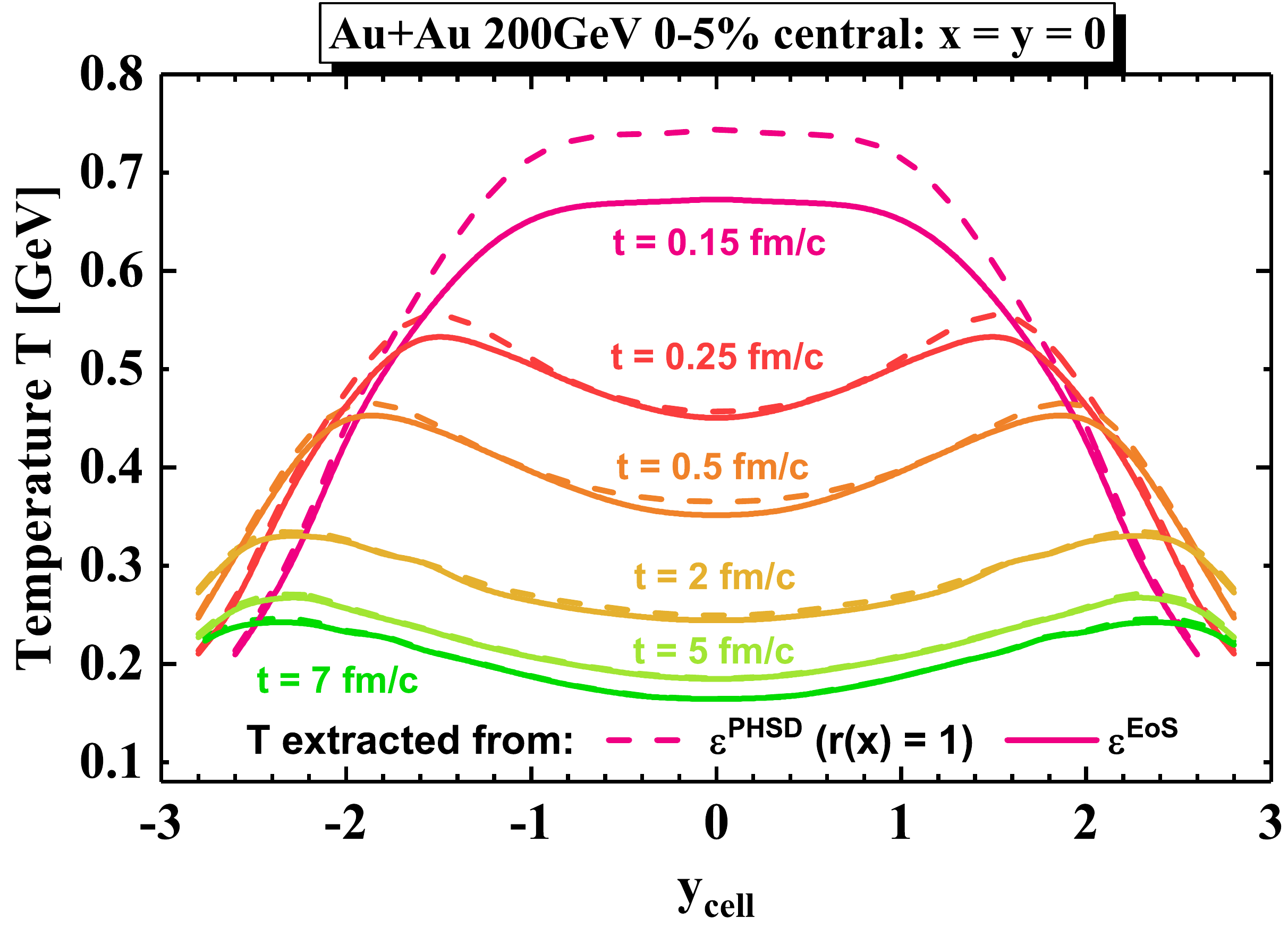}
	\caption{(Color online) The temperature profile (for the central cell) as a function of the cell rapidity  $y_{\text{cell}}$ and different times (from 0.15 to 7 fm/c) for a 5\% central Au + Au collision at $\sqrt{s_{NN}}$ = 200 GeV from PHSD. The dashed lines result when extracting the temperature directly from the energy density of the central cell in PHSD $\epsilon^{\text{PHSD}}$ while the solid lines refer to the extraction from the equation of state $\epsilon^{\text{EoS}}$ (see text).}
	\label{fig-Pressure_200AGeV}
\end{figure}

\begin{figure*}[t]
	\centering
	\includegraphics[width=0.3\linewidth]{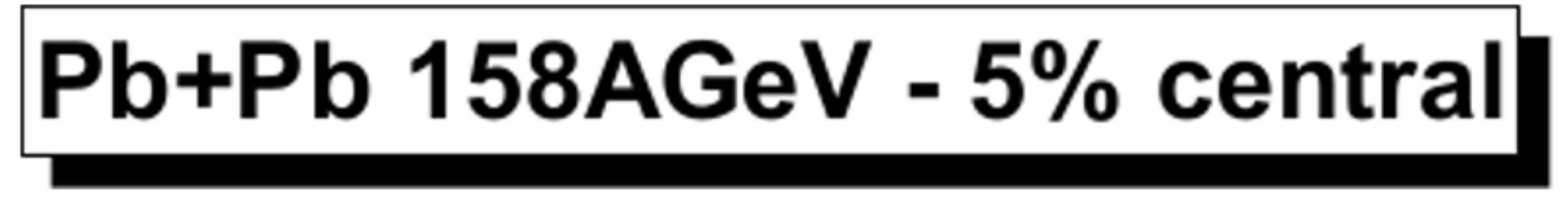}
	\includegraphics[width=1.\linewidth]{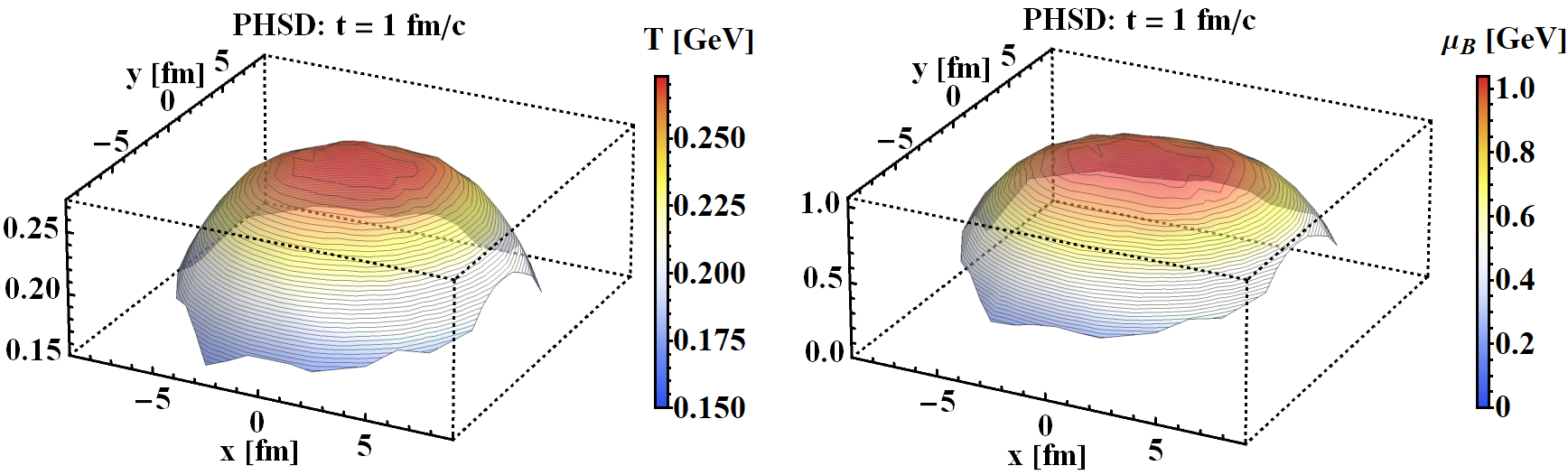}
	\includegraphics[width=1.\linewidth]{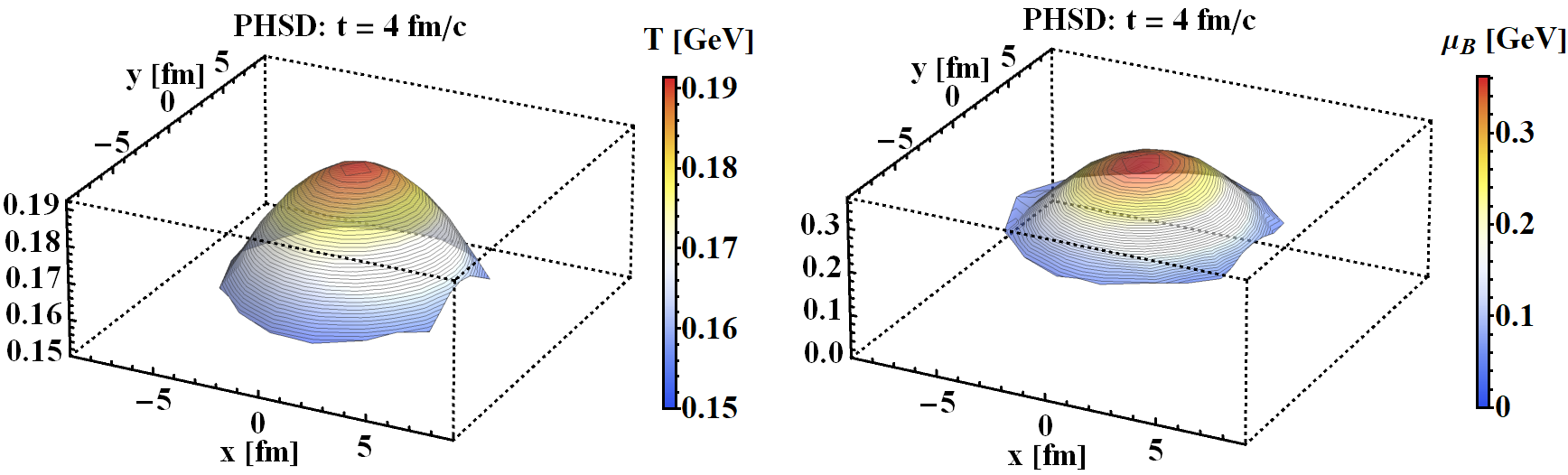}
	\caption{(Color online) (Left) The temperature profile in $(x,y)$ at midrapidity ($|y_{\text{cell}}| < $  1) at 1 and 4 fm/c after the initial collision in case of a 5\% central Pb + Pb collision at 158 A GeV from PHSD. (Right) The profile in the chemical potential $\mu_B$ in $(x,y)$ at midrapidity ($|y_{\text{cell}}| < $  1) for different times from 0.5 to 6 fm/c for the same collision.}
	\label{fig-Eval-T-muB-158AGeV-xy}
\end{figure*}

We now turn to the evaluation of $T$ and $\mu_B$ in actual PHSD
simulations for $A + A$ collisions. As an example for our results, we show in Fig. \ref{fig-muBoT_158AGeV} the
ratio $\mu_B/T$ as a function of the cell rapidity $y_{\text{cell}}$
at different times (from 1 to 6 fm/c) for 5\% central Pb + Pb
collisions at 158 A GeV. The largest ratios are seen for all times
for rapidities closer to projectile and target rapidities (cf. Ref. \cite{Li:2018ini}),
while at midrapidity this ratio is initially high (at $t$ = 1 fm/c) but drops to $\mu_B/T \sim 2$ at $t$ = 2 fm/c and remains approximately constant afterward. We mention that this profile is very close to that calculated in the
hydrodynamics + hadronic transport approach by Denicol {\itshape  et al.}
(Fig. 5 in Ref. \cite{Denicol:2018wdp}) and also shows an increase of $\mu_B/T$ with increasing $|y|$.

Figure \ref{fig-Pressure_200AGeV} shows the temperature profile (for the central cell) as a function of the cell rapidity  $y_{\text{cell}}$ and different times (from 0.15 to 7 fm/c) for a 5\% central Au + Au collision at $\sqrt{s_{NN}}$ = 200 GeV. This temperature profile initially ($t$ = 0.15 fm/c) has a broad maximum at midrapidity but for $t>$ 0.25 fm/c slight maxima at $|y_{\text{cell}}| \approx$ 1.5 appear which move to higher cell rapidity with increasing time while the average temperature drops rapidly in time. The lowest temperatures (at midrapidity), however, are still on the level of 250 MeV at $t$ = 2 fm/c, i.e., well above the critical temperature $T_c$. We note in passing that the temperature profiles from two different extraction methods, directly from the energy density of the cell in PHSD $\epsilon^{\text{PHSD}}$ (dashed lines) --- setting $r(x)$ = 1 in Eq. (\ref{iso}) --- and from the equation of state $\epsilon^{\text{EoS}}$ (solid lines) are practically the same for $t\ge$ 0.25 fm/c and provide an idea about the accuracy of our extraction method.

\begin{figure*}[h!]
	\centering
	\begin{tabular}{cc}
		\includegraphics[width=0.5\linewidth]{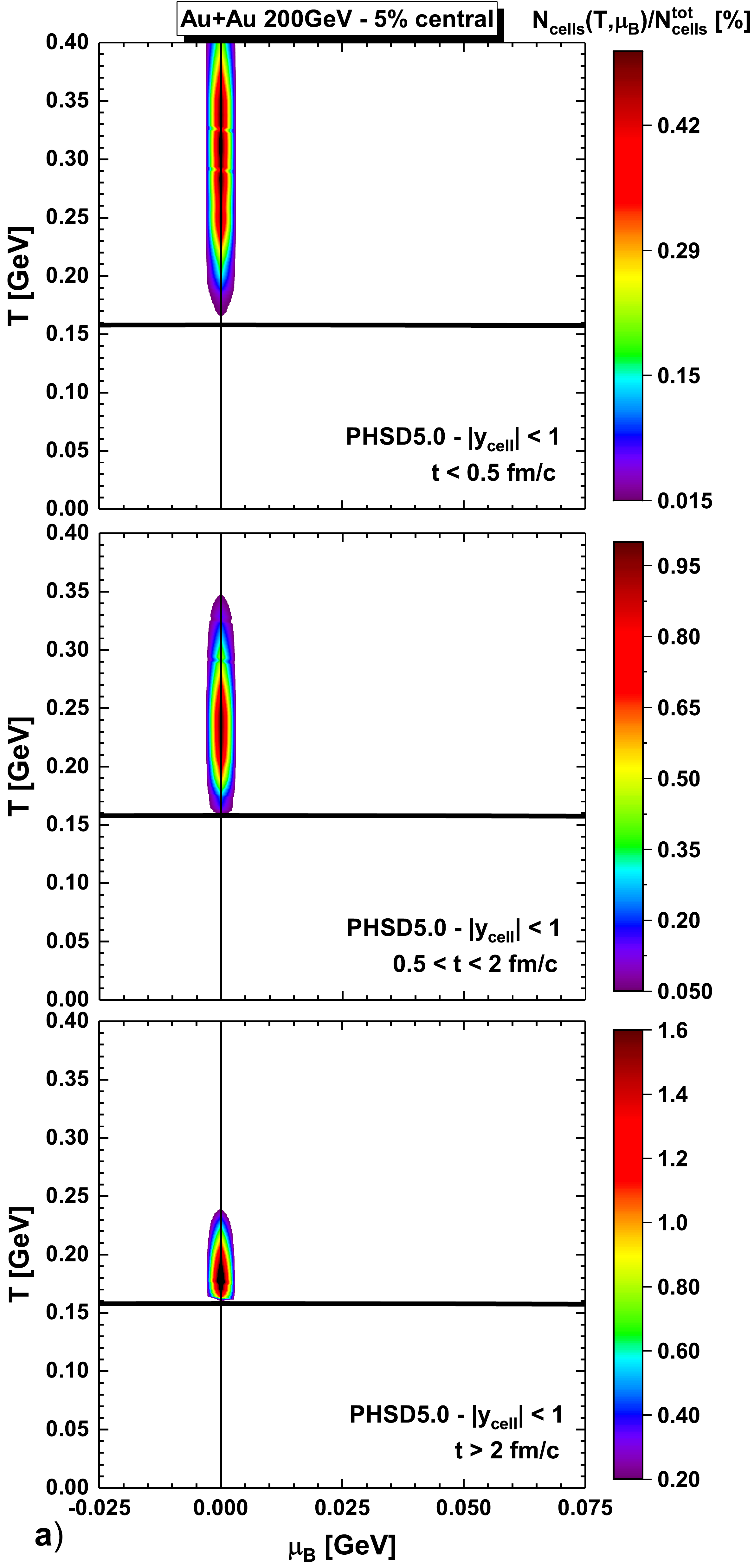} &
		\includegraphics[width=0.5\linewidth]{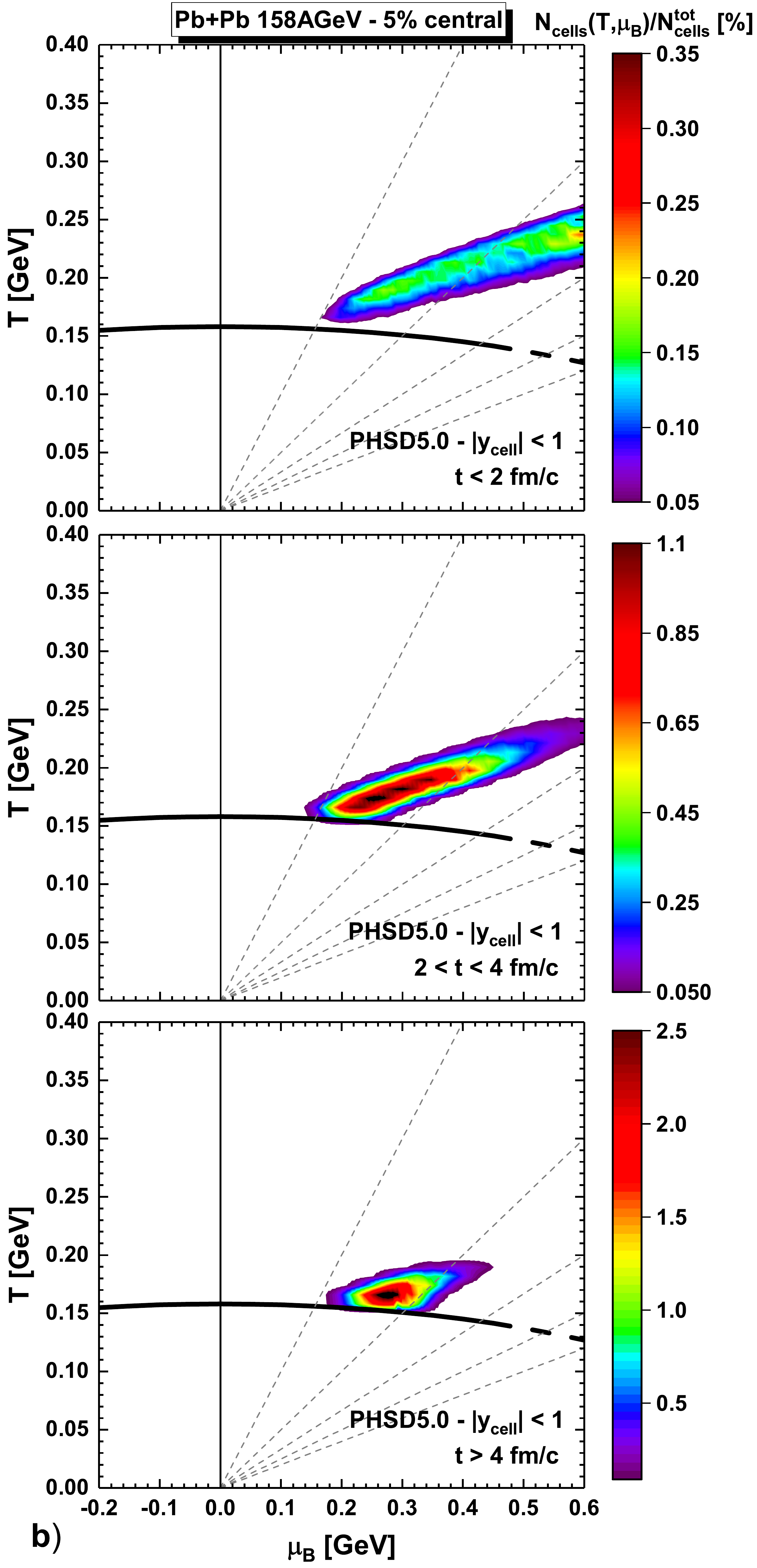}
	\end{tabular}
	\caption{(Color online) Distributions in $T$ and $\mu_B$ as extracted from the DQPM equation of state in a PHSD simulation of a central Au + Au collision at $\sqrt{s_{NN}}$ = 200 GeV (a) and of a central Pb + Pb collision at 158 A GeV (b) for cells with a temperature $T > T_c(\mu_B)$. The scale corresponds to the number of cells in the PHSD event in the considered bin in $T-\mu_B$ divided by the total number of cells in the corresponding time window (see legend). The solid black line is the DQPM phase boundary for orientation, the gray dashed lines indicate ratios of $\mu_B/T$ ranging from 1 to 5, while the vertical line stands for $\mu_B$ = 0.}
	\label{fig-Eval-T-muB-200GeV-158AGeV}
\end{figure*}

We now focus on the space-time distribution of the extracted temperatures $T$ (left column) and chemical potentials $\mu_B$ (right column) for a 5\% central collision of Pb + Pb at 158 A GeV from PHSD as shown in Fig. \ref{fig-Eval-T-muB-158AGeV-xy}.
These distributions correspond to the transverse plane ($x,y$) --- orthogonal to the beam direction --- at midrapi\-dity ($|y_{\text{cell}}| < $  1), i.e., in the center of the collision zone. For early times, we find the temperatures of the fireball to be well above $T_c$ practically everywhere, with a maximum in the center. Then the fireball expands in space with time while the temperature (and thus the QGP region) drops accordingly. Indeed, at $t$ = 4 fm/c the temperature is above $T_c$ only in the very central region, whereas on the outside hadronization already occurred. This is in line with the common picture of fireball expansion and hadronization. However, the profile in the chemical potential $\mu_B$ (right column) shows that the chemical potential $\mu_B$ is very large for early times in the whole fireball and drops to values of around $\approx 0.2 - 0.3$ in the hot QGP zone for $t \approx$ 4 fm/c.

We close this section by visualizing the time evolution of the distribution in $T$ and $\mu_B$ for cells having a temperature $T > T_c(\mu_B)$ at midrapidity ($|y_{\text{cell}}| <$ 1) for 5\% central heavy-ion collisions. Figure \ref{fig-Eval-T-muB-200GeV-158AGeV} (a) shows this distribution for a Au + Au collision at $\sqrt{s_{NN}}$ = 200 GeV from PHSD for times $t <$ 0.5 fm/c, 0.5 fm/c $< t <$ 2 fm/c, and $t>$ 2 fm/c. The scale corresponds to the number of cells in the PHSD event in the considered bin in $T-\mu_B$ divided by the total number of cells in the corresponding time window while the solid black line is the DQPM phase boundary for orientation. At the very early times $t <$ 0.5 fm/c, the distribution peaks at $T \approx$ 0.3 GeV and is concentrated around $\mu_B \approx 0$. For times 0.5 fm/c $< t <$ 2 fm/c, the average temperature has dropped to about 0.24 GeV, and for later times ($t >$ 2 fm/c), the distribution peaks at an average temperature slightly above $T_c$.
Note that a negative $\mu_B$ implies that there are more antiquarks (antibaryons) than quarks (baryons) in the individual cell. This time evolution of the distribution at the top RHIC energy matches well-known expectations.

Figure \ref{fig-Eval-T-muB-200GeV-158AGeV} (b) shows the distribution in $T$ and $\mu_B$ for cells at midrapidity ($|y_{\text{cell}}| <$ 1) in case of Pb + Pb collisions at 158 A GeV from PHSD for times $t<$ 2 fm/c, 2 fm/c $<t<$ 4 fm/c, and $t>$ 4 fm/c. For early times $t<$ 2 fm/c, the distribution peaks at a temperature of about 0.25 GeV and a sizable chemical potential of about 0.6 GeV, while for times in the interval 2 fm/c $<t<$ 4 fm/c, the maximum has dropped already to an average temperature $\sim$ 0.18 GeV and a chemical potential of about 0.3 GeV. For later times $t>$ 4 fm/c, the distribution (above $T_c$) essentially stays around $\mu_B \approx 0.25$ GeV. We mention that the values of $\mu_B$ probed around the transition temperature $T_c$ in the PHSD are in accordance with the expectation from statistical models which for central Pb + Pb collisions at 158 A\ GeV quote a value of $\mu_B = 0.2489$ GeV \cite{Cleymans:2005xv}. Furthermore, the trajectory of the fireball in the ($T, \mu_B$) plane resembles the isentropic trajectories shown in Ref. \cite{Gunther:2017sxn}, which for this energy corresponds approximately to a fixed ratio of entropy over baryon density of $s/n_B \approx 40$ \cite{Reiter:1998uq} and a fixed ratio of $\mu_B/T \approx 2$ (see Fig. \ref{fig-muBoT_158AGeV}). 


\section{Observables from relativistic nucleus-nucleus collisions}
\label{Section7}

As mentioned above, the PHSD transport
approach~\cite{Cassing:2009vt,Bratkovskaya:2011wp} is a microscopic covariant dynamical
model for strongly interacting systems formulated on the basis of
Kadanoff-Baym equations \cite{Cassing:2008nn}
for Green's functions in phase-space representation (in first-order
gradient expansion beyond the quasiparticle approximation). The
approach consistently describes the full evolution of a relativistic
heavy-ion collision from the initial hard scatterings and string
formation through the dynamical deconfinement phase transition to
the strongly interacting quark-gluon plasma (sQGP) as well as
hadronization and the subsequent interactions in the expanding
hadronic phase as in the hadron-string-dynamics (HSD) transport
approach \cite{Cassing:1999es}. Note that at lower bombarding energies --- without any partonic phase ---
the PHSD approach merges to the HSD approach with only hadronic and string degrees of freedom.
Since we only look for modifications in the partonic sector --- cf. Secs. \ref{Section3} and \ref{Section4} --- we
do not further specify the hadronic sector and refer the reader to Refs. \cite{Cassing:1999es,Cassing:2015owa,Palmese:2016rtq} for details. We recall that in the PHSD4.0 version, the partonic cross sections are parametrized as a function of the energy density to comply with the individual widths of quarks, antiquarks, and gluons (cf. Ref. \cite{Ozvenchuk:2012fn}), while the parton masses are parametrized as a function of the scalar density (cf. Ref. \cite{Bratkovskaya:2011wp}).

\subsection{AGS-SPS energies}

We start with lower and intermediate energies covered experimentally by the AGS (BNL) and SPS (CERN) with a focus on central Au + Au or Pb + Pb collisions. We will compare results for the ``bulk'' observables (rapidity distributions and $p_T$ or $m_T$ spectra) from PHSD calculations based on the default DQPM parameters (PHSD4.0) with the new PHSD5.0 including the differential cross sections from Sec. \ref{Section3} for the individual partonic channels at finite $T$ and $\mu_B$ as well as the parton masses $M_i(T,\mu_B)$ from Eqs. (\ref{Mg9}) and (\ref{Mq9}). A comparison to the available experimental data is included (for orientation) but not discussed explicitly since this has been done in earlier work in detail \cite{Cassing:2015owa,Palmese:2016rtq,Linnyk:2015rco}.

\begin{figure}[h!]
	\centering
	\includegraphics[width=0.92\columnwidth]{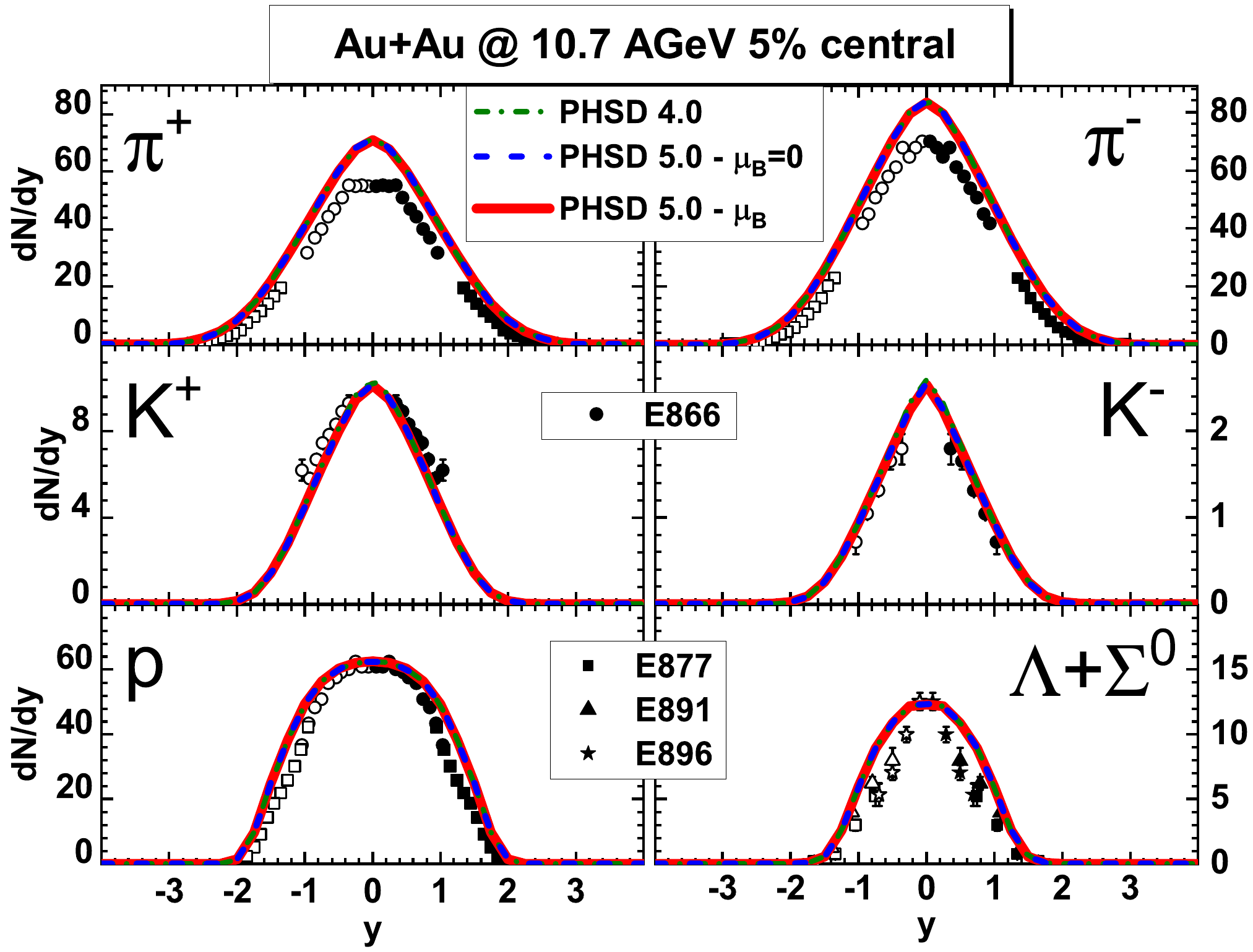}
	\caption{(Color online) The rapidity distributions for  5\% central Au + Au collisions at 10.7 A GeV for PHSD4.0 (green dot-dashed lines), PHSD5.0 with partonic cross sections and parton masses calculated for $\mu_B$ = 0 (blue dashed lines), and with cross sections and parton masses evaluated at the actual chemical potential $\mu_B$ in each individual space-time cell (red lines) in comparison to the experimental data from the E866 \cite{Akiba:1996xf}, E877 \cite{Lacasse:1996gb}, E891 \cite{Ahmad:1991nv}, E877 \cite{Barrette:2000cb}, and E896 \cite{Albergo:2002tn} Collaborations. All PHSD results are the same within the linewidth.}
	\label{fig-dNdy-11AGeV}
\end{figure}

\begin{figure}[h!]
	\centering
	\includegraphics[width=\columnwidth]{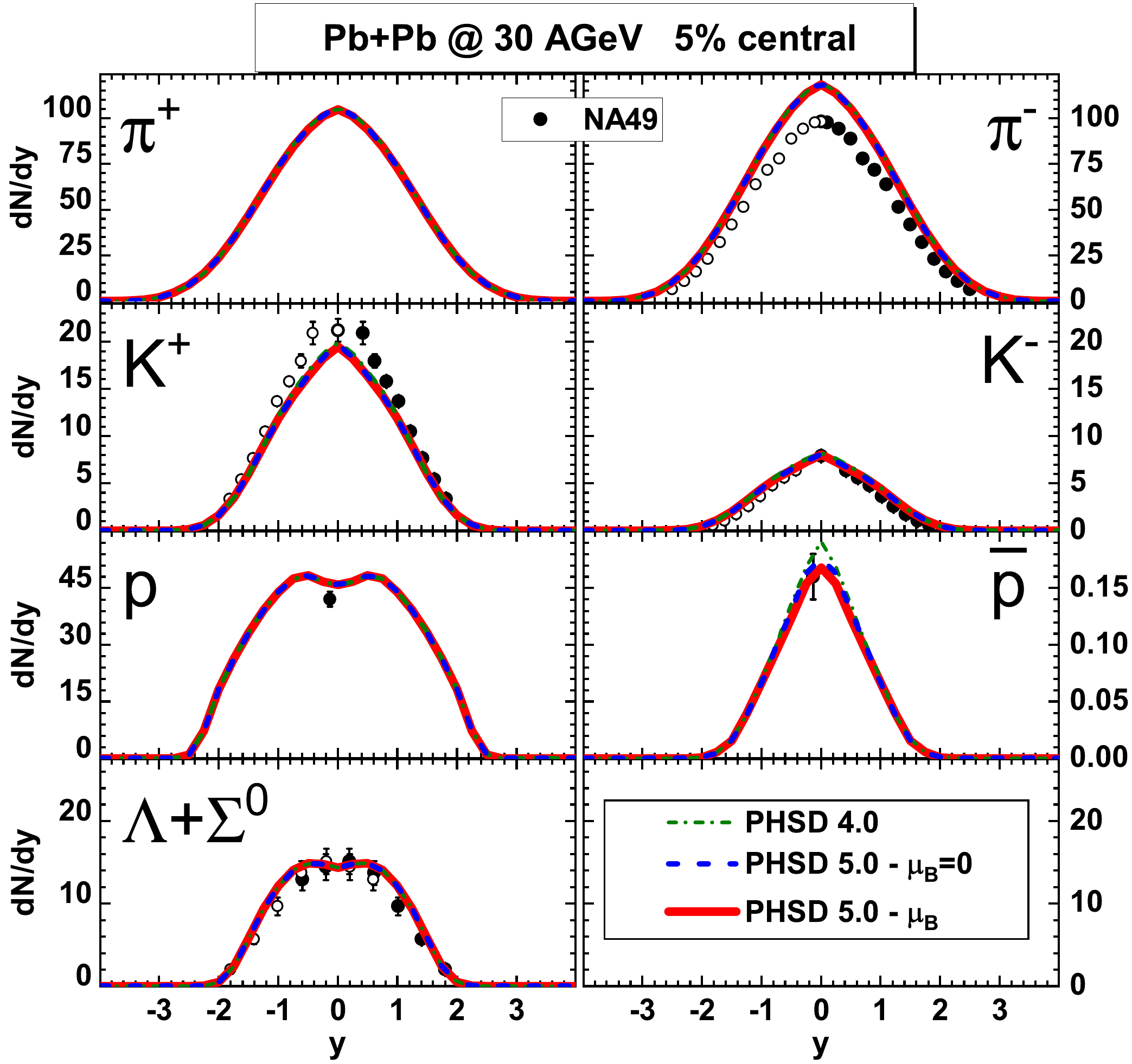}
	\caption{(Color online) The rapidity distributions for 5\% central Pb + Pb collisions at 30 A GeV for PHSD4.0 (green dot-dashed lines), PHSD5.0 with partonic cross sections and parton masses calculated for $\mu_B$ = 0 (blue dashed lines) and with cross sections and parton masses evaluated at the actual chemical potential $\mu_B$ in each individual space-time cell (red lines) in comparison to the experimental data from the NA49 Collaboration \cite{Alt:2006dk,Alt:2007aa,Alt:2008qm}. All PHSD results are practically the same within the linewidth.}
	\label{fig-dNdy-30AGeV}
\end{figure}

\begin{figure}[h!]
	\centering
	\includegraphics[width=\columnwidth]{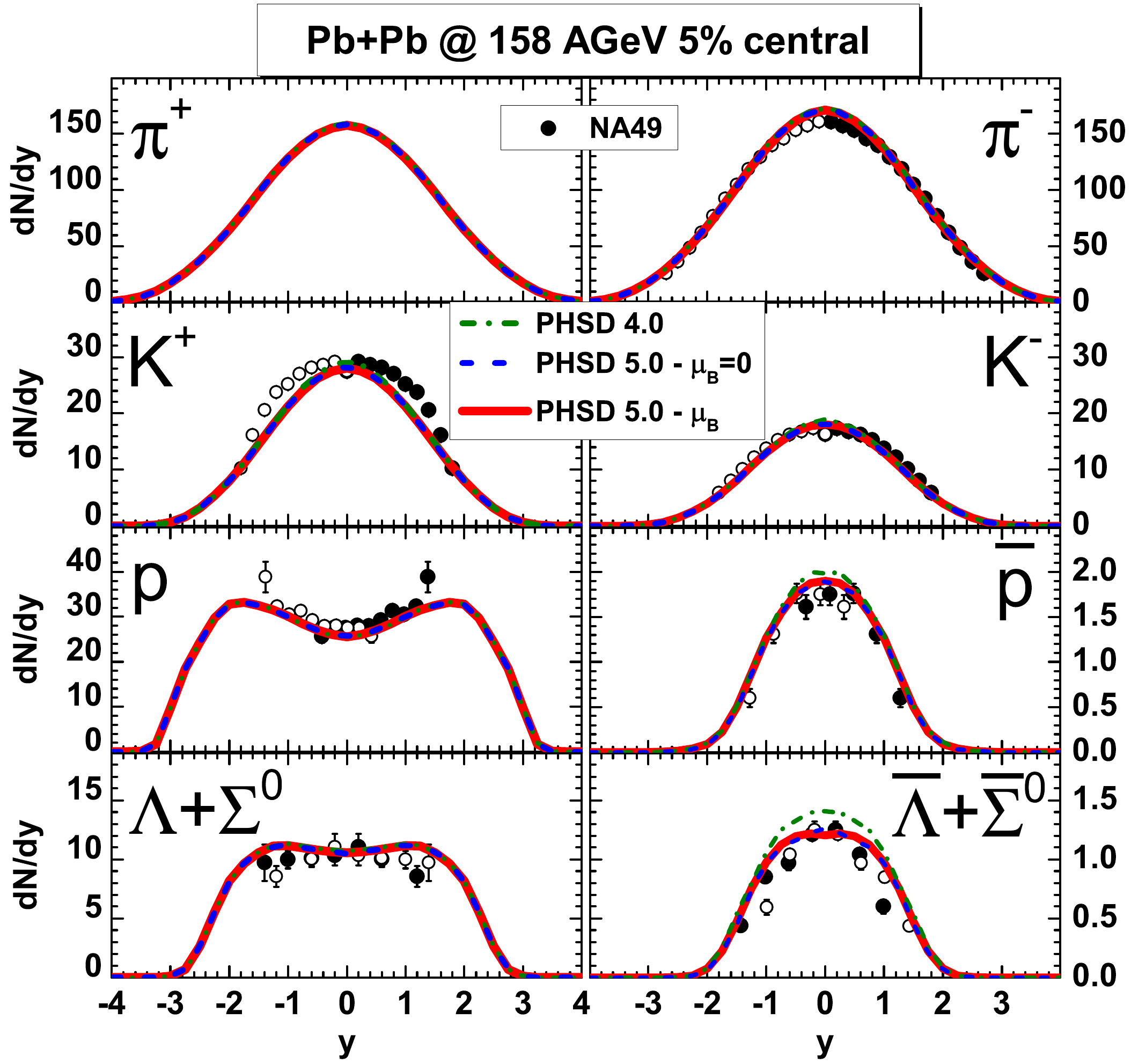}
	\caption{(Color online) The rapidity distributions for 5\% central Pb + Pb collisions at 158 A GeV for PHSD4.0 (green dot-dashed lines), PHSD5.0 with partonic cross sections and parton masses calculated for $\mu_B$ = 0 (blue dashed lines), and with cross sections and parton masses evaluated at the actual chemical potential $\mu_B$ in each individual space-time cell (red lines) in comparison to the experimental data from the NA49 Collaboration \cite{Afanasiev:2002mx,Anticic:2003ux,Anticic:2010mp,Anticic:2012ay}. All PHSD results are the same within the linewidth except for the antibaryons. }
	\label{fig-dNdy-158AGeV}
\end{figure}

Figure \ref{fig-dNdy-11AGeV} displays the actual results for hadronic rapidity distributions in the case of 5\% central Au + Au collisions at 10.7 A GeV for PHSD4.0 (green dot-dashed lines), PHSD5.0 with partonic cross sections and parton masses calculated for $\mu_B$ = 0 (blue dashed lines), and with cross sections and parton masses evaluated at the actual chemical potential $\mu_B$ in each individual space-time cell (red lines) in comparison to the experimental data from the E866 \cite{Akiba:1996xf}, E877 \cite{Lacasse:1996gb}, E891 \cite{Ahmad:1991nv}, E877 \cite{Barrette:2000cb}, and E896 \cite{Albergo:2002tn} Collaborations. Here, we focus on the most abundant hadrons, i.e., pions, kaons, protons, and neutral hyperons. We note in passing that the effects of chiral symmetry restoration are incorporated as in Refs. \cite{Cassing:2015owa,Palmese:2016rtq} since this was found to be mandatory to achieve a reasonable description of the strangeness degrees of freedom reflected in the kaon and neutral hyperon dynamics. As seen from Fig. \ref{fig-dNdy-11AGeV}, there is no difference in rapidity distributions for all the hadron species from the different versions of PHSD within linewidth, which implies that there is no sensitivity to the new partonic differential cross sections and parton masses employed. One could argue that this result might be due to the low amount of QGP produced at this energy, but the different PHSD calculations for 5\% central Pb + Pb collisions at 30 A GeV in Fig. \ref{fig-dNdy-30AGeV} for the hadronic rapidity distributions do not provide a different picture. Only when stepping up to the top SPS energy of 158 A GeV can one identify a small difference in the antibaryon sector (${\bar p}$, ${\bar \Lambda} + {\bar \Sigma}^0$) in the case of 5\% central Pb + Pb collisions (cf. Fig. \ref{fig-dNdy-158AGeV}).

\begin{figure}[h!]
	\centering
	\includegraphics[width=0.95\columnwidth]{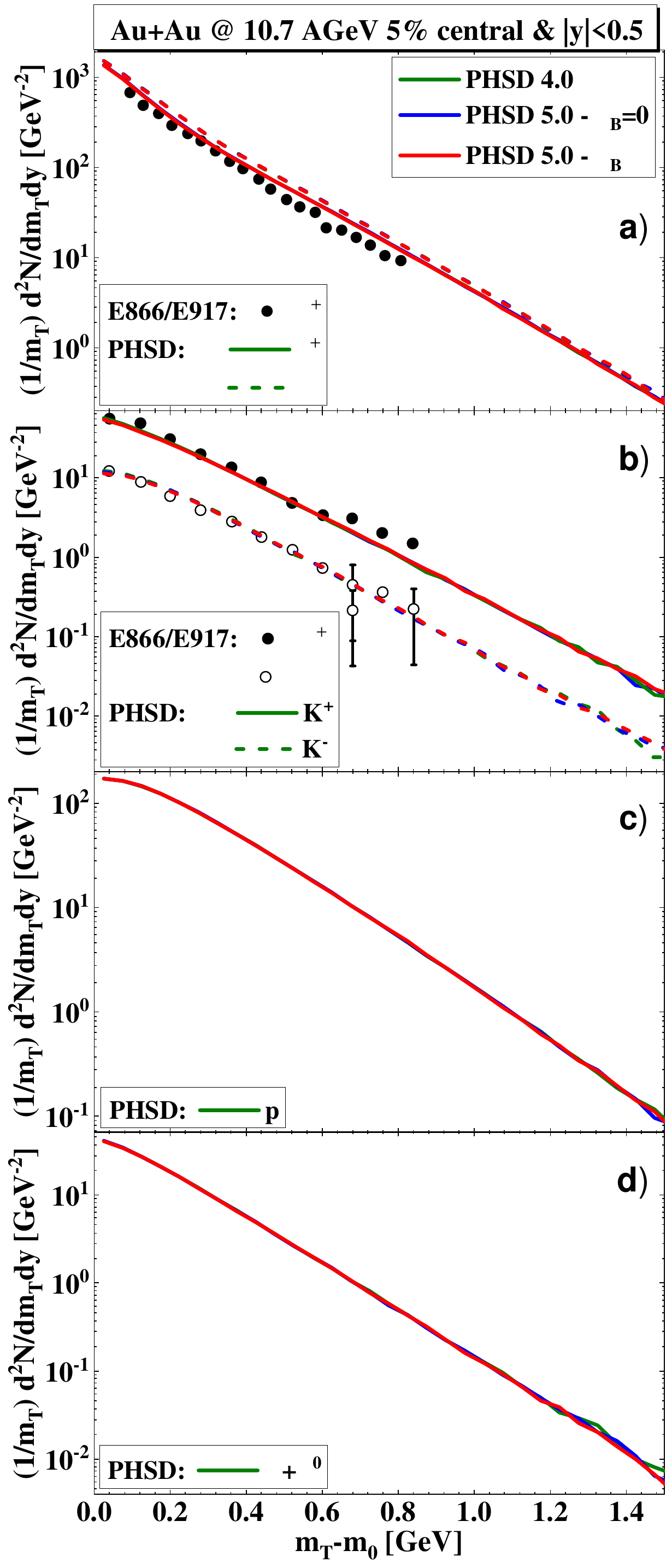}
	\caption{(Color online) The transverse momentum distributions for  5\% central Pb + Pb collisions at 11 A GeV and midrapidity ($|y| < $ 0.5) for PHSD4.0 (green lines), PHSD5.0 with partonic cross sections and parton masses calculated for $\mu_B$ = 0 (blue lines), and with cross sections and parton masses evaluated at the actual chemical potential $\mu_B$ in each individual space-time cell (red lines) in comparison to the experimental data from the E917 and E866 Collaborations \cite{Ahle:1999uy,Ahle:2000wq}.}
	\label{fig-dNdpT-11AGeV}
\end{figure}

\begin{figure}[h!]
	\centering
	\includegraphics[width=0.97\columnwidth]{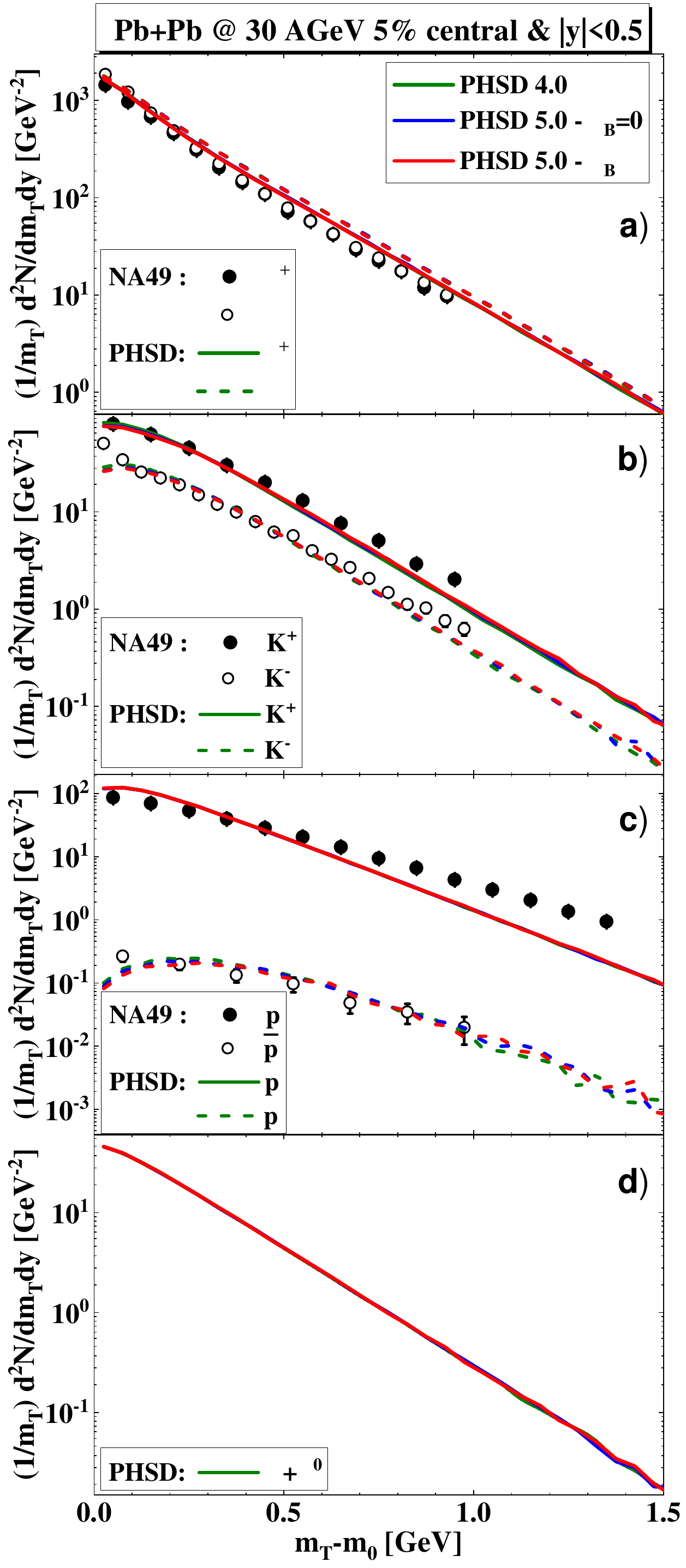}
	\caption{(Color online) The transverse momentum distributions for  5\% central Pb + Pb collisions at 30 A GeV and midrapidity ($|y| < $ 0.5) for PHSD4.0 (green lines), PHSD5.0 with partonic cross sections and parton masses calculated for $\mu_B$ = 0 (blue lines), and with cross sections and parton masses evaluated at the actual chemical potential $\mu_B$ in each individual space-time cell (red lines) in comparison to the experimental data from the NA49 Collaboration \cite{Alt:2006dk,Alt:2007aa,Alt:2008qm}.}
	\label{fig-dNdpT-30AGeV}
\end{figure}

\begin{figure}[h!]
	\centering
	\includegraphics[width=0.91\columnwidth]{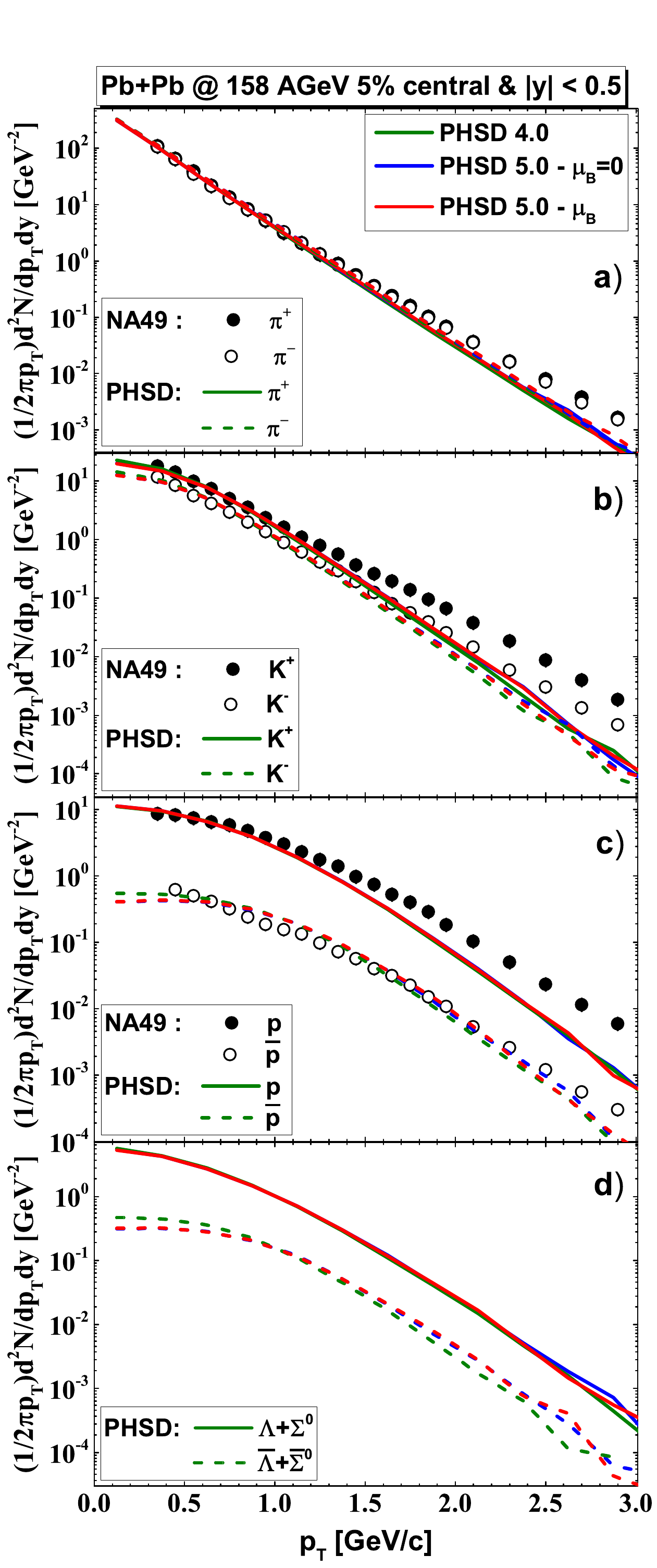}
	\caption{(Color online) The transverse momentum distributions for  5\% central Pb + Pb collisions at 158 A GeV and midrapidity ($|y| < $ 0.5) for PHSD4.0 (green lines), PHSD5.0 with partonic cross sections and parton masses calculated for $\mu_B$ = 0 (blue lines), and with cross sections and parton masses evaluated at the actual chemical potential $\mu_B$ in each individual space-time cell (red lines) in comparison to the experimental data from the NA49 Collaboration \cite{Afanasiev:2002mx,Anticic:2003ux,Anticic:2010mp}.)}
	\label{fig-dNdydNdpt-158AGeV}
\end{figure}

According to the studies above, there is apparently no sizable sensitivity in the hadronic rapidity distributions to the actual differential partonic cross sections, but one has to explore the transverse dynamics
in addition. To this end, we show in Figs. \ref{fig-dNdpT-30AGeV} and \ref{fig-dNdydNdpt-158AGeV} the transverse momentum distributions for  5\% central Pb + Pb collisions at 158 A GeV and midrapidity ($|y| < $ 0.5) for PHSD4.0 (green lines), PHSD5.0 with partonic cross sections and parton masses calculated for $\mu_B$ = 0 (blue lines), and with cross sections and parton masses evaluated at the actual chemical potential $\mu_B$ in each individual space-time cell (red lines) in comparison to the experimental data from the NA49 Collaboration \cite{Alt:2006dk,Alt:2007aa,Alt:2008qm,Afanasiev:2002mx,Anticic:2003ux,Anticic:2010mp}. Here, the solid lines stand for positively charged particles while the dashed lines display the results for negatively charged particles. We find that at 30 A GeV there is practically no change in the $p_T$ spectra for all PHSD versions; only at 158 A GeV do tiny changes in the $p_T$ spectra become visible for transverse momenta above about 2.5 GeV/c. We mention for completeness that again for 10.7 A GeV Au + Au collisions we do not find any changes also in the $p_T$ spectra within the linewidth (cf. Fig. \ref{fig-dNdpT-11AGeV}). Apparently, the space-time volume of the partonic phase is too small at AGS and SPS energies even in central Pb + Pb collisions such that one has practically no sensitivity to the microscopic collisional details in the partonic phase. However, this might change for ultra-relativistic collision systems where the QGP phase becomes dominant.

\subsection{RHIC energies}
\begin{figure}[h!]
	\centering
	\includegraphics[width=\columnwidth]{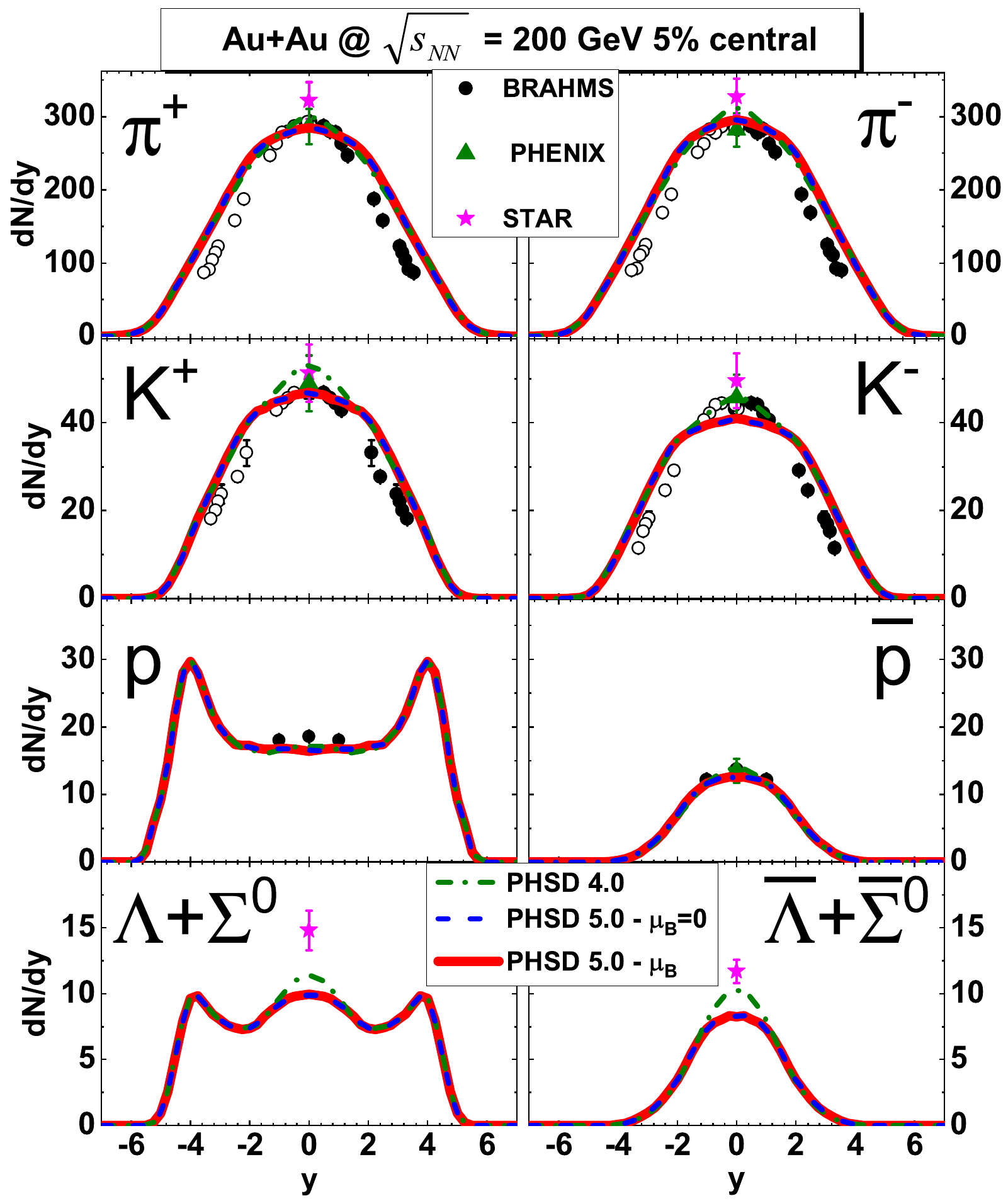}
	\caption{(Color online) The rapidity distributions for  5\% central Au + Au collisions at $\sqrt{s_{NN}}$ = 200 GeV for PHSD4.0 (green dot-dashed lines), PHSD5.0 with partonic cross sections and parton masses calculated for $\mu_B$ = 0 (blue dashed lines) and with cross sections and parton masses evaluated at the actual chemical potential $\mu_B$ in each individual space-time cell (red lines) in comparison to the experimental data from the BRAHMS \cite{Bearden:2004yx,Arsene:2005mr}, PHENIX \cite{Adler:2003cb} and STAR \cite{Agakishiev:2011ar} Collaborations.}
	\label{fig-dNdy-200GeV}
\end{figure}

As demonstrated in Ref. \cite{Bratkovskaya:2011wp}, one expects a dominantly partonic phase in central Au + Au collisions at $\sqrt{s_{NN}}$ = 200 GeV especially when gating on midrapidity. However, the differences between PHSD4.0 and PHSD5.0 (with and without $\mu_B$ dependence) in the hadronic rapidity distributions for 5\% central Au+Au collisions turn out to be rather small for mesons ($\pi^\pm,K^\pm$) and also for baryons and antibaryons ($p,\bar{p},\Lambda+\Sigma^0,\bar{\Lambda}+\bar{\Sigma}^0$) (cf. Fig. \ref{fig-dNdy-200GeV}) such that no robust conclusion on the partonic collisional dynamics can be drawn even in this case.

\begin{figure}[h!]
	\centering
	\includegraphics[width=0.96\columnwidth]{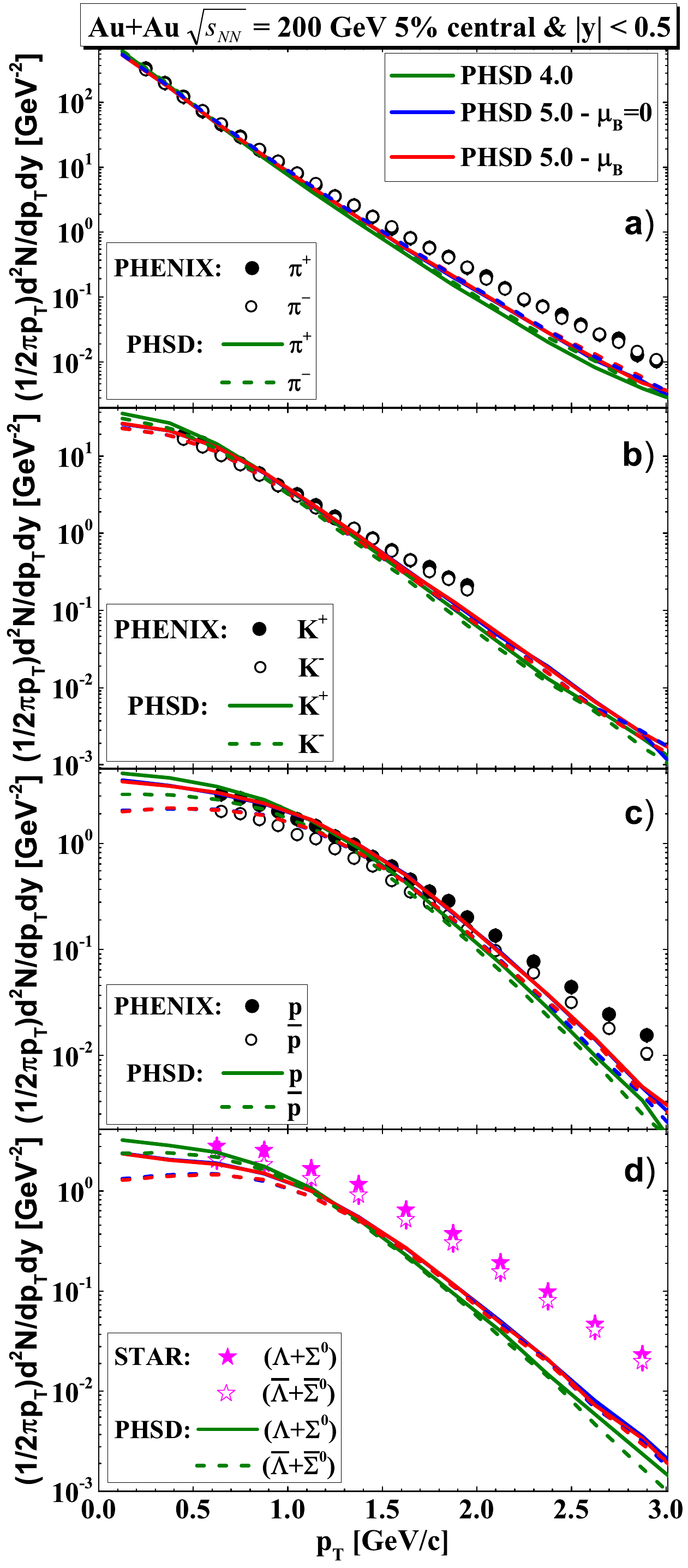}
	\caption{(Color online) The transverse momentum distributions for  5\% central Au+Au collisions at $\sqrt{s_{NN}}$ = 200 GeV and midrapidity ($|y| < $ 0.5) for PHSD4.0 (green lines), PHSD5.0 with partonic cross sections and parton masses calculated for $\mu_B$ = 0 (blue lines), and with cross sections and parton masses evaluated at the actual chemical potential $\mu_B$ in each individual space-time cell (red lines) in comparison to the experimental data from the PHENIX \cite{Adler:2003cb} and STAR \cite{Agakishiev:2011ar} Collaborations.}
	\label{fig-dNdpT-200GeV}
\end{figure}

This also holds true for the transverse momentum distributions at midrapidity ($|y| < $ 0.5) for these collisions when comparing the results from the different PHSD versions with each other and the data from the PHENIX \cite{Adler:2003cb} and STAR \cite{Agakishiev:2011ar} Collaborations in Fig. \ref{fig-dNdpT-200GeV}. Only for high transverse momenta can small differences be seen with the tendency to improve the description of the data in the novel versions of PHSD5.0 with the microscopic differential partonic cross sections.

\subsection{Asymmetric systems}
Since the central collisions of the heavy systems (Au + Au or Pb + Pb) only provide information on the total partonic reaction rate and not details of the partonic collisional dynamics, one has to explore asymmetric heavy-ion collisions --- such as C + Au or Cu + Au --- in addition in order find out a possible sensitivity to the partonic collisions. To this end, we have performed a systematic study of 5\% C + Au and Cu + Au collisions at bombarding energies from AGS to top RHIC energies for the ``bulk'' observables within the different PHSD versions. We note that (without explicit representation) we did not find any difference at 10.7 and 30 A GeV as in the case of the heavy symmetric systems for the hadronic rapidity distributions and transverse momentum spectra at midrapidity. For Cu + Au, the actual results --- with regard to the differences
between PHSD4.0 and PHSD5.0 --- at all bombarding energies
turned out to be very similar to the central Au + Au or Pb + Pb
collisions such that an explicit representation is discarded. Only
in the case of 5\% C + Au reactions at top SPS and top RHIC energies have some
differences been found, which will be discussed in the
following.

\begin{figure}[h!]
	\centering
	\includegraphics[width=\columnwidth]{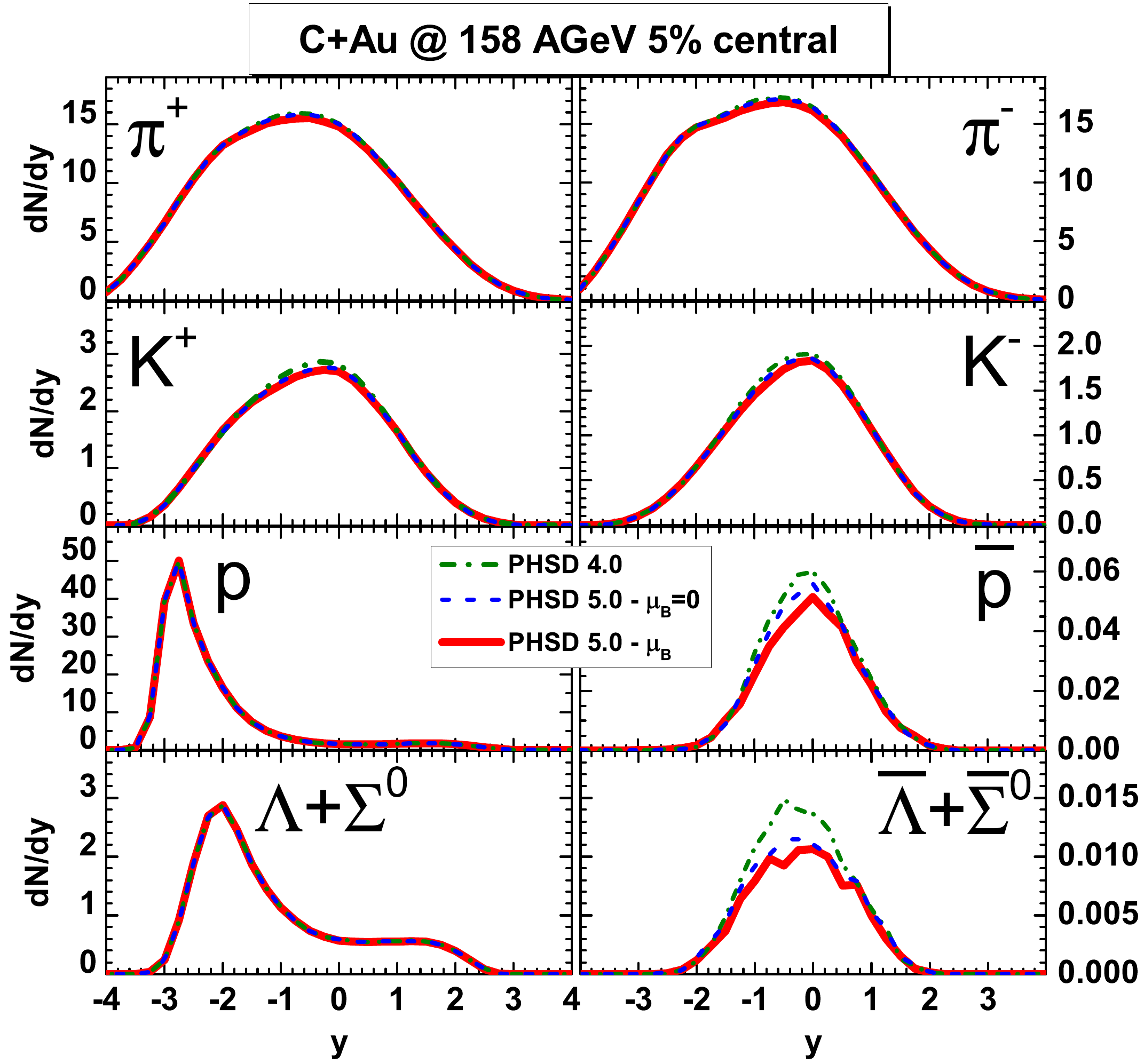}
	\caption{(Color online) The rapidity distributions for 5\% central C + Au collisions at 158 A GeV for PHSD4.0 (green dot-dashed lines), PHSD5.0 with partonic cross sections and parton masses calculated for $\mu_B$ = 0 (blue dashed lines), and with cross sections and parton masses evaluated at the actual chemical potential $\mu_B$ in each individual space-time cell (red lines).}
	\label{dNdy-158GeV-CAu}
\end{figure}

\begin{figure}[h!]
	\centering
	\includegraphics[width=\columnwidth]{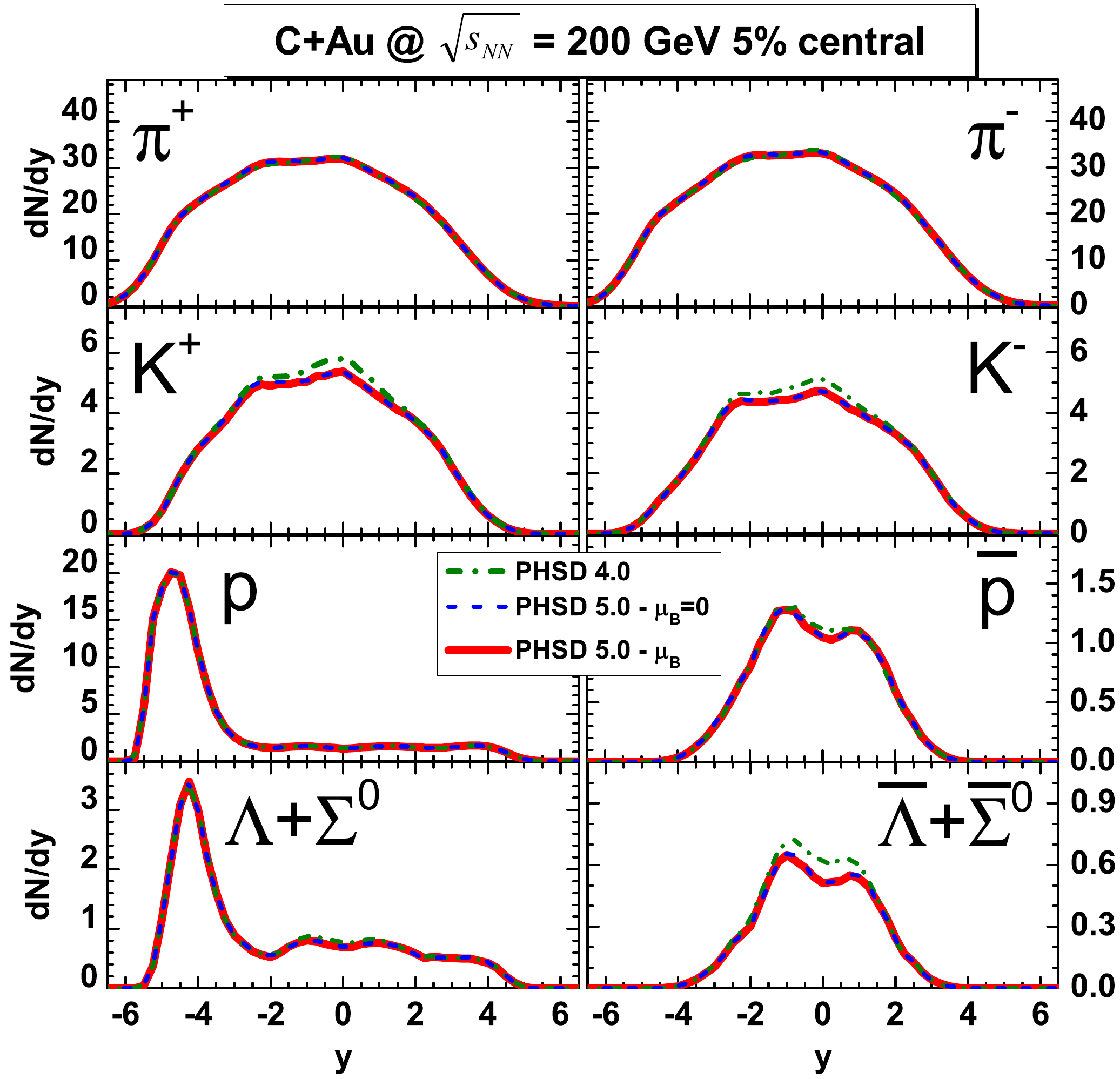}
	\caption{(Color online) The rapidity distributions for 5\% central C + Au collisions at $\sqrt{s_{NN}}$ = 200 GeV for PHSD4.0 (green dot-dashed lines), PHSD5.0 with partonic cross sections and parton masses calculated for $\mu_B$ = 0 (blue dashed lines) and with cross sections and parton masses evaluated at the actual chemical potential $\mu_B$ in each individual space-time cell (red lines).}
	\label{dNdy-200GeV-CAu}
\end{figure}

\begin{figure}[h!]
	\centering
	\includegraphics[width=0.95\columnwidth]{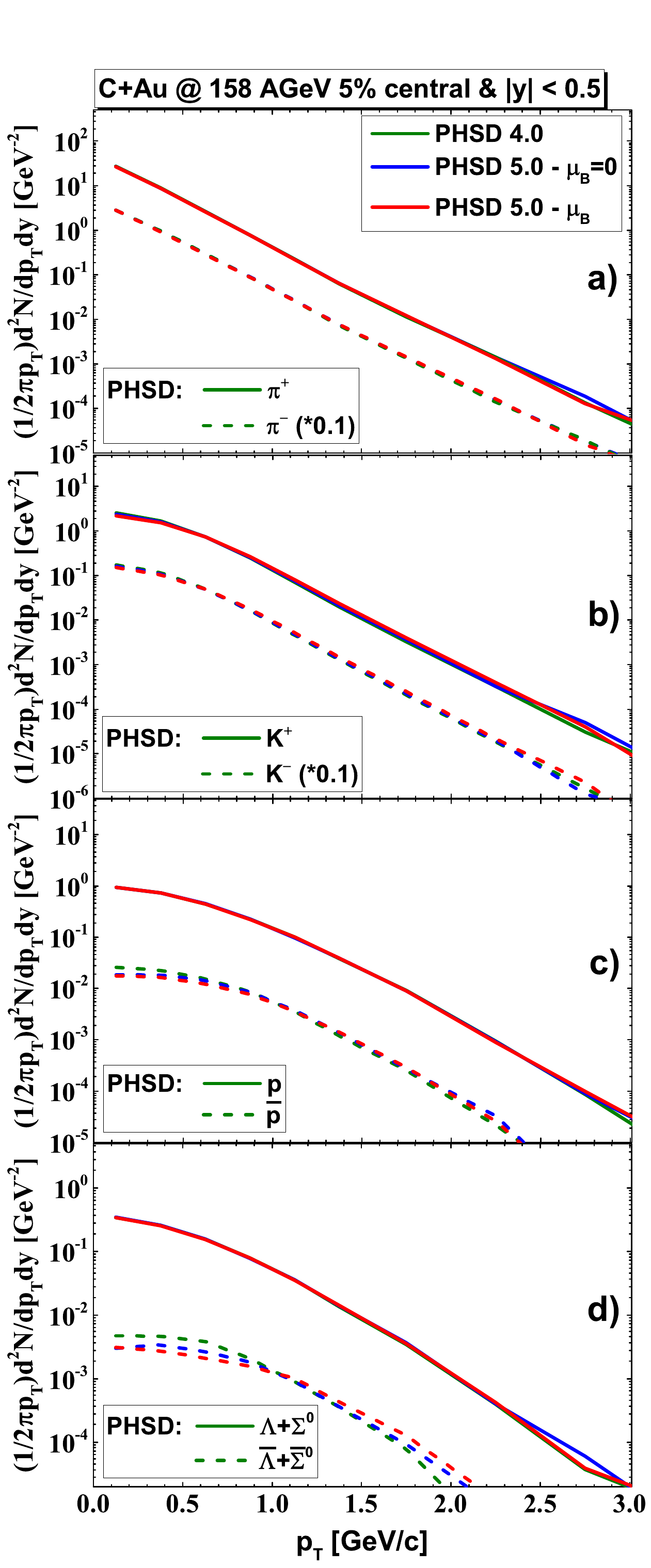}
	\caption{(Color online) The transverse momentum distributions for 5\% central C + Au collisions at 158 A GeV and midrapidity ($|y| < $ 0.5) for PHSD4.0 (green lines), PHSD5.0 with partonic cross sections and parton masses calculated for $\mu_B$ = 0 (blue lines), and with cross sections and parton masses evaluated at the actual chemical potential $\mu_B$ in each individual space-time cell (red lines).}
	\label{dNdpT-158GeV-CAu}
\end{figure}

\begin{figure}[h!]
	\centering
	\includegraphics[width=0.95\columnwidth]{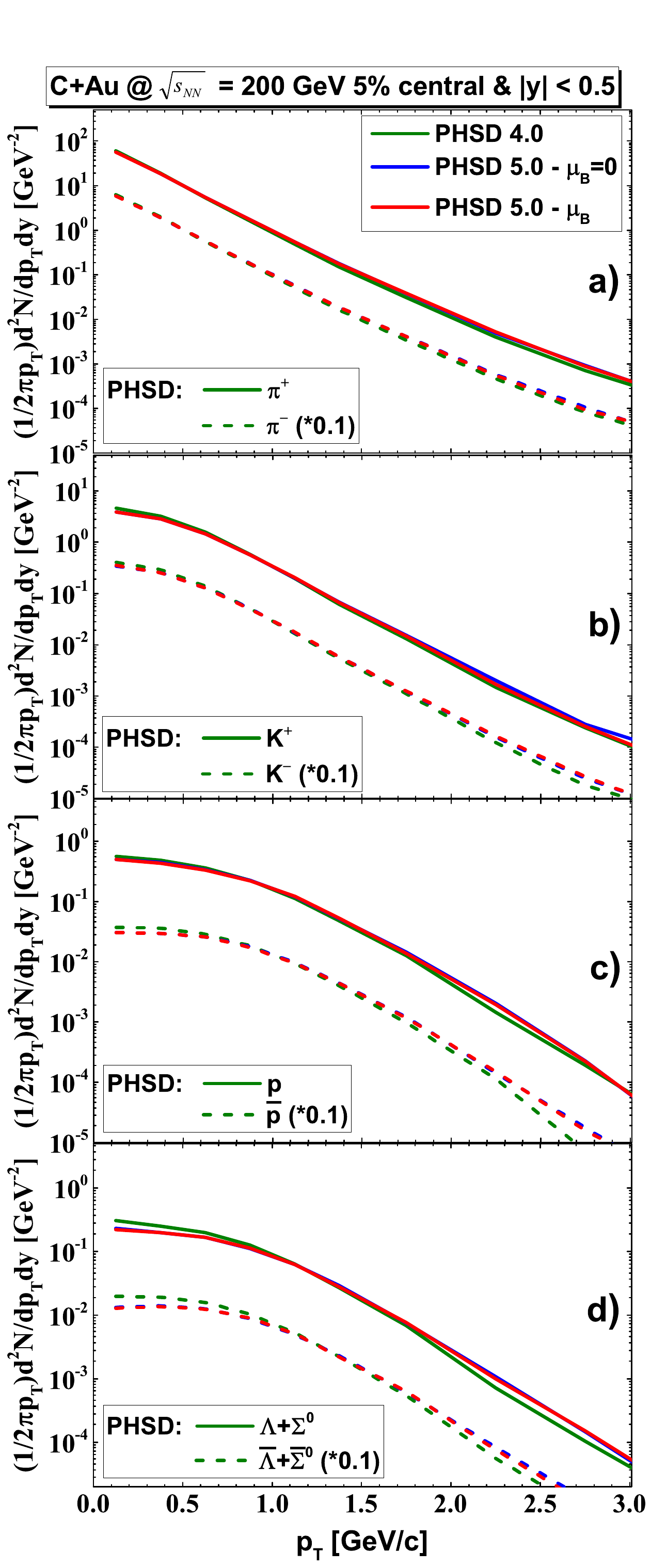}
	\caption{(Color online) The transverse momentum distributions for  5\% central C + Au collisions at $\sqrt{s_{NN}}$ = 200 GeV and midrapidity ($|y| < $ 0.5) for PHSD4.0 (green lines), PHSD5.0 with partonic cross sections and parton masses calculated for $\mu_B$ = 0 (blue lines), and with cross sections and parton masses evaluated at the actual chemical potential $\mu_B$ in each individual space-time cell (red lines).}
	\label{dNdpT-200GeV-CAu}
\end{figure}

The rapidity distributions of hadrons for 5\% central C + Au collisions are displayed in Figs. \ref{dNdy-158GeV-CAu} and  \ref{dNdy-200GeV-CAu} for 158 A GeV and $\sqrt{s_{NN}}$ = 200 GeV, respectively. Note that the rapidity distributions are no longer symmetric in rapidity $y$ but enhanced for $y <$ 0 (Au-going side). There is no change of the pion and baryon distributions at both energies for the different PHSD versions as in the case of the heavy symmetric systems while tiny differences can again be seen in the antibaryon spectra. However, in the case of C + Au, now there is also a small signal in the kaon rapidity distributions which is more pronounced at $\sqrt{s_{NN}}$ = 200 GeV. This suggests that the strangeness degree of freedom might be explored in very asymmetric systems to obtain additional information on the partonic scattering dynamics.

The transverse momentum spectra of hadrons at midrapidity (for C + Au) are shown in Figs. \ref{dNdpT-158GeV-CAu} and  \ref{dNdpT-200GeV-CAu} for 158 A GeV and $\sqrt{s_{NN}}$ = 200 GeV, respectively.
There is practically no difference in the PHSD4.0 and PHSD5.0 results for pions, kaons, protons, and antiprotons and only a very small signal in the antihyperons can be identified. Nevertheless, our results for this very asymmetric system can be considered as predictions for the production of the most abundant hadron species at top SPS and RHIC energies.

\section{Summary}
\label{Section8}

In this work, we have extended the PHSD transport approach (PHSD4.0 \cite{Linnyk:2015rco,Palmese:2016rtq}) to incorporate differential ``off-shell cross sections'' for all binary partonic channels that are based on the same effective propagators and couplings as employed in the QGP equation of state and the parton propagation. To this end, we have recalled the extraction of the partonic masses and the coupling $g^2$ from lattice QCD data (within the DQPM) and calculated the partonic differential cross sections as a function of $T$ and $\mu_B$ for the leading tree-level diagrams (cf. the \hyperref[Appendix]{Appendixes}). Furthermore, in Sec. \ref{Section4}, we have used these differential cross sections to evaluate partonic scattering rates for fixed $T$ and $\mu_B$ as well as to compute the ratio of the shear viscosity $\eta$ to entropy density $s$ within the Kubo formalism in comparison to calculations from lQCD.
It turns out that the ratio $\eta/s$ calculated with the partonic scattering rates in the relaxation-time approximation is very similar to the original result from the DQPM and to lQCD results such that the present extension of the approach does not lead to different partonic transport properties except for temperatures close to $T_c$. 
We recall that the novel PHSD version (PHSD5.0) is practically parameter free in the partonic sector since the effective coupling (squared) is determined by a fit to the scaled entropy density from lQCD. The dynamical masses for quarks and gluons then are fixed by the HTL expressions. The interaction rate in the timelike sector is, furthermore, calculated in leading order employing the DQPM propagators and coupling. 


When implementing the differential cross sections and parton masses into the PHSD5.0 approach, one has to specify the Lagrange parameters $T$ and $\mu_B$ in each computational cell in space-time. This has been done by employing a state-of-the-art lattice QCD equation of state \cite{Gunther:2017sxn} and a diagonalization of the energy-momentum tensor from PHSD as described in Sec. \ref{Section5}. Detailed results for $T$ and $\mu_B$ have been presented for central collisions of Pb + Pb at $\sqrt{s_{NN}}$ = 17.3 and Au + Au at $\sqrt{s_{NN}}$ = 200 GeV in the ($T, \mu_B$) plane as a function of reaction time. It turns out that the evolution of the QGP phase from the PHSD approximately follows the expectation from the isentropic trajectory $s/n_B \approx 40$ at $\sqrt{s_{NN}}$ = 17.3 for $T$ above $T_c$ while at the top RHIC energy the distribution in $T$ and $\mu_B$ spreads around zero for all reaction times considered.

In Sec. \ref{Section6} we then have calculated 5\% central Au + Au (or Pb + Pb) collisions and compared the results for hadronic rapidity distributions and transverse momentum spectra (at midrapidity) from the previous PHSD4.0 with the novel version PHSD5.0 (with and without the explicit dependence of the partonic differential cross sections and parton masses on $\mu_B$).  No differences for all the hadron ``bulk'' observables from the various PHSD versions have been found at AGS and FAIR/NICA energies within linewidth, which implies that there is no sensitivity to the new partonic differential cross sections employed. Only in the case of the kaons and the antibaryons $\bar{p}$ and $\bar{\Lambda} + \bar{\Sigma}^0$ could a small difference between PHSD4.0 and PHSD5.0 be seen at top SPS and top RHIC energies; however, there was no clear difference between the PHSD5.0 calculations with partonic cross sections for $\mu_B$ = 0  and actual $\mu_B$ in the local cells. When considering very asymmetric collisions of C + Au, a small sensitivity to the partonic scatterings was found in the kaon and antibaryon rapidity distributions, too.
However, it will be very hard to extract a robust signal experimentally.

Our findings can be understood as follows:
The fact that we find only small traces
of the $\mu_B$ dependence of
partonic scattering dynamics
in heavy-ion ``bulk'' observables --- although the
differential cross sections and parton masses clearly depend on $\mu_B$ --- implies that one needs a sizable partonic density and large space-time QGP volume to explore the dynamics in the QGP phase. These conditions are only fulfilled at high bombarding energies (top SPS, RHIC energies) where, however, $\mu_B$ is rather low. On the other hand, decreasing the bombarding energy to FAIR-NICA energies and thus increasing $\mu_B$ lead to collisions that are dominated by the hadronic phase where the extraction of information about the parton dynamics will be rather complicated based on ``bulk'' observables. Further investigations of other observables (such as flow coefficients $v_n$ of particles and antiparticles, fluctuations, and correlations) might contain more visible ``$\mu_B$ traces'' from the QGP phase and will be the subject of a forthcoming study.

\section*{Acknowledgements}
The authors acknowledge inspiring discussions with J. Aichelin, H. Berrehrah, C. Ratti, and T. Steinert. This
work was supported  by the LOEWE center ``HIC for FAIR''. Furthermore, P.M., L.O.,
and E.B. acknowledge support by the Deutsche Forschungsgemeinschaft (DFG, German Research Foundation) through Grant CRC-TR 211 ``Strong-interaction matter under extreme conditions'', Project No. 315477589 - TRR 211. O.S. acknowledges support from HGS-HIRe for FAIR; L.O. and E.B. thank the COST Action THOR, CA15213.
The computational resources have been provided by the LOEWE Center for Scientific Computing.


\begin{widetext}

\appendix

\label{Appendix}

\section{Matrix elements for $qq' \rightarrow qq'$ scattering}
\label{AppendixA}

Here, we give the details on the calculation of the matrix elements used to evaluate the DQPM partonic cross sections which are based on Refs. \cite{Cutler:1977qm,Bengtsson:1984yx,Berrehrah:2013mua}. We recall that the Mandelstam variables are given by the momenta as $s=(k_i+p_i)^2=(k_f+p_f)^2$, $t=(k_i-k_f)^2=(p_i-p_f)^2$, $u=(k_i-p_f)^2=(p_i-k_f)^2$. The generators of SU(3) associated with QCD are denoted by the matrices $T^a = \lambda^a/2$ with $a$ being the gluon color and $\lambda^a$ the Gell-Mann matrices \cite{GellMann:1962xb}. The Lie algebra formed by the generators $T^a$ is given by the commutation relation $\left[ T^a,T^b \right] = if^{abc}T^c$ where $f^{abc}$ are the structure constants. We refer the reader to Ref. \cite{Schwartz:2013pla} where all the rules for calculating the color factors in the following calculations are given in detail. The $\gamma$ matrices are denoted by $\gamma^\mu$ and the Dirac spinors are $u$ for particles and $v$ for antiparticles. The final analytical expressions for the matrix elements used in the PHSD code were evaluated using FeynCalc \cite{MERTIG1991345,SHTABOVENKO2016432}.

\begin{figure*}[h!]
	\centering
	\includegraphics[width=0.75\linewidth]{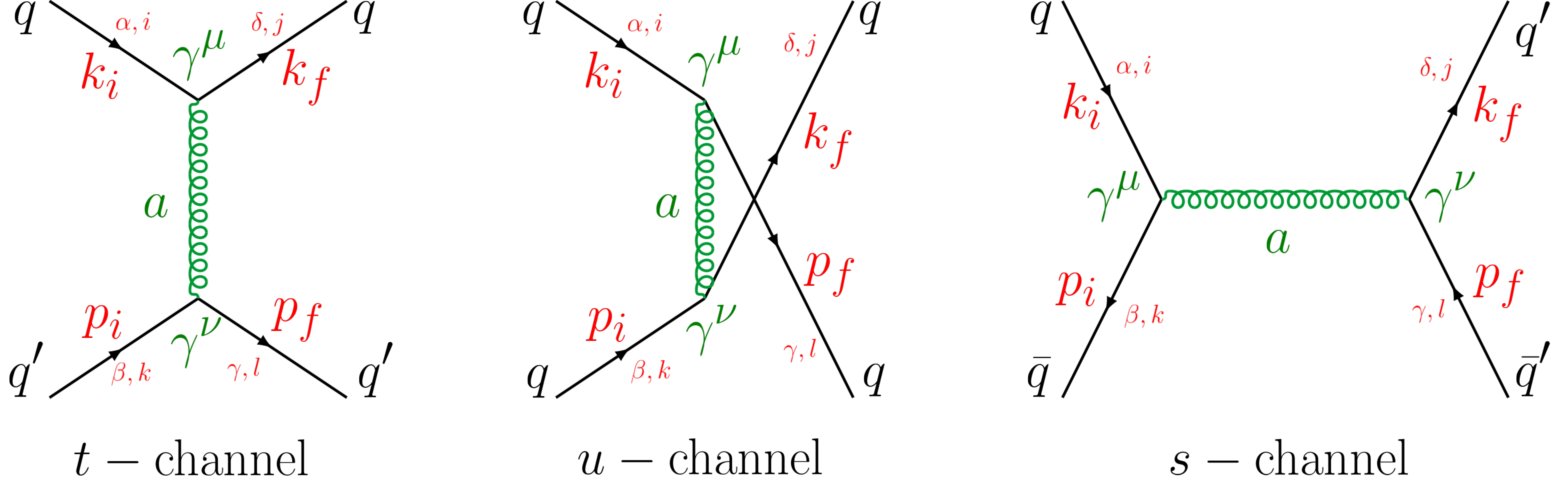}
	\caption{Leading-order Feynman diagrams for the $q q' \rightarrow q q' $ and $q \bar{q} \rightarrow q' \bar{q}' $ processes. The initial and final 4-momenta are $k_i$ and $p_i$, and $k_f$ and $p_f$, respectively. The indices ${\color{Red} i,j,k,l}=$ 1--3 denote the quark colors, ${\color{ForestGreen}a}=$ 1--8 denote the gluon colors, while the quark flavor is indicated by the indices ${\color{Red}\alpha,\beta,\delta,\gamma}=u,d,s,...$.}
	\label{fig-feynman1}
\end{figure*}

The invariant matrix elements corresponding to Fig. \ref{fig-feynman1} are given by the following expressions:

\begin{equation}
i  \mathcal{M}_t (q^i_\alpha q^k_\beta \rightarrow q^j_\delta q^l_\gamma)  = \delta_{\alpha \delta} \ \delta_{\beta \gamma}\ \bar{u}^j_\delta(k_f) (-ig\gamma^\mu T^a_{ij}) u^i_\alpha(k_i) \left[ -i \frac{g_{\mu\nu}-(q^t_\mu q^t_\nu)/M_g^2}{(k_f-k_i)^2-M_g^2+2i\gamma_g \omega_t} \right]  \bar{u}^l_\gamma(p_f) (-ig\gamma^\nu T^a_{kl}) u^k_\beta(p_i) ,
\end{equation}

\begin{equation}
i \mathcal{M}_u (q^i_\alpha q^k_\beta \rightarrow q^j_\delta q^l_\gamma)  = - \delta_{\alpha \beta} \ \delta_{\alpha \delta} \ \delta_{\beta \gamma}\ \bar{u}^j_\delta(k_f) (-ig\gamma^\nu T^a_{kj}) u^k_\beta(p_i) \left[ -i \frac{g_{\mu\nu}-(q^u_\mu q^u_\nu)/M_g^2}{(p_f-k_i)^2-M_g^2+2i\gamma_g \omega_u} \right]  \bar{u}^l_\gamma(p_f) (-ig\gamma^\mu T^a_{il}) u^i_\alpha(k_i) ,
\end{equation}

\begin{equation}
i \mathcal{M}_s (q^i_\alpha q^k_\beta \rightarrow q^j_\delta q^l_\gamma)  = - \delta_{\alpha \bar{\beta}}\ \delta_{\delta \bar{\gamma}}\ \bar{u}^j_\delta(k_f) (-ig\gamma^\nu T^a_{lj}) v^l_\gamma(p_f) \left[ -i \frac{g_{\mu\nu}-(q^s_\mu q^s_\nu)/M_g^2}{(k_i+p_i)^2-M_g^2+2i\gamma_g \omega_s} \right]  \bar{v}^k_\beta(p_i) (-ig\gamma^\mu T^a_{ik}) u^i_\alpha(k_i) ,
\end{equation}
~\\
where the energy of the exchanged gluon is $\omega_t = |k^0_f - k^0_i|$, $\omega_u = |p_f^0-k_i^0|$, and $\omega_s = |k_i^0+p_i^0|$, and its momentum denoted by $q_t^\mu = (k_f-k_i)^\mu$, $q_u^\mu = (p_i-k_f)^\mu$, and $q_s^\mu = (k_i+p_i)^\mu$. We note here that the matrix element $\mathcal{M}_t$ given above only corresponds to a $q$--$q$ elastic scattering, but the final contribution from $t$-channel diagrams to $|\overline{\mathcal{M}}|^2$ is found to be the same whether a quark scatters with a quark or antiquark since the amplitudes are averaged (summed) over initial (final) partons as:

\begin{align}
|\bar{\mathcal{M}} (q_\alpha q_\beta \rightarrow q_\delta q_\gamma)|^2 =\ & \frac{1}{d^2_q} \sum_{\text{color}} \sum_{\text{spin}}  | \mathcal{M}_t+\mathcal{M}_u+\mathcal{M}_s|^2 \nonumber \\
=\ & \frac{1}{3 \cross 3} \sum_{\text{color}} \frac{1}{2 \cross 2} \sum_{\text{spin}} \left( |\mathcal{M}_t|^2 + |\mathcal{M}_u|^2 + 2 \Re \left[ \mathcal{M}_t \mathcal{M}^\star_u \right] + 2 \Re \left[ \mathcal{M}_t \mathcal{M}^\star_s \right] \right) .
\end{align}

\section{Matrix elements for $gq \rightarrow gq$ scattering}
\label{AppendixB}

\begin{figure}[h!]
	\centering
	\includegraphics[width=0.75\columnwidth]{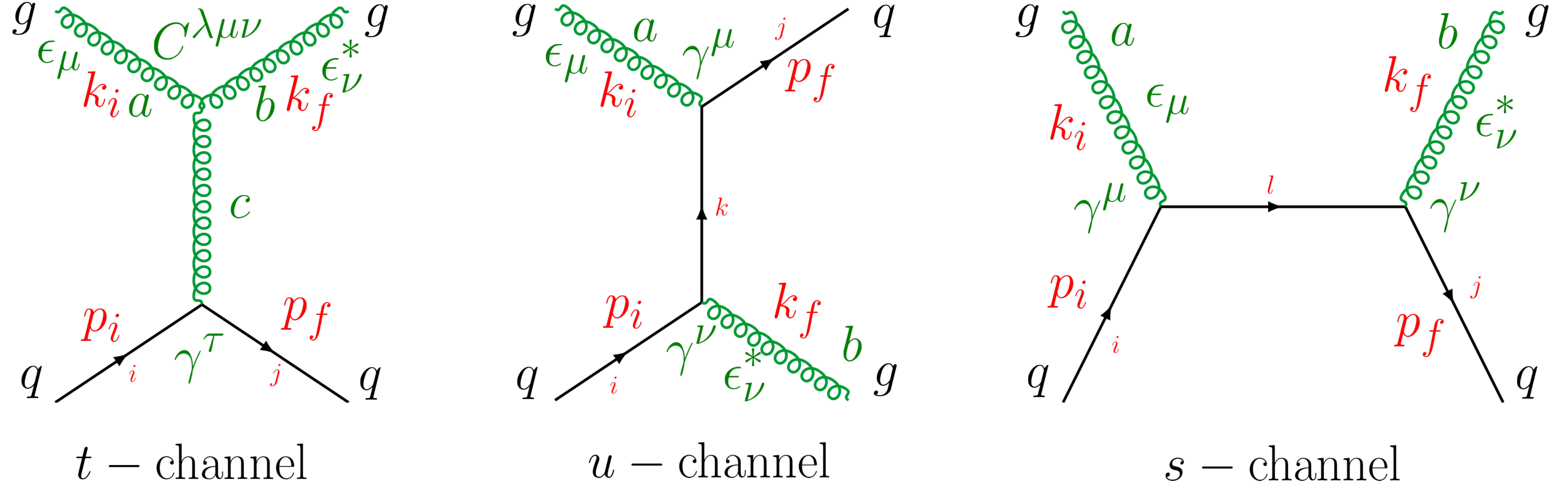}
	\caption{Leading-order Feynman diagrams for the $g q \rightarrow g q $ processes. The initial and final 4-momenta are $k_i$ and $p_i$, and $k_f$ and $p_f$, respectively. The indices ${\color{Red} i,j,k,l}=$ 1--3 denote the quark colors, ${\color{ForestGreen}a,b,c}=$ 1--8 indicate the gluon colors, while the quark flavor is indicated by the indices ${\color{Red}\alpha,\beta,\delta,\gamma}=u,d,s,...$.}
	\label{fig-feynman2}
\end{figure}

The invariant matrix elements corresponding to the Feynman diagrams in Fig. \ref{fig-feynman2} are given by the following expressions:

\begin{equation}
i\mathcal{M}_t (g^a q^i \rightarrow g^b q^j)  = (\epsilon^\star_{b,f})_\nu \left(-gf^{cab} C^{\lambda \mu \nu}(k_i-k_f,-k_i,k_f)\right) (\epsilon_{a,i})_\mu \left[ -i \frac{g_{\lambda \tau}-(q^t_\lambda q^t_\tau)/M_g^2}{(k_f-k_i)^2-M_g^2+2i\gamma_g \omega_t} \right]  \bar{u}^j(p_f) (-ig\gamma^\tau T^c_{ij}) u^i(p_i) ,
\end{equation}

\begin{equation}
i\mathcal{M}_u (g^a q^i \rightarrow g^b q^j) =  \bar{u}^j(p_f) (-ig\gamma^\mu T^a_{kj}) (\epsilon_{a,i})_\mu \left[ i \frac{\slashed{q}^u+M_q}{u-M_q^2+2i\gamma_q \omega_u} \right]  (\epsilon^\star_{b,f})_\nu (-ig\gamma^\nu T^b_{ik}) u^i(p_i) ,
\end{equation}

\begin{equation}
i\mathcal{M}_s (g^a q^i \rightarrow g^b q^j) =  \bar{u}^j(p_f) (-ig\gamma^\nu T^b_{lj}) (\epsilon^\star_{b,f})_\nu \left[ i \frac{\slashed{q}^s+M_q}{s-M_q^2+2i\gamma_q \omega_s} \right]  (\epsilon_{a,i})_\mu (-ig\gamma^\mu T^a_{il}) u^i(p_i) .
\end{equation}
~\\
with the 3-gluon vertex $C^{\lambda \mu \nu}(q_1,q_2,q_3) = \left[ (q_1-q_2)^\nu g^{\lambda \mu} + (q_2-q_3)^\lambda g^{\mu \nu} + (q_3-q_1)^\mu g^{\lambda \nu} \right]$ and the momentum of the exchanged gluon $q_t^\mu = (k_f-k_i)^\mu$ in the $t$-channel. \\

In the case of a massive gluon, the sum over polarizations is given by [in accordance with the denominator of the propagator in Eq. (\ref{propg})]
\begin{equation}
\sum_{\text{pol.}} (\epsilon_i)_\mu (\epsilon^\star_i)_{\mu'} = - g_{\mu\mu'} + \frac{(k_i)_{\mu}(k_i)_{\mu'}}{(M_i)^2_g} ,
\end{equation}

The invariant matrix element squared averaged (summed) over the initial (final) partons is:

\begin{align}
|\overline{\mathcal{M}}  (g q \rightarrow g q)|^2 & = \frac{1}{d_g d_q} \sum_{\text{color}}   \sum_{\text{spin}}  | \mathcal{M}_t+\mathcal{M}_u+\mathcal{M}_s|^2  \\
& = \frac{1}{8 \cross 3} \sum_{\text{color}} \frac{1}{2\times 2} \sum_{\text{spin}} \left( |\mathcal{M}_t|^2 + |\mathcal{M}_u|^2 + |\mathcal{M}_s|^2 + 2\Re \left[ \mathcal{M}_t \mathcal{M}^\star_u \right] + 2\Re \left[ \mathcal{M}_t \mathcal{M}^\star_s \right] + 2\Re \left[ \mathcal{M}_u \mathcal{M}^\star_s \right] \right) \nonumber .
\end{align}

\section{Matrix elements for $gg \rightarrow gg$ scattering}
\label{AppendixC}

\begin{figure}[h!]
	\centering
	\includegraphics[width=0.85\columnwidth]{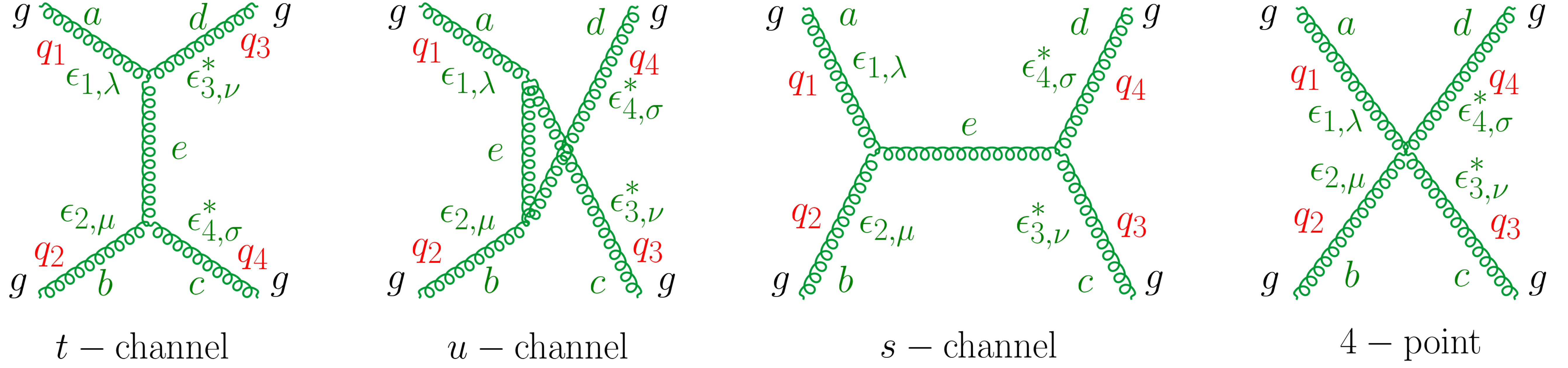}
	\caption{Leading-order Feynman diagrams for the $g g \rightarrow g g $ scatterings. The initial and final 4-momenta are $k_i$ and $p_i$, and $k_f$ and $p_f$, respectively. The indices ${\color{ForestGreen}a,b,c,d,e}=$1--8 denote the gluon colors.}
	\label{fig-feynman3}
\end{figure}

The invariant matrix elements corresponding to the Feynman diagrams in Fig. \ref{fig-feynman3} are given by the following expressions:

\begin{align}
i \mathcal{M}_t (g^a g^b \rightarrow g^c g^d)  = & (\epsilon^\star_{d,4})_\sigma \left(-gf^{ead} C^{\tau \lambda \sigma}(q_1-q_4,-q_1,q_4)\right) (\epsilon_{a,1})_\lambda \nonumber \\
\left[ -i \frac{g_{\tau \tau'}-(q^t_\tau q^t_{\tau'})/M_g^2}{(q_4-q_1)^2-M_g^2+2i\gamma_g \omega_t} \right]  & (\epsilon^\star_{c,3})_\nu \left(-gf^{ecb} C^{\tau' \nu \mu}(-q_3+q_2,q_3,-q_2)\right) (\epsilon_{b,2})_\mu ,
\end{align}
\begin{align}
i \mathcal{M}_s (g^a g^b \rightarrow g^c g^d)  = & (\epsilon^\star_{d,4})_\sigma \left(-gf^{edc} C^{\tau' \sigma \nu}(-q_4-q_3,q_4,q_3)\right) (\epsilon^\star_{c,3})_\nu \nonumber \\
\left[ -i \frac{g_{\tau \tau'}-(q^s_\tau q^s_{\tau'})/M_g^2}{(q_1+q_2)^2-M_g^2+2i\gamma_g \omega_s} \right]  & (\epsilon_{b,2})_\mu \left(-gf^{eba} C^{\tau \mu \lambda}(q_2+q_1,-q_2,-q_1)\right) (\epsilon_{a,1})_\lambda
\end{align}
\begin{align}
i \mathcal{M}_u (g^a g^b \rightarrow g^c g^d)  = & (\epsilon^\star_{d,4})_\sigma \left(-gf^{edb} C^{\tau' \sigma \mu}(-q_4+q_2,q_4,-q_2)\right) (\epsilon_{b,2})_\mu \nonumber \\
\left[ -i \frac{g_{\tau \tau'}-(q^u_\tau q^u_{\tau'})/M_g^2}{(q_2-q_4)^2-M_g^2+2i\gamma_g \omega_u} \right]  & (\epsilon^\star_{c,3})_\nu \left(-gf^{eac} C^{\tau \lambda \nu}(q_1-q_3,-q_1,q_3)\right) (\epsilon_{a,1})_\lambda ,
\end{align}
~\\
with the 3-gluon vertex $C^{\lambda \mu \nu}(q_1,q_2,q_3) = \left[ (q_1-q_2)^\nu g^{\lambda \mu} + (q_2-q_3)^\lambda g^{\mu \nu} + (q_3-q_1)^\mu g^{\lambda \nu} \right]$ and the momentum of the exchanged gluon $q_t^\mu = (q_4-q_1)^\mu$, $q_s^\mu = (q_1+q_2)^\mu$, and  $q_u^\mu = (q_2-q_4)^\mu$. \\

The 4-point invariant amplitude is given by (according to Refs. \cite{Cutler:1977qm,Bengtsson:1984yx})
\begin{align}
i\mathcal{M}_4 (g^a g^b \rightarrow g^c g^d) & = -ig^2 \left[ f^{abe} f^{cde} (g^{\lambda \nu}g^{\mu \sigma}-g^{\lambda \sigma}g^{\mu \nu}) + f^{ace} f^{bde} (g^{\lambda \mu}g^{\nu \sigma}-g^{\lambda \sigma}g^{\nu \mu}) \right. \nonumber \\
& \left. + f^{ade} f^{cbe} (g^{\lambda \mu}g^{\sigma \nu}-g^{\lambda \nu}g^{\sigma \mu})\right] (\epsilon^\star_{d,4})_\sigma (\epsilon^\star_{c,3})_\nu (\epsilon_{b,2})_\mu  (\epsilon_{a,1})_\lambda , \nonumber \\
\mathcal{M}_4 (g^a g^b \rightarrow g^c g^d) & = f^{abe} f^{cde} \mathcal{M}^s_4 +  f^{ace} f^{bde} \mathcal{M}^u_4 + f^{ade} f^{cbe} \mathcal{M}^t_4 .
\end{align}
~\\
The invariant matrix element squared averaged (summed) over the initial (final) gluons is
\begin{align}
|\overline{\mathcal{M}}  (g g \rightarrow g g)|^2 = &\ \frac{1}{d^2_g} \sum_{\text{color}}   \sum_{\text{spin}}  |\mathcal{M}  (g^a g^b \rightarrow g^c g^d)|^2 \\
= & \frac{1}{8 \cross 8} \sum_{\text{color}} \frac{1}{2 \cross 2} \sum_{\text{spin}} \left( |\mathcal{M}_t|^2 + |\mathcal{M}_s|^2 + |\mathcal{M}_u|^2 + |\mathcal{M}_4|^2 + 2\Re \left[ \mathcal{M}_t \mathcal{M}^\star_s \right] + 2\Re \left[ \mathcal{M}_t \mathcal{M}^\star_u \right] \right. \nonumber \\
& \left. + 2\Re \left[ \mathcal{M}_u \mathcal{M}^\star_s \right] + 2\Re \left[ \mathcal{M}_t \mathcal{M}^\star_4 \right] + 2\Re \left[ \mathcal{M}_s \mathcal{M}^\star_4 \right] + + 2\Re \left[ \mathcal{M}_u \mathcal{M}^\star_4 \right] \right) \nonumber .
\end{align}

\end{widetext}


\bibliographystyle{aipnum4-1}
\bibliography{References}

\begin{thebibliography}{133}%
\makeatletter
\providecommand \@ifxundefined [1]{%
 \@ifx{#1\undefined}
}%
\providecommand \@ifnum [1]{%
 \ifnum #1\expandafter \@firstoftwo
 \else \expandafter \@secondoftwo
 \fi
}%
\providecommand \@ifx [1]{%
 \ifx #1\expandafter \@firstoftwo
 \else \expandafter \@secondoftwo
 \fi
}%
\providecommand \natexlab [1]{#1}%
\providecommand \enquote  [1]{``#1''}%
\providecommand \bibnamefont  [1]{#1}%
\providecommand \bibfnamefont [1]{#1}%
\providecommand \citenamefont [1]{#1}%
\providecommand \href@noop [0]{\@secondoftwo}%
\providecommand \href [0]{\begingroup \@sanitize@url \@href}%
\providecommand \@href[1]{\@@startlink{#1}\@@href}%
\providecommand \@@href[1]{\endgroup#1\@@endlink}%
\providecommand \@sanitize@url [0]{\catcode `\\12\catcode `\$12\catcode
  `\&12\catcode `\#12\catcode `\^12\catcode `\_12\catcode `\%12\relax}%
\providecommand \@@startlink[1]{}%
\providecommand \@@endlink[0]{}%
\providecommand \url  [0]{\begingroup\@sanitize@url \@url }%
\providecommand \@url [1]{\endgroup\@href {#1}{\urlprefix }}%
\providecommand \urlprefix  [0]{URL }%
\providecommand \Eprint [0]{\href }%
\providecommand \doibase [0]{http://dx.doi.org/}%
\providecommand \selectlanguage [0]{\@gobble}%
\providecommand \bibinfo  [0]{\@secondoftwo}%
\providecommand \bibfield  [0]{\@secondoftwo}%
\providecommand \translation [1]{[#1]}%
\providecommand \BibitemOpen [0]{}%
\providecommand \bibitemStop [0]{}%
\providecommand \bibitemNoStop [0]{.\EOS\space}%
\providecommand \EOS [0]{\spacefactor3000\relax}%
\providecommand \BibitemShut  [1]{\csname bibitem#1\endcsname}%
\let\auto@bib@innerbib\@empty
\bibitem [{\citenamefont {Schwinger}(1961)}]{Schwinger:1960qe}%
  \BibitemOpen
  \bibfield  {author} {\bibinfo {author} {\bibfnamefont {J.~S.}\ \bibnamefont
  {Schwinger}},\ }\href {\doibase 10.1063/1.1703727} {\bibfield  {journal}
  {\bibinfo  {journal} {J. Math. Phys.}\ }\textbf {\bibinfo {volume} {2}},\
  \bibinfo {pages} {407} (\bibinfo {year} {1961})}\BibitemShut {NoStop}%
\bibitem [{\citenamefont {Bakshi}\ and\ \citenamefont
  {Mahanthappa}(1963)}]{Bakshi:1963bn}%
  \BibitemOpen
  \bibfield  {author} {\bibinfo {author} {\bibfnamefont {P.~M.}\ \bibnamefont
  {Bakshi}}\ and\ \bibinfo {author} {\bibfnamefont {K.~T.}\ \bibnamefont
  {Mahanthappa}},\ }\href {\doibase 10.1063/1.1703879} {\bibfield  {journal}
  {\bibinfo  {journal} {J. Math. Phys.}\ }\textbf {\bibinfo {volume} {4}},\
  \bibinfo {pages} {12} (\bibinfo {year} {1963})}\BibitemShut {NoStop}%
\bibitem [{\citenamefont {Keldysh}(1964)}]{Keldysh:1964ud}%
  \BibitemOpen
  \bibfield  {author} {\bibinfo {author} {\bibfnamefont {L.~V.}\ \bibnamefont
  {Keldysh}},\ }\href@noop {} {\bibfield  {journal} {\bibinfo  {journal} {Zh.
  Eksp. Teor. Fiz.}\ }\textbf {\bibinfo {volume} {47}},\ \bibinfo {pages}
  {1515} (\bibinfo {year} {1964})},\ \bibinfo {note} {[Sov. Phys.
  JETP20,1018(1965)]}\BibitemShut {NoStop}%
\bibitem [{\citenamefont {Craig}(1968)}]{Cr68}%
  \BibitemOpen
  \bibfield  {author} {\bibinfo {author} {\bibfnamefont {R.~A.}\ \bibnamefont
  {Craig}},\ }\href {\doibase 10.1063/1.1664616} {\bibfield  {journal}
  {\bibinfo  {journal} {Journal of Mathematical Physics}\ }\textbf {\bibinfo
  {volume} {9}},\ \bibinfo {pages} {605} (\bibinfo {year} {1968})},\ \Eprint
  {http://arxiv.org/abs/https://doi.org/10.1063/1.1664616}
  {https://doi.org/10.1063/1.1664616} \BibitemShut {NoStop}%
\bibitem [{\citenamefont {Bonitz}(1998)}]{Bonitz}%
  \BibitemOpen
  \bibfield  {author} {\bibinfo {author} {\bibfnamefont {M.}~\bibnamefont
  {Bonitz}},\ }\href {https://books.google.de/books?id=B7ON7nhWcDIC} {\emph
  {\bibinfo {title} {Quantum Kinetic Theory}}},\ Teubner Texte zur Physik\
  (\bibinfo  {publisher} {Vieweg+Teubner Verlag, Stuttgart, Leipzig},\ \bibinfo
  {year} {1998})\BibitemShut {NoStop}%
\bibitem [{\citenamefont {Kadanoff}\ and\ \citenamefont
  {Baym}(1962)}]{KadanoffBaym}%
  \BibitemOpen
  \bibfield  {author} {\bibinfo {author} {\bibfnamefont {L.~P.}\ \bibnamefont
  {Kadanoff}}\ and\ \bibinfo {author} {\bibfnamefont {G.}~\bibnamefont
  {Baym}},\ }\href@noop {} {\emph {\bibinfo {title} {{Quantum Statistical
  mechanics}}}}\ (\bibinfo  {publisher} {W. A. Benjamin, Inc., New York},\
  \bibinfo {year} {1962})\BibitemShut {NoStop}%
\bibitem [{\citenamefont {DuBois}(1967)}]{dubois1967lectures}%
  \BibitemOpen
  \bibfield  {author} {\bibinfo {author} {\bibfnamefont {D.}~\bibnamefont
  {DuBois}},\ }\href@noop {} {\bibfield  {journal} {\bibinfo  {journal} {{\it
  Lectures in Theoretical Physics}, Brittin, W.E. (ed.), London: Gordon and
  Breach}\ } (\bibinfo {year} {1967})}\BibitemShut {NoStop}%
\bibitem [{\citenamefont {Danielewicz}(1984)}]{Danielewicz:1982ca}%
  \BibitemOpen
  \bibfield  {author} {\bibinfo {author} {\bibfnamefont {P.}~\bibnamefont
  {Danielewicz}},\ }\href {\doibase 10.1016/0003-4916(84)90093-9} {\bibfield
  {journal} {\bibinfo  {journal} {Annals Phys.}\ }\textbf {\bibinfo {volume}
  {152}},\ \bibinfo {pages} {305} (\bibinfo {year} {1984})}\BibitemShut
  {NoStop}%
\bibitem [{\citenamefont {Chou}\ \emph {et~al.}(1985)\citenamefont {Chou},
  \citenamefont {Su}, \citenamefont {Hao},\ and\ \citenamefont
  {Yu}}]{Chou:1984es}%
  \BibitemOpen
  \bibfield  {author} {\bibinfo {author} {\bibfnamefont {K.-c.}\ \bibnamefont
  {Chou}}, \bibinfo {author} {\bibfnamefont {Z.-b.}\ \bibnamefont {Su}},
  \bibinfo {author} {\bibfnamefont {B.-l.}\ \bibnamefont {Hao}}, \ and\
  \bibinfo {author} {\bibfnamefont {L.}~\bibnamefont {Yu}},\ }\href {\doibase
  10.1016/0370-1573(85)90136-X} {\bibfield  {journal} {\bibinfo  {journal}
  {Phys. Rept.}\ }\textbf {\bibinfo {volume} {118}},\ \bibinfo {pages} {1}
  (\bibinfo {year} {1985})}\BibitemShut {NoStop}%
\bibitem [{\citenamefont {Rammer}\ and\ \citenamefont
  {Smith}(1986)}]{Rammer:1986zz}%
  \BibitemOpen
  \bibfield  {author} {\bibinfo {author} {\bibfnamefont {J.}~\bibnamefont
  {Rammer}}\ and\ \bibinfo {author} {\bibfnamefont {H.}~\bibnamefont {Smith}},\
  }\href {\doibase 10.1103/RevModPhys.58.323} {\bibfield  {journal} {\bibinfo
  {journal} {Rev. Mod. Phys.}\ }\textbf {\bibinfo {volume} {58}},\ \bibinfo
  {pages} {323} (\bibinfo {year} {1986})}\BibitemShut {NoStop}%
\bibitem [{\citenamefont {Calzetta}\ and\ \citenamefont
  {Hu}(1988)}]{Calzetta:1986cq}%
  \BibitemOpen
  \bibfield  {author} {\bibinfo {author} {\bibfnamefont {E.}~\bibnamefont
  {Calzetta}}\ and\ \bibinfo {author} {\bibfnamefont {B.~L.}\ \bibnamefont
  {Hu}},\ }\href {\doibase 10.1103/PhysRevD.37.2878} {\bibfield  {journal}
  {\bibinfo  {journal} {Phys. Rev.}\ }\textbf {\bibinfo {volume} {D37}},\
  \bibinfo {pages} {2878} (\bibinfo {year} {1988})}\BibitemShut {NoStop}%
\bibitem [{\citenamefont {Haug}\ and\ \citenamefont {Jauho}(1998)}]{Haug}%
  \BibitemOpen
  \bibfield  {author} {\bibinfo {author} {\bibfnamefont {H.}~\bibnamefont
  {Haug}}\ and\ \bibinfo {author} {\bibfnamefont {A.}~\bibnamefont {Jauho}},\
  }\href {https://books.google.de/books?id=i8\_hQgAACAAJ} {\emph {\bibinfo
  {title} {Quantum Kinetics in Transport and Optics of Semiconductors}}},\
  Springer Series in Solid-State Sciences\ (\bibinfo  {publisher} {Springer
  Berlin Heidelberg},\ \bibinfo {year} {1998})\BibitemShut {NoStop}%
\bibitem [{\citenamefont {Bezzerides}\ and\ \citenamefont
  {DuBois}(1972)}]{BEZZERIDES197210}%
  \BibitemOpen
  \bibfield  {author} {\bibinfo {author} {\bibfnamefont {B.}~\bibnamefont
  {Bezzerides}}\ and\ \bibinfo {author} {\bibfnamefont {D.}~\bibnamefont
  {DuBois}},\ }\href {\doibase https://doi.org/10.1016/0003-4916(72)90329-6}
  {\bibfield  {journal} {\bibinfo  {journal} {Annals of Physics}\ }\textbf
  {\bibinfo {volume} {70}},\ \bibinfo {pages} {10 } (\bibinfo {year}
  {1972})}\BibitemShut {NoStop}%
\bibitem [{\citenamefont {Botermans}\ and\ \citenamefont
  {Malfliet}(1990)}]{Botermans:1990qi}%
  \BibitemOpen
  \bibfield  {author} {\bibinfo {author} {\bibfnamefont {W.}~\bibnamefont
  {Botermans}}\ and\ \bibinfo {author} {\bibfnamefont {R.}~\bibnamefont
  {Malfliet}},\ }\href {\doibase 10.1016/0370-1573(90)90174-Z} {\bibfield
  {journal} {\bibinfo  {journal} {Phys. Rept.}\ }\textbf {\bibinfo {volume}
  {198}},\ \bibinfo {pages} {115} (\bibinfo {year} {1990})}\BibitemShut
  {NoStop}%
\bibitem [{\citenamefont {Mrowczynski}\ and\ \citenamefont
  {Danielewicz}(1990)}]{Mrowczynski:1989bu}%
  \BibitemOpen
  \bibfield  {author} {\bibinfo {author} {\bibfnamefont {S.}~\bibnamefont
  {Mrowczynski}}\ and\ \bibinfo {author} {\bibfnamefont {P.}~\bibnamefont
  {Danielewicz}},\ }\href {\doibase 10.1016/0550-3213(90)90194-I} {\bibfield
  {journal} {\bibinfo  {journal} {Nucl. Phys.}\ }\textbf {\bibinfo {volume}
  {B342}},\ \bibinfo {pages} {345} (\bibinfo {year} {1990})}\BibitemShut
  {NoStop}%
\bibitem [{\citenamefont {Makhlin}\ and\ \citenamefont
  {Surdutovich}(1998)}]{Makhlin:1998zi}%
  \BibitemOpen
  \bibfield  {author} {\bibinfo {author} {\bibfnamefont {A.}~\bibnamefont
  {Makhlin}}\ and\ \bibinfo {author} {\bibfnamefont {E.}~\bibnamefont
  {Surdutovich}},\ }\href {\doibase 10.1103/PhysRevC.58.389} {\bibfield
  {journal} {\bibinfo  {journal} {Phys. Rev.}\ }\textbf {\bibinfo {volume}
  {C58}},\ \bibinfo {pages} {389} (\bibinfo {year} {1998})},\ \Eprint
  {http://arxiv.org/abs/hep-ph/9803364} {arXiv:hep-ph/9803364 [hep-ph]}
  \BibitemShut {NoStop}%
\bibitem [{\citenamefont {Makhlin}(1995)}]{Makhlin:1994ew}%
  \BibitemOpen
  \bibfield  {author} {\bibinfo {author} {\bibfnamefont {A.}~\bibnamefont
  {Makhlin}},\ }\href {\doibase 10.1103/PhysRevC.52.995} {\bibfield  {journal}
  {\bibinfo  {journal} {Phys. Rev.}\ }\textbf {\bibinfo {volume} {C52}},\
  \bibinfo {pages} {995} (\bibinfo {year} {1995})},\ \Eprint
  {http://arxiv.org/abs/hep-ph/9412363} {arXiv:hep-ph/9412363 [hep-ph]}
  \BibitemShut {NoStop}%
\bibitem [{\citenamefont {Geiger}(1996)}]{Geiger:1995ak}%
  \BibitemOpen
  \bibfield  {author} {\bibinfo {author} {\bibfnamefont {K.}~\bibnamefont
  {Geiger}},\ }\href {\doibase 10.1103/PhysRevD.54.949} {\bibfield  {journal}
  {\bibinfo  {journal} {Phys. Rev.}\ }\textbf {\bibinfo {volume} {D54}},\
  \bibinfo {pages} {949} (\bibinfo {year} {1996})},\ \Eprint
  {http://arxiv.org/abs/hep-ph/9507365} {arXiv:hep-ph/9507365 [hep-ph]}
  \BibitemShut {NoStop}%
\bibitem [{\citenamefont {Geiger}(1997)}]{Geiger:1996ym}%
  \BibitemOpen
  \bibfield  {author} {\bibinfo {author} {\bibfnamefont {K.}~\bibnamefont
  {Geiger}},\ }\href {\doibase 10.1103/PhysRevD.56.2665} {\bibfield  {journal}
  {\bibinfo  {journal} {Phys. Rev.}\ }\textbf {\bibinfo {volume} {D56}},\
  \bibinfo {pages} {2665} (\bibinfo {year} {1997})},\ \Eprint
  {http://arxiv.org/abs/hep-ph/9611400} {arXiv:hep-ph/9611400 [hep-ph]}
  \BibitemShut {NoStop}%
\bibitem [{\citenamefont {Brown}\ and\ \citenamefont
  {Danielewicz}(1998)}]{Brown:1998zx}%
  \BibitemOpen
  \bibfield  {author} {\bibinfo {author} {\bibfnamefont {D.~A.}\ \bibnamefont
  {Brown}}\ and\ \bibinfo {author} {\bibfnamefont {P.}~\bibnamefont
  {Danielewicz}},\ }\href {\doibase 10.1103/PhysRevD.58.094003} {\bibfield
  {journal} {\bibinfo  {journal} {Phys. Rev.}\ }\textbf {\bibinfo {volume}
  {D58}},\ \bibinfo {pages} {094003} (\bibinfo {year} {1998})},\ \Eprint
  {http://arxiv.org/abs/nucl-th/9802015} {arXiv:nucl-th/9802015 [nucl-th]}
  \BibitemShut {NoStop}%
\bibitem [{\citenamefont {Blaizot}\ and\ \citenamefont
  {Iancu}(1999)}]{Blaizot:1999xk}%
  \BibitemOpen
  \bibfield  {author} {\bibinfo {author} {\bibfnamefont {J.-P.}\ \bibnamefont
  {Blaizot}}\ and\ \bibinfo {author} {\bibfnamefont {E.}~\bibnamefont
  {Iancu}},\ }\href {\doibase 10.1016/S0550-3213(99)00341-7} {\bibfield
  {journal} {\bibinfo  {journal} {Nucl. Phys.}\ }\textbf {\bibinfo {volume}
  {B557}},\ \bibinfo {pages} {183} (\bibinfo {year} {1999})},\ \Eprint
  {http://arxiv.org/abs/hep-ph/9903389} {arXiv:hep-ph/9903389 [hep-ph]}
  \BibitemShut {NoStop}%
\bibitem [{\citenamefont {Cassing}\ and\ \citenamefont
  {Juchem}(2000{\natexlab{a}})}]{Cassing:1999wx}%
  \BibitemOpen
  \bibfield  {author} {\bibinfo {author} {\bibfnamefont {W.}~\bibnamefont
  {Cassing}}\ and\ \bibinfo {author} {\bibfnamefont {S.}~\bibnamefont
  {Juchem}},\ }\href {\doibase 10.1016/S0375-9474(99)00393-0} {\bibfield
  {journal} {\bibinfo  {journal} {Nucl. Phys.}\ }\textbf {\bibinfo {volume}
  {A665}},\ \bibinfo {pages} {377} (\bibinfo {year} {2000}{\natexlab{a}})},\
  \Eprint {http://arxiv.org/abs/nucl-th/9903070} {arXiv:nucl-th/9903070
  [nucl-th]} \BibitemShut {NoStop}%
\bibitem [{\citenamefont {Cassing}\ and\ \citenamefont
  {Juchem}(2000{\natexlab{b}})}]{Cassing:1999mh}%
  \BibitemOpen
  \bibfield  {author} {\bibinfo {author} {\bibfnamefont {W.}~\bibnamefont
  {Cassing}}\ and\ \bibinfo {author} {\bibfnamefont {S.}~\bibnamefont
  {Juchem}},\ }\href {\doibase 10.1016/S0375-9474(00)00050-6} {\bibfield
  {journal} {\bibinfo  {journal} {Nucl. Phys.}\ }\textbf {\bibinfo {volume}
  {A672}},\ \bibinfo {pages} {417} (\bibinfo {year} {2000}{\natexlab{b}})},\
  \Eprint {http://arxiv.org/abs/nucl-th/9910052} {arXiv:nucl-th/9910052
  [nucl-th]} \BibitemShut {NoStop}%
\bibitem [{\citenamefont {Ivanov}, \citenamefont {Knoll},\ and\ \citenamefont
  {Voskresensky}(1999)}]{Ivanov:1998nv}%
  \BibitemOpen
  \bibfield  {author} {\bibinfo {author} {\bibfnamefont {{\relax Yu}.~B.}\
  \bibnamefont {Ivanov}}, \bibinfo {author} {\bibfnamefont {J.}~\bibnamefont
  {Knoll}}, \ and\ \bibinfo {author} {\bibfnamefont {D.~N.}\ \bibnamefont
  {Voskresensky}},\ }\href {\doibase 10.1016/S0375-9474(99)00313-9} {\bibfield
  {journal} {\bibinfo  {journal} {Nucl. Phys.}\ }\textbf {\bibinfo {volume}
  {A657}},\ \bibinfo {pages} {413} (\bibinfo {year} {1999})},\ \Eprint
  {http://arxiv.org/abs/hep-ph/9807351} {arXiv:hep-ph/9807351 [hep-ph]}
  \BibitemShut {NoStop}%
\bibitem [{\citenamefont {Knoll}, \citenamefont {Ivanov},\ and\ \citenamefont
  {Voskresensky}(2001)}]{Knoll:2001jx}%
  \BibitemOpen
  \bibfield  {author} {\bibinfo {author} {\bibfnamefont {J.}~\bibnamefont
  {Knoll}}, \bibinfo {author} {\bibfnamefont {{\relax Yu}.~B.}\ \bibnamefont
  {Ivanov}}, \ and\ \bibinfo {author} {\bibfnamefont {D.~N.}\ \bibnamefont
  {Voskresensky}},\ }\href {\doibase 10.1006/aphy.2001.6185} {\bibfield
  {journal} {\bibinfo  {journal} {Annals Phys.}\ }\textbf {\bibinfo {volume}
  {293}},\ \bibinfo {pages} {126} (\bibinfo {year} {2001})},\ \Eprint
  {http://arxiv.org/abs/nucl-th/0102044} {arXiv:nucl-th/0102044 [nucl-th]}
  \BibitemShut {NoStop}%
\bibitem [{\citenamefont {Cassing}(2009)}]{Cassing:2008nn}%
  \BibitemOpen
  \bibfield  {author} {\bibinfo {author} {\bibfnamefont {W.}~\bibnamefont
  {Cassing}},\ }\href {\doibase 10.1140/epjst/e2009-00959-x} {\bibfield
  {journal} {\bibinfo  {journal} {Eur. Phys. J. ST}\ }\textbf {\bibinfo
  {volume} {168}},\ \bibinfo {pages} {3} (\bibinfo {year} {2009})},\ \Eprint
  {http://arxiv.org/abs/0808.0715} {arXiv:0808.0715 [nucl-th]} \BibitemShut
  {NoStop}%
\bibitem [{\citenamefont {Weldon}(1983)}]{Weldon:1983jn}%
  \BibitemOpen
  \bibfield  {author} {\bibinfo {author} {\bibfnamefont {H.~A.}\ \bibnamefont
  {Weldon}},\ }\href {\doibase 10.1103/PhysRevD.28.2007} {\bibfield  {journal}
  {\bibinfo  {journal} {Phys. Rev.}\ }\textbf {\bibinfo {volume} {D28}},\
  \bibinfo {pages} {2007} (\bibinfo {year} {1983})}\BibitemShut {NoStop}%
\bibitem [{\citenamefont {Lebedev}\ and\ \citenamefont
  {Smilga}(1990)}]{Lebedev:1989ev}%
  \BibitemOpen
  \bibfield  {author} {\bibinfo {author} {\bibfnamefont {V.~V.}\ \bibnamefont
  {Lebedev}}\ and\ \bibinfo {author} {\bibfnamefont {A.~V.}\ \bibnamefont
  {Smilga}},\ }\href {\doibase 10.1016/0003-4916(90)90225-D} {\bibfield
  {journal} {\bibinfo  {journal} {Annals Phys.}\ }\textbf {\bibinfo {volume}
  {202}},\ \bibinfo {pages} {229} (\bibinfo {year} {1990})}\BibitemShut
  {NoStop}%
\bibitem [{\citenamefont {Braaten}\ and\ \citenamefont
  {Pisarski}(1990)}]{Braaten:1990it}%
  \BibitemOpen
  \bibfield  {author} {\bibinfo {author} {\bibfnamefont {E.}~\bibnamefont
  {Braaten}}\ and\ \bibinfo {author} {\bibfnamefont {R.~D.}\ \bibnamefont
  {Pisarski}},\ }\href {\doibase 10.1103/PhysRevD.42.2156} {\bibfield
  {journal} {\bibinfo  {journal} {Phys. Rev.}\ }\textbf {\bibinfo {volume}
  {D42}},\ \bibinfo {pages} {2156} (\bibinfo {year} {1990})}\BibitemShut
  {NoStop}%
\bibitem [{\citenamefont {Braaten}\ and\ \citenamefont
  {Pisarski}(1992)}]{Braaten:1992gd}%
  \BibitemOpen
  \bibfield  {author} {\bibinfo {author} {\bibfnamefont {E.}~\bibnamefont
  {Braaten}}\ and\ \bibinfo {author} {\bibfnamefont {R.~D.}\ \bibnamefont
  {Pisarski}},\ }\href {\doibase 10.1103/PhysRevD.46.1829} {\bibfield
  {journal} {\bibinfo  {journal} {Phys. Rev.}\ }\textbf {\bibinfo {volume}
  {D46}},\ \bibinfo {pages} {1829} (\bibinfo {year} {1992})}\BibitemShut
  {NoStop}%
\bibitem [{\citenamefont {Pisarski}(1993)}]{Pisarski:1993rf}%
  \BibitemOpen
  \bibfield  {author} {\bibinfo {author} {\bibfnamefont {R.~D.}\ \bibnamefont
  {Pisarski}},\ }\href {\doibase 10.1103/PhysRevD.47.5589} {\bibfield
  {journal} {\bibinfo  {journal} {Phys. Rev.}\ }\textbf {\bibinfo {volume}
  {D47}},\ \bibinfo {pages} {5589} (\bibinfo {year} {1993})}\BibitemShut
  {NoStop}%
\bibitem [{\citenamefont {Jeon}(1995)}]{Jeon:1994if}%
  \BibitemOpen
  \bibfield  {author} {\bibinfo {author} {\bibfnamefont {S.}~\bibnamefont
  {Jeon}},\ }\href {\doibase 10.1103/PhysRevD.52.3591} {\bibfield  {journal}
  {\bibinfo  {journal} {Phys. Rev.}\ }\textbf {\bibinfo {volume} {D52}},\
  \bibinfo {pages} {3591} (\bibinfo {year} {1995})},\ \Eprint
  {http://arxiv.org/abs/hep-ph/9409250} {arXiv:hep-ph/9409250 [hep-ph]}
  \BibitemShut {NoStop}%
\bibitem [{\citenamefont {Wang}\ and\ \citenamefont
  {Heinz}(1996)}]{Wang:1995qf}%
  \BibitemOpen
  \bibfield  {author} {\bibinfo {author} {\bibfnamefont {E.-k.}\ \bibnamefont
  {Wang}}\ and\ \bibinfo {author} {\bibfnamefont {U.~W.}\ \bibnamefont
  {Heinz}},\ }\href {\doibase 10.1103/PhysRevD.53.899} {\bibfield  {journal}
  {\bibinfo  {journal} {Phys. Rev.}\ }\textbf {\bibinfo {volume} {D53}},\
  \bibinfo {pages} {899} (\bibinfo {year} {1996})},\ \Eprint
  {http://arxiv.org/abs/hep-ph/9509333} {arXiv:hep-ph/9509333 [hep-ph]}
  \BibitemShut {NoStop}%
\bibitem [{\citenamefont {Thoma}(1994)}]{Thoma:1993vs}%
  \BibitemOpen
  \bibfield  {author} {\bibinfo {author} {\bibfnamefont {M.~H.}\ \bibnamefont
  {Thoma}},\ }\href {\doibase 10.1103/PhysRevD.49.451} {\bibfield  {journal}
  {\bibinfo  {journal} {Phys. Rev.}\ }\textbf {\bibinfo {volume} {D49}},\
  \bibinfo {pages} {451} (\bibinfo {year} {1994})},\ \Eprint
  {http://arxiv.org/abs/hep-ph/9308257} {arXiv:hep-ph/9308257 [hep-ph]}
  \BibitemShut {NoStop}%
\bibitem [{\citenamefont {Liu}\ and\ \citenamefont {Rapp}(2018)}]{Liu:2017qah}%
  \BibitemOpen
  \bibfield  {author} {\bibinfo {author} {\bibfnamefont {S.~Y.~F.}\
  \bibnamefont {Liu}}\ and\ \bibinfo {author} {\bibfnamefont {R.}~\bibnamefont
  {Rapp}},\ }\href {\doibase 10.1103/PhysRevC.97.034918} {\bibfield  {journal}
  {\bibinfo  {journal} {Phys. Rev.}\ }\textbf {\bibinfo {volume} {C97}},\
  \bibinfo {pages} {034918} (\bibinfo {year} {2018})},\ \Eprint
  {http://arxiv.org/abs/1711.03282} {arXiv:1711.03282 [nucl-th]} \BibitemShut
  {NoStop}%
\bibitem [{\citenamefont {Vanderheyden}\ and\ \citenamefont
  {Baym}(1998)}]{Vanderheyden:1998ph}%
  \BibitemOpen
  \bibfield  {author} {\bibinfo {author} {\bibfnamefont {B.}~\bibnamefont
  {Vanderheyden}}\ and\ \bibinfo {author} {\bibfnamefont {G.}~\bibnamefont
  {Baym}},\ }\href {\doibase 10.1023/B:JOSS.0000033166.37520.ae} {\bibfield
  {journal} {\bibinfo  {journal} {J. Stat. Phys.}\ } (\bibinfo {year} {1998}),\
  10.1023/B:JOSS.0000033166.37520.ae},\ \bibinfo {note} {[J. Statist.
  Phys.93,843(1998)]},\ \Eprint {http://arxiv.org/abs/hep-ph/9803300}
  {arXiv:hep-ph/9803300 [hep-ph]} \BibitemShut {NoStop}%
\bibitem [{\citenamefont {Bernard}\ \emph {et~al.}(2005)\citenamefont
  {Bernard}, \citenamefont {Burch}, \citenamefont {Gregory}, \citenamefont
  {Toussaint}, \citenamefont {DeTar}, \citenamefont {Osborn}, \citenamefont
  {Gottlieb}, \citenamefont {Heller},\ and\ \citenamefont
  {Sugar}}]{Bernard:2004je}%
  \BibitemOpen
  \bibfield  {author} {\bibinfo {author} {\bibfnamefont {C.}~\bibnamefont
  {Bernard}}, \bibinfo {author} {\bibfnamefont {T.}~\bibnamefont {Burch}},
  \bibinfo {author} {\bibfnamefont {E.~B.}\ \bibnamefont {Gregory}}, \bibinfo
  {author} {\bibfnamefont {D.}~\bibnamefont {Toussaint}}, \bibinfo {author}
  {\bibfnamefont {C.~E.}\ \bibnamefont {DeTar}}, \bibinfo {author}
  {\bibfnamefont {J.}~\bibnamefont {Osborn}}, \bibinfo {author} {\bibfnamefont
  {S.}~\bibnamefont {Gottlieb}}, \bibinfo {author} {\bibfnamefont {U.~M.}\
  \bibnamefont {Heller}}, \ and\ \bibinfo {author} {\bibfnamefont
  {R.}~\bibnamefont {Sugar}} (\bibinfo {collaboration} {MILC}),\ }\href
  {\doibase 10.1103/PhysRevD.71.034504} {\bibfield  {journal} {\bibinfo
  {journal} {Phys. Rev.}\ }\textbf {\bibinfo {volume} {D71}},\ \bibinfo {pages}
  {034504} (\bibinfo {year} {2005})},\ \Eprint
  {http://arxiv.org/abs/hep-lat/0405029} {arXiv:hep-lat/0405029 [hep-lat]}
  \BibitemShut {NoStop}%
\bibitem [{\citenamefont {Aoki}\ \emph {et~al.}(2006)\citenamefont {Aoki},
  \citenamefont {Endrodi}, \citenamefont {Fodor}, \citenamefont {Katz},\ and\
  \citenamefont {Szabo}}]{Aoki:2006we}%
  \BibitemOpen
  \bibfield  {author} {\bibinfo {author} {\bibfnamefont {Y.}~\bibnamefont
  {Aoki}}, \bibinfo {author} {\bibfnamefont {G.}~\bibnamefont {Endrodi}},
  \bibinfo {author} {\bibfnamefont {Z.}~\bibnamefont {Fodor}}, \bibinfo
  {author} {\bibfnamefont {S.~D.}\ \bibnamefont {Katz}}, \ and\ \bibinfo
  {author} {\bibfnamefont {K.~K.}\ \bibnamefont {Szabo}},\ }\href {\doibase
  10.1038/nature05120} {\bibfield  {journal} {\bibinfo  {journal} {Nature}\
  }\textbf {\bibinfo {volume} {443}},\ \bibinfo {pages} {675} (\bibinfo {year}
  {2006})},\ \Eprint {http://arxiv.org/abs/hep-lat/0611014}
  {arXiv:hep-lat/0611014 [hep-lat]} \BibitemShut {NoStop}%
\bibitem [{\citenamefont {Bazavov}\ \emph {et~al.}(2012)\citenamefont
  {Bazavov}, \citenamefont {Bhattacharya}, \citenamefont {Cheng}, \citenamefont
  {DeTar}, \citenamefont {Ding}, \citenamefont {Gottlieb}, \citenamefont
  {Gupta}, \citenamefont {Hegde}, \citenamefont {Heller}, \citenamefont
  {Karsch} \emph {et~al.}}]{Bazavov:2011nk}%
  \BibitemOpen
  \bibfield  {author} {\bibinfo {author} {\bibfnamefont {A.}~\bibnamefont
  {Bazavov}}, \bibinfo {author} {\bibfnamefont {T.}~\bibnamefont
  {Bhattacharya}}, \bibinfo {author} {\bibfnamefont {M.}~\bibnamefont {Cheng}},
  \bibinfo {author} {\bibfnamefont {C.}~\bibnamefont {DeTar}}, \bibinfo
  {author} {\bibfnamefont {H.-T.}\ \bibnamefont {Ding}}, \bibinfo {author}
  {\bibfnamefont {S.}~\bibnamefont {Gottlieb}}, \bibinfo {author}
  {\bibfnamefont {R.}~\bibnamefont {Gupta}}, \bibinfo {author} {\bibfnamefont
  {P.}~\bibnamefont {Hegde}}, \bibinfo {author} {\bibfnamefont {U.~M.}\
  \bibnamefont {Heller}}, \bibinfo {author} {\bibfnamefont {F.}~\bibnamefont
  {Karsch}},  \emph {et~al.} (\bibinfo {collaboration} {HotQCD
  Collaboration}),\ }\href {\doibase 10.1103/PhysRevD.85.054503} {\bibfield
  {journal} {\bibinfo  {journal} {Phys. Rev. D}\ }\textbf {\bibinfo {volume}
  {85}},\ \bibinfo {pages} {054503} (\bibinfo {year} {2012})}\BibitemShut
  {NoStop}%
\bibitem [{\citenamefont {Fischer}\ and\ \citenamefont
  {Luecker}(2013)}]{Fischer:2012vc}%
  \BibitemOpen
  \bibfield  {author} {\bibinfo {author} {\bibfnamefont {C.~S.}\ \bibnamefont
  {Fischer}}\ and\ \bibinfo {author} {\bibfnamefont {J.}~\bibnamefont
  {Luecker}},\ }\href {\doibase 10.1016/j.physletb.2012.11.054} {\bibfield
  {journal} {\bibinfo  {journal} {Phys. Lett.}\ }\textbf {\bibinfo {volume}
  {B718}},\ \bibinfo {pages} {1036} (\bibinfo {year} {2013})},\ \Eprint
  {http://arxiv.org/abs/1206.5191} {arXiv:1206.5191 [hep-ph]} \BibitemShut
  {NoStop}%
\bibitem [{\citenamefont {Fischer}(2019)}]{Fischer:2018sdj}%
  \BibitemOpen
  \bibfield  {author} {\bibinfo {author} {\bibfnamefont {C.~S.}\ \bibnamefont
  {Fischer}},\ }\href {\doibase 10.1016/j.ppnp.2019.01.002} {\bibfield
  {journal} {\bibinfo  {journal} {Prog. Part. Nucl. Phys.}\ }\textbf {\bibinfo
  {volume} {105}},\ \bibinfo {pages} {1} (\bibinfo {year} {2019})},\ \Eprint
  {http://arxiv.org/abs/1810.12938} {arXiv:1810.12938 [hep-ph]} \BibitemShut
  {NoStop}%
\bibitem [{\citenamefont {Senger}\ \emph {et~al.}(2011)\citenamefont {Senger},
  \citenamefont {Bratkovskaya}, \citenamefont {Andronic}, \citenamefont
  {Averbeck}, \citenamefont {Bellwied}, \citenamefont {Friese}, \citenamefont
  {Fuchs}, \citenamefont {Knoll}, \citenamefont {Randrup},\ and\ \citenamefont
  {Steinheimer}}]{Senger:2011zza}%
  \BibitemOpen
  \bibfield  {author} {\bibinfo {author} {\bibfnamefont {P.}~\bibnamefont
  {Senger}}, \bibinfo {author} {\bibfnamefont {E.}~\bibnamefont
  {Bratkovskaya}}, \bibinfo {author} {\bibfnamefont {A.}~\bibnamefont
  {Andronic}}, \bibinfo {author} {\bibfnamefont {R.}~\bibnamefont {Averbeck}},
  \bibinfo {author} {\bibfnamefont {R.}~\bibnamefont {Bellwied}}, \bibinfo
  {author} {\bibfnamefont {V.}~\bibnamefont {Friese}}, \bibinfo {author}
  {\bibfnamefont {C.}~\bibnamefont {Fuchs}}, \bibinfo {author} {\bibfnamefont
  {J.}~\bibnamefont {Knoll}}, \bibinfo {author} {\bibfnamefont
  {J.}~\bibnamefont {Randrup}}, \ and\ \bibinfo {author} {\bibfnamefont
  {J.}~\bibnamefont {Steinheimer}},\ }\href {\doibase
  10.1007/978-3-642-13293-3_6} {\bibfield  {journal} {\bibinfo  {journal}
  {Lect. Notes Phys.}\ }\textbf {\bibinfo {volume} {814}},\ \bibinfo {pages}
  {681} (\bibinfo {year} {2011})}\BibitemShut {NoStop}%
\bibitem [{\citenamefont {Ruggieri}\ \emph {et~al.}(2014)\citenamefont
  {Ruggieri}, \citenamefont {Oliva}, \citenamefont {Castorina}, \citenamefont
  {Gatto},\ and\ \citenamefont {Greco}}]{Ruggieri:2014bqa}%
  \BibitemOpen
  \bibfield  {author} {\bibinfo {author} {\bibfnamefont {M.}~\bibnamefont
  {Ruggieri}}, \bibinfo {author} {\bibfnamefont {L.}~\bibnamefont {Oliva}},
  \bibinfo {author} {\bibfnamefont {P.}~\bibnamefont {Castorina}}, \bibinfo
  {author} {\bibfnamefont {R.}~\bibnamefont {Gatto}}, \ and\ \bibinfo {author}
  {\bibfnamefont {V.}~\bibnamefont {Greco}},\ }\href {\doibase
  10.1016/j.physletb.2014.05.073} {\bibfield  {journal} {\bibinfo  {journal}
  {Phys. Lett.}\ }\textbf {\bibinfo {volume} {B734}},\ \bibinfo {pages} {255}
  (\bibinfo {year} {2014})},\ \Eprint {http://arxiv.org/abs/1402.0737}
  {arXiv:1402.0737 [hep-ph]} \BibitemShut {NoStop}%
\bibitem [{\citenamefont {Mohanty}(2011)}]{Mohanty:2011nm}%
  \BibitemOpen
  \bibfield  {author} {\bibinfo {author} {\bibfnamefont {B.}~\bibnamefont
  {Mohanty}} (\bibinfo {collaboration} {STAR}),\ }\href {\doibase
  10.1088/0954-3899/38/12/124023} {\bibfield  {journal} {\bibinfo  {journal}
  {J. Phys.}\ }\textbf {\bibinfo {volume} {G38}},\ \bibinfo {pages} {124023}
  (\bibinfo {year} {2011})},\ \Eprint {http://arxiv.org/abs/1106.5902}
  {arXiv:1106.5902 [nucl-ex]} \BibitemShut {NoStop}%
\bibitem [{\citenamefont {Kumar}(2011)}]{Kumar:2011us}%
  \BibitemOpen
  \bibfield  {author} {\bibinfo {author} {\bibfnamefont {L.}~\bibnamefont
  {Kumar}} (\bibinfo {collaboration} {STAR}),\ }\href {\doibase
  10.1088/0954-3899/38/12/124145} {\bibfield  {journal} {\bibinfo  {journal}
  {J. Phys.}\ }\textbf {\bibinfo {volume} {G38}},\ \bibinfo {pages} {124145}
  (\bibinfo {year} {2011})},\ \Eprint {http://arxiv.org/abs/1106.6071}
  {arXiv:1106.6071 [nucl-ex]} \BibitemShut {NoStop}%
\bibitem [{\citenamefont {Cassing}\ \emph {et~al.}(2016)\citenamefont
  {Cassing}, \citenamefont {Palmese}, \citenamefont {Moreau},\ and\
  \citenamefont {Bratkovskaya}}]{Cassing:2015owa}%
  \BibitemOpen
  \bibfield  {author} {\bibinfo {author} {\bibfnamefont {W.}~\bibnamefont
  {Cassing}}, \bibinfo {author} {\bibfnamefont {A.}~\bibnamefont {Palmese}},
  \bibinfo {author} {\bibfnamefont {P.}~\bibnamefont {Moreau}}, \ and\ \bibinfo
  {author} {\bibfnamefont {E.~L.}\ \bibnamefont {Bratkovskaya}},\ }\href
  {\doibase 10.1103/PhysRevC.93.014902} {\bibfield  {journal} {\bibinfo
  {journal} {Phys. Rev.}\ }\textbf {\bibinfo {volume} {C93}},\ \bibinfo {pages}
  {014902} (\bibinfo {year} {2016})},\ \Eprint
  {http://arxiv.org/abs/1510.04120} {arXiv:1510.04120 [nucl-th]} \BibitemShut
  {NoStop}%
\bibitem [{\citenamefont {Palmese}\ \emph {et~al.}(2016)\citenamefont
  {Palmese}, \citenamefont {Cassing}, \citenamefont {Seifert}, \citenamefont
  {Steinert}, \citenamefont {Moreau},\ and\ \citenamefont
  {Bratkovskaya}}]{Palmese:2016rtq}%
  \BibitemOpen
  \bibfield  {author} {\bibinfo {author} {\bibfnamefont {A.}~\bibnamefont
  {Palmese}}, \bibinfo {author} {\bibfnamefont {W.}~\bibnamefont {Cassing}},
  \bibinfo {author} {\bibfnamefont {E.}~\bibnamefont {Seifert}}, \bibinfo
  {author} {\bibfnamefont {T.}~\bibnamefont {Steinert}}, \bibinfo {author}
  {\bibfnamefont {P.}~\bibnamefont {Moreau}}, \ and\ \bibinfo {author}
  {\bibfnamefont {E.~L.}\ \bibnamefont {Bratkovskaya}},\ }\href {\doibase
  10.1103/PhysRevC.94.044912} {\bibfield  {journal} {\bibinfo  {journal} {Phys.
  Rev.}\ }\textbf {\bibinfo {volume} {C94}},\ \bibinfo {pages} {044912}
  (\bibinfo {year} {2016})},\ \Eprint {http://arxiv.org/abs/1607.04073}
  {arXiv:1607.04073 [nucl-th]} \BibitemShut {NoStop}%
\bibitem [{\citenamefont {Shen}\ and\ \citenamefont
  {Schenke}(2019)}]{Shen:2018pty}%
  \BibitemOpen
  \bibfield  {author} {\bibinfo {author} {\bibfnamefont {C.}~\bibnamefont
  {Shen}}\ and\ \bibinfo {author} {\bibfnamefont {B.}~\bibnamefont {Schenke}},\
  }\bibfield  {booktitle} {\emph {\bibinfo {booktitle} {{Proceedings, 27th
  International Conference on Ultrarelativistic Nucleus-Nucleus Collisions
  (Quark Matter 2018): Venice, Italy, May 14-19, 2018}}},\ }\href {\doibase
  10.1016/j.nuclphysa.2018.08.007} {\bibfield  {journal} {\bibinfo  {journal}
  {Nucl. Phys.}\ }\textbf {\bibinfo {volume} {A982}},\ \bibinfo {pages} {411}
  (\bibinfo {year} {2019})},\ \Eprint {http://arxiv.org/abs/1807.05141}
  {arXiv:1807.05141 [nucl-th]} \BibitemShut {NoStop}%
\bibitem [{\citenamefont {Li}\ and\ \citenamefont {Shen}(2018)}]{Li:2018fow}%
  \BibitemOpen
  \bibfield  {author} {\bibinfo {author} {\bibfnamefont {M.}~\bibnamefont
  {Li}}\ and\ \bibinfo {author} {\bibfnamefont {C.}~\bibnamefont {Shen}},\
  }\href {\doibase 10.1103/PhysRevC.98.064908} {\bibfield  {journal} {\bibinfo
  {journal} {Phys. Rev.}\ }\textbf {\bibinfo {volume} {C98}},\ \bibinfo {pages}
  {064908} (\bibinfo {year} {2018})},\ \Eprint
  {http://arxiv.org/abs/1809.04034} {arXiv:1809.04034 [nucl-th]} \BibitemShut
  {NoStop}%
\bibitem [{\citenamefont {Ivanov}, \citenamefont {Russkikh},\ and\
  \citenamefont {Toneev}(2006)}]{Ivanov:2005yw}%
  \BibitemOpen
  \bibfield  {author} {\bibinfo {author} {\bibfnamefont {{\relax Yu}.~B.}\
  \bibnamefont {Ivanov}}, \bibinfo {author} {\bibfnamefont {V.~N.}\
  \bibnamefont {Russkikh}}, \ and\ \bibinfo {author} {\bibfnamefont {V.~D.}\
  \bibnamefont {Toneev}},\ }\href {\doibase 10.1103/PhysRevC.73.044904}
  {\bibfield  {journal} {\bibinfo  {journal} {Phys. Rev.}\ }\textbf {\bibinfo
  {volume} {C73}},\ \bibinfo {pages} {044904} (\bibinfo {year} {2006})},\
  \Eprint {http://arxiv.org/abs/nucl-th/0503088} {arXiv:nucl-th/0503088
  [nucl-th]} \BibitemShut {NoStop}%
\bibitem [{\citenamefont {Ivanov}\ and\ \citenamefont
  {Blaschke}(2015)}]{Ivanov:2015vna}%
  \BibitemOpen
  \bibfield  {author} {\bibinfo {author} {\bibfnamefont {{\relax Yu}.~B.}\
  \bibnamefont {Ivanov}}\ and\ \bibinfo {author} {\bibfnamefont
  {D.}~\bibnamefont {Blaschke}},\ }\href {\doibase 10.1103/PhysRevC.92.024916}
  {\bibfield  {journal} {\bibinfo  {journal} {Phys. Rev.}\ }\textbf {\bibinfo
  {volume} {C92}},\ \bibinfo {pages} {024916} (\bibinfo {year} {2015})},\
  \Eprint {http://arxiv.org/abs/1504.03992} {arXiv:1504.03992 [nucl-th]}
  \BibitemShut {NoStop}%
\bibitem [{\citenamefont {Denicol}\ \emph {et~al.}(2018)\citenamefont
  {Denicol}, \citenamefont {Gale}, \citenamefont {Jeon}, \citenamefont
  {Monnai}, \citenamefont {Schenke},\ and\ \citenamefont
  {Shen}}]{Denicol:2018wdp}%
  \BibitemOpen
  \bibfield  {author} {\bibinfo {author} {\bibfnamefont {G.~S.}\ \bibnamefont
  {Denicol}}, \bibinfo {author} {\bibfnamefont {C.}~\bibnamefont {Gale}},
  \bibinfo {author} {\bibfnamefont {S.}~\bibnamefont {Jeon}}, \bibinfo {author}
  {\bibfnamefont {A.}~\bibnamefont {Monnai}}, \bibinfo {author} {\bibfnamefont
  {B.}~\bibnamefont {Schenke}}, \ and\ \bibinfo {author} {\bibfnamefont
  {C.}~\bibnamefont {Shen}},\ }\href {\doibase 10.1103/PhysRevC.98.034916}
  {\bibfield  {journal} {\bibinfo  {journal} {Phys. Rev.}\ }\textbf {\bibinfo
  {volume} {C98}},\ \bibinfo {pages} {034916} (\bibinfo {year} {2018})},\
  \Eprint {http://arxiv.org/abs/1804.10557} {arXiv:1804.10557 [nucl-th]}
  \BibitemShut {NoStop}%
\bibitem [{\citenamefont {Petersen}\ \emph {et~al.}(2008)\citenamefont
  {Petersen}, \citenamefont {Steinheimer}, \citenamefont {Burau}, \citenamefont
  {Bleicher},\ and\ \citenamefont {St{\"o}cker}}]{Petersen:2008dd}%
  \BibitemOpen
  \bibfield  {author} {\bibinfo {author} {\bibfnamefont {H.}~\bibnamefont
  {Petersen}}, \bibinfo {author} {\bibfnamefont {J.}~\bibnamefont
  {Steinheimer}}, \bibinfo {author} {\bibfnamefont {G.}~\bibnamefont {Burau}},
  \bibinfo {author} {\bibfnamefont {M.}~\bibnamefont {Bleicher}}, \ and\
  \bibinfo {author} {\bibfnamefont {H.}~\bibnamefont {St{\"o}cker}},\ }\href
  {\doibase 10.1103/PhysRevC.78.044901} {\bibfield  {journal} {\bibinfo
  {journal} {Phys. Rev.}\ }\textbf {\bibinfo {volume} {C78}},\ \bibinfo {pages}
  {044901} (\bibinfo {year} {2008})},\ \Eprint {http://arxiv.org/abs/0806.1695}
  {arXiv:0806.1695 [nucl-th]} \BibitemShut {NoStop}%
\bibitem [{\citenamefont {Karpenko}(2019)}]{Karpenko:2018xam}%
  \BibitemOpen
  \bibfield  {author} {\bibinfo {author} {\bibfnamefont {I.}~\bibnamefont
  {Karpenko}},\ }\href {\doibase 10.5506/APhysPolB.50.141} {\bibfield
  {journal} {\bibinfo  {journal} {Acta Phys. Polon.}\ }\textbf {\bibinfo
  {volume} {B50}},\ \bibinfo {pages} {141} (\bibinfo {year} {2019})},\ \Eprint
  {http://arxiv.org/abs/1805.11998} {arXiv:1805.11998 [nucl-th]} \BibitemShut
  {NoStop}%
\bibitem [{\citenamefont {Li}\ and\ \citenamefont
  {Kapusta}(2019)}]{Li:2018ini}%
  \BibitemOpen
  \bibfield  {author} {\bibinfo {author} {\bibfnamefont {M.}~\bibnamefont
  {Li}}\ and\ \bibinfo {author} {\bibfnamefont {J.~I.}\ \bibnamefont
  {Kapusta}},\ }\href {\doibase 10.1103/PhysRevC.99.014906} {\bibfield
  {journal} {\bibinfo  {journal} {Phys. Rev.}\ }\textbf {\bibinfo {volume}
  {C99}},\ \bibinfo {pages} {014906} (\bibinfo {year} {2019})},\ \Eprint
  {http://arxiv.org/abs/1808.05751} {arXiv:1808.05751 [nucl-th]} \BibitemShut
  {NoStop}%
\bibitem [{\citenamefont {Cassing}\ and\ \citenamefont
  {Bratkovskaya}(2009)}]{Cassing:2009vt}%
  \BibitemOpen
  \bibfield  {author} {\bibinfo {author} {\bibfnamefont {W.}~\bibnamefont
  {Cassing}}\ and\ \bibinfo {author} {\bibfnamefont {E.~L.}\ \bibnamefont
  {Bratkovskaya}},\ }\href {\doibase 10.1016/j.nuclphysa.2009.09.007}
  {\bibfield  {journal} {\bibinfo  {journal} {Nucl. Phys.}\ }\textbf {\bibinfo
  {volume} {A831}},\ \bibinfo {pages} {215} (\bibinfo {year} {2009})},\ \Eprint
  {http://arxiv.org/abs/0907.5331} {arXiv:0907.5331 [nucl-th]} \BibitemShut
  {NoStop}%
\bibitem [{\citenamefont {Linnyk}, \citenamefont {Bratkovskaya},\ and\
  \citenamefont {Cassing}(2016)}]{Linnyk:2015rco}%
  \BibitemOpen
  \bibfield  {author} {\bibinfo {author} {\bibfnamefont {O.}~\bibnamefont
  {Linnyk}}, \bibinfo {author} {\bibfnamefont {E.~L.}\ \bibnamefont
  {Bratkovskaya}}, \ and\ \bibinfo {author} {\bibfnamefont {W.}~\bibnamefont
  {Cassing}},\ }\href {\doibase 10.1016/j.ppnp.2015.12.003} {\bibfield
  {journal} {\bibinfo  {journal} {Prog. Part. Nucl. Phys.}\ }\textbf {\bibinfo
  {volume} {87}},\ \bibinfo {pages} {50} (\bibinfo {year} {2016})},\ \Eprint
  {http://arxiv.org/abs/1512.08126} {arXiv:1512.08126 [nucl-th]} \BibitemShut
  {NoStop}%
\bibitem [{\citenamefont {Ozvenchuk}\ \emph
  {et~al.}(2013{\natexlab{a}})\citenamefont {Ozvenchuk}, \citenamefont
  {Linnyk}, \citenamefont {Gorenstein}, \citenamefont {Bratkovskaya},\ and\
  \citenamefont {Cassing}}]{Ozvenchuk:2012fn}%
  \BibitemOpen
  \bibfield  {author} {\bibinfo {author} {\bibfnamefont {V.}~\bibnamefont
  {Ozvenchuk}}, \bibinfo {author} {\bibfnamefont {O.}~\bibnamefont {Linnyk}},
  \bibinfo {author} {\bibfnamefont {M.~I.}\ \bibnamefont {Gorenstein}},
  \bibinfo {author} {\bibfnamefont {E.~L.}\ \bibnamefont {Bratkovskaya}}, \
  and\ \bibinfo {author} {\bibfnamefont {W.}~\bibnamefont {Cassing}},\ }\href
  {\doibase 10.1103/PhysRevC.87.024901} {\bibfield  {journal} {\bibinfo
  {journal} {Phys. Rev.}\ }\textbf {\bibinfo {volume} {C87}},\ \bibinfo {pages}
  {024901} (\bibinfo {year} {2013}{\natexlab{a}})},\ \Eprint
  {http://arxiv.org/abs/1203.4734} {arXiv:1203.4734 [nucl-th]} \BibitemShut
  {NoStop}%
\bibitem [{\citenamefont {Linnyk}\ \emph {et~al.}(2013)\citenamefont {Linnyk},
  \citenamefont {Cassing}, \citenamefont {Manninen}, \citenamefont
  {Bratkovskaya}, \citenamefont {Gossiaux}, \citenamefont {Aichelin},
  \citenamefont {Song},\ and\ \citenamefont {Ko}}]{Linnyk:2012pu}%
  \BibitemOpen
  \bibfield  {author} {\bibinfo {author} {\bibfnamefont {O.}~\bibnamefont
  {Linnyk}}, \bibinfo {author} {\bibfnamefont {W.}~\bibnamefont {Cassing}},
  \bibinfo {author} {\bibfnamefont {J.}~\bibnamefont {Manninen}}, \bibinfo
  {author} {\bibfnamefont {E.~L.}\ \bibnamefont {Bratkovskaya}}, \bibinfo
  {author} {\bibfnamefont {P.~B.}\ \bibnamefont {Gossiaux}}, \bibinfo {author}
  {\bibfnamefont {J.}~\bibnamefont {Aichelin}}, \bibinfo {author}
  {\bibfnamefont {T.}~\bibnamefont {Song}}, \ and\ \bibinfo {author}
  {\bibfnamefont {C.~M.}\ \bibnamefont {Ko}},\ }\href {\doibase
  10.1103/PhysRevC.87.014905} {\bibfield  {journal} {\bibinfo  {journal} {Phys.
  Rev.}\ }\textbf {\bibinfo {volume} {C87}},\ \bibinfo {pages} {014905}
  (\bibinfo {year} {2013})},\ \Eprint {http://arxiv.org/abs/1208.1279}
  {arXiv:1208.1279 [nucl-th]} \BibitemShut {NoStop}%
\bibitem [{\citenamefont {Bratkovskaya}\ \emph {et~al.}(2011)\citenamefont
  {Bratkovskaya}, \citenamefont {Cassing}, \citenamefont {Konchakovski},\ and\
  \citenamefont {Linnyk}}]{Bratkovskaya:2011wp}%
  \BibitemOpen
  \bibfield  {author} {\bibinfo {author} {\bibfnamefont {E.~L.}\ \bibnamefont
  {Bratkovskaya}}, \bibinfo {author} {\bibfnamefont {W.}~\bibnamefont
  {Cassing}}, \bibinfo {author} {\bibfnamefont {V.~P.}\ \bibnamefont
  {Konchakovski}}, \ and\ \bibinfo {author} {\bibfnamefont {O.}~\bibnamefont
  {Linnyk}},\ }\href {\doibase 10.1016/j.nuclphysa.2011.03.003} {\bibfield
  {journal} {\bibinfo  {journal} {Nucl. Phys.}\ }\textbf {\bibinfo {volume}
  {A856}},\ \bibinfo {pages} {162} (\bibinfo {year} {2011})},\ \Eprint
  {http://arxiv.org/abs/1101.5793} {arXiv:1101.5793 [nucl-th]} \BibitemShut
  {NoStop}%
\bibitem [{\citenamefont {Konchakovski}, \citenamefont {Cassing},\ and\
  \citenamefont {Toneev}(2015)}]{Konchakovski:2014fya}%
  \BibitemOpen
  \bibfield  {author} {\bibinfo {author} {\bibfnamefont {V.~P.}\ \bibnamefont
  {Konchakovski}}, \bibinfo {author} {\bibfnamefont {W.}~\bibnamefont
  {Cassing}}, \ and\ \bibinfo {author} {\bibfnamefont {V.~D.}\ \bibnamefont
  {Toneev}},\ }\href {\doibase 10.1088/0954-3899/42/5/055106} {\bibfield
  {journal} {\bibinfo  {journal} {J. Phys.}\ }\textbf {\bibinfo {volume}
  {G42}},\ \bibinfo {pages} {055106} (\bibinfo {year} {2015})},\ \Eprint
  {http://arxiv.org/abs/1411.5534} {arXiv:1411.5534 [nucl-th]} \BibitemShut
  {NoStop}%
\bibitem [{\citenamefont {Konchakovski}\ \emph
  {et~al.}(2012{\natexlab{a}})\citenamefont {Konchakovski}, \citenamefont
  {Bratkovskaya}, \citenamefont {Cassing}, \citenamefont {Toneev},\ and\
  \citenamefont {Voronyuk}}]{Konchakovski:2011qa}%
  \BibitemOpen
  \bibfield  {author} {\bibinfo {author} {\bibfnamefont {V.~P.}\ \bibnamefont
  {Konchakovski}}, \bibinfo {author} {\bibfnamefont {E.~L.}\ \bibnamefont
  {Bratkovskaya}}, \bibinfo {author} {\bibfnamefont {W.}~\bibnamefont
  {Cassing}}, \bibinfo {author} {\bibfnamefont {V.~D.}\ \bibnamefont {Toneev}},
  \ and\ \bibinfo {author} {\bibfnamefont {V.}~\bibnamefont {Voronyuk}},\
  }\href {\doibase 10.1103/PhysRevC.85.011902} {\bibfield  {journal} {\bibinfo
  {journal} {Phys. Rev.}\ }\textbf {\bibinfo {volume} {C85}},\ \bibinfo {pages}
  {011902} (\bibinfo {year} {2012}{\natexlab{a}})},\ \Eprint
  {http://arxiv.org/abs/1109.3039} {arXiv:1109.3039 [nucl-th]} \BibitemShut
  {NoStop}%
\bibitem [{\citenamefont {Konchakovski}\ \emph
  {et~al.}(2012{\natexlab{b}})\citenamefont {Konchakovski}, \citenamefont
  {Bratkovskaya}, \citenamefont {Cassing}, \citenamefont {Toneev},
  \citenamefont {Voloshin},\ and\ \citenamefont
  {Voronyuk}}]{Konchakovski:2012yg}%
  \BibitemOpen
  \bibfield  {author} {\bibinfo {author} {\bibfnamefont {V.~P.}\ \bibnamefont
  {Konchakovski}}, \bibinfo {author} {\bibfnamefont {E.~L.}\ \bibnamefont
  {Bratkovskaya}}, \bibinfo {author} {\bibfnamefont {W.}~\bibnamefont
  {Cassing}}, \bibinfo {author} {\bibfnamefont {V.~D.}\ \bibnamefont {Toneev}},
  \bibinfo {author} {\bibfnamefont {S.~A.}\ \bibnamefont {Voloshin}}, \ and\
  \bibinfo {author} {\bibfnamefont {V.}~\bibnamefont {Voronyuk}},\ }\href
  {\doibase 10.1103/PhysRevC.85.044922} {\bibfield  {journal} {\bibinfo
  {journal} {Phys. Rev.}\ }\textbf {\bibinfo {volume} {C85}},\ \bibinfo {pages}
  {044922} (\bibinfo {year} {2012}{\natexlab{b}})},\ \Eprint
  {http://arxiv.org/abs/1201.3320} {arXiv:1201.3320 [nucl-th]} \BibitemShut
  {NoStop}%
\bibitem [{\citenamefont {Konchakovski}\ \emph {et~al.}(2014)\citenamefont
  {Konchakovski}, \citenamefont {Cassing}, \citenamefont {Ivanov},\ and\
  \citenamefont {Toneev}}]{Konchakovski:2014gda}%
  \BibitemOpen
  \bibfield  {author} {\bibinfo {author} {\bibfnamefont {V.~P.}\ \bibnamefont
  {Konchakovski}}, \bibinfo {author} {\bibfnamefont {W.}~\bibnamefont
  {Cassing}}, \bibinfo {author} {\bibfnamefont {{\relax Yu}.~B.}\ \bibnamefont
  {Ivanov}}, \ and\ \bibinfo {author} {\bibfnamefont {V.~D.}\ \bibnamefont
  {Toneev}},\ }\href {\doibase 10.1103/PhysRevC.90.014903} {\bibfield
  {journal} {\bibinfo  {journal} {Phys. Rev.}\ }\textbf {\bibinfo {volume}
  {C90}},\ \bibinfo {pages} {014903} (\bibinfo {year} {2014})},\ \Eprint
  {http://arxiv.org/abs/1404.2765} {arXiv:1404.2765 [nucl-th]} \BibitemShut
  {NoStop}%
\bibitem [{\citenamefont {Berrehrah}\ \emph
  {et~al.}(2016{\natexlab{a}})\citenamefont {Berrehrah}, \citenamefont
  {Cassing}, \citenamefont {Bratkovskaya},\ and\ \citenamefont
  {Steinert}}]{Berrehrah:2015vhe}%
  \BibitemOpen
  \bibfield  {author} {\bibinfo {author} {\bibfnamefont {H.}~\bibnamefont
  {Berrehrah}}, \bibinfo {author} {\bibfnamefont {W.}~\bibnamefont {Cassing}},
  \bibinfo {author} {\bibfnamefont {E.}~\bibnamefont {Bratkovskaya}}, \ and\
  \bibinfo {author} {\bibfnamefont {T.}~\bibnamefont {Steinert}},\ }\href
  {\doibase 10.1103/PhysRevC.93.044914} {\bibfield  {journal} {\bibinfo
  {journal} {Phys. Rev.}\ }\textbf {\bibinfo {volume} {C93}},\ \bibinfo {pages}
  {044914} (\bibinfo {year} {2016}{\natexlab{a}})},\ \Eprint
  {http://arxiv.org/abs/1512.06909} {arXiv:1512.06909 [hep-ph]} \BibitemShut
  {NoStop}%
\bibitem [{\citenamefont {Berrehrah}\ \emph
  {et~al.}(2016{\natexlab{b}})\citenamefont {Berrehrah}, \citenamefont
  {Bratkovskaya}, \citenamefont {Steinert},\ and\ \citenamefont
  {Cassing}}]{Berrehrah:2016vzw}%
  \BibitemOpen
  \bibfield  {author} {\bibinfo {author} {\bibfnamefont {H.}~\bibnamefont
  {Berrehrah}}, \bibinfo {author} {\bibfnamefont {E.}~\bibnamefont
  {Bratkovskaya}}, \bibinfo {author} {\bibfnamefont {T.}~\bibnamefont
  {Steinert}}, \ and\ \bibinfo {author} {\bibfnamefont {W.}~\bibnamefont
  {Cassing}},\ }\href {\doibase 10.1142/S0218301316420039} {\bibfield
  {journal} {\bibinfo  {journal} {Int. J. Mod. Phys.}\ }\textbf {\bibinfo
  {volume} {E25}},\ \bibinfo {pages} {1642003} (\bibinfo {year}
  {2016}{\natexlab{b}})},\ \Eprint {http://arxiv.org/abs/1605.02371}
  {arXiv:1605.02371 [hep-ph]} \BibitemShut {NoStop}%
\bibitem [{\citenamefont {Adler}\ \emph {et~al.}(2003)\citenamefont {Adler}
  \emph {et~al.}}]{Adler:2003kt}%
  \BibitemOpen
  \bibfield  {author} {\bibinfo {author} {\bibfnamefont {S.~S.}\ \bibnamefont
  {Adler}} \emph {et~al.} (\bibinfo {collaboration} {PHENIX}),\ }\href
  {\doibase 10.1103/PhysRevLett.91.182301} {\bibfield  {journal} {\bibinfo
  {journal} {Phys. Rev. Lett.}\ }\textbf {\bibinfo {volume} {91}},\ \bibinfo
  {pages} {182301} (\bibinfo {year} {2003})},\ \Eprint
  {http://arxiv.org/abs/nucl-ex/0305013} {arXiv:nucl-ex/0305013 [nucl-ex]}
  \BibitemShut {NoStop}%
\bibitem [{\citenamefont {Berrehrah}\ \emph {et~al.}(2014)\citenamefont
  {Berrehrah}, \citenamefont {Bratkovskaya}, \citenamefont {Cassing},
  \citenamefont {Gossiaux}, \citenamefont {Aichelin},\ and\ \citenamefont
  {Bleicher}}]{Berrehrah:2013mua}%
  \BibitemOpen
  \bibfield  {author} {\bibinfo {author} {\bibfnamefont {H.}~\bibnamefont
  {Berrehrah}}, \bibinfo {author} {\bibfnamefont {E.}~\bibnamefont
  {Bratkovskaya}}, \bibinfo {author} {\bibfnamefont {W.}~\bibnamefont
  {Cassing}}, \bibinfo {author} {\bibfnamefont {P.~B.}\ \bibnamefont
  {Gossiaux}}, \bibinfo {author} {\bibfnamefont {J.}~\bibnamefont {Aichelin}},
  \ and\ \bibinfo {author} {\bibfnamefont {M.}~\bibnamefont {Bleicher}},\
  }\href {\doibase 10.1103/PhysRevC.89.054901} {\bibfield  {journal} {\bibinfo
  {journal} {Phys. Rev.}\ }\textbf {\bibinfo {volume} {C89}},\ \bibinfo {pages}
  {054901} (\bibinfo {year} {2014})},\ \Eprint {http://arxiv.org/abs/1308.5148}
  {arXiv:1308.5148 [hep-ph]} \BibitemShut {NoStop}%
\bibitem [{\citenamefont {Berrehrah}\ \emph
  {et~al.}(2015{\natexlab{a}})\citenamefont {Berrehrah}, \citenamefont
  {Bratkovskaya}, \citenamefont {Cassing},\ and\ \citenamefont
  {Marty}}]{Berrehrah:2014ysa}%
  \BibitemOpen
  \bibfield  {author} {\bibinfo {author} {\bibfnamefont {H.}~\bibnamefont
  {Berrehrah}}, \bibinfo {author} {\bibfnamefont {E.}~\bibnamefont
  {Bratkovskaya}}, \bibinfo {author} {\bibfnamefont {W.}~\bibnamefont
  {Cassing}}, \ and\ \bibinfo {author} {\bibfnamefont {R.}~\bibnamefont
  {Marty}},\ }\href {\doibase 10.1088/1742-6596/612/1/012050} {\bibfield
  {journal} {\bibinfo  {journal} {J. Phys. Conf. Ser.}\ }\textbf {\bibinfo
  {volume} {612}},\ \bibinfo {pages} {012050} (\bibinfo {year}
  {2015}{\natexlab{a}})},\ \Eprint {http://arxiv.org/abs/1412.1017}
  {arXiv:1412.1017 [hep-ph]} \BibitemShut {NoStop}%
\bibitem [{\citenamefont {Berrehrah}\ \emph
  {et~al.}(2015{\natexlab{b}})\citenamefont {Berrehrah}, \citenamefont
  {Bratkovskaya}, \citenamefont {Cassing}, \citenamefont {Gossiaux},\ and\
  \citenamefont {Aichelin}}]{Berrehrah:2015ywa}%
  \BibitemOpen
  \bibfield  {author} {\bibinfo {author} {\bibfnamefont {H.}~\bibnamefont
  {Berrehrah}}, \bibinfo {author} {\bibfnamefont {E.}~\bibnamefont
  {Bratkovskaya}}, \bibinfo {author} {\bibfnamefont {W.}~\bibnamefont
  {Cassing}}, \bibinfo {author} {\bibfnamefont {P.~B.}\ \bibnamefont
  {Gossiaux}}, \ and\ \bibinfo {author} {\bibfnamefont {J.}~\bibnamefont
  {Aichelin}},\ }\href {\doibase 10.1103/PhysRevC.91.054902} {\bibfield
  {journal} {\bibinfo  {journal} {Phys. Rev.}\ }\textbf {\bibinfo {volume}
  {C91}},\ \bibinfo {pages} {054902} (\bibinfo {year} {2015}{\natexlab{b}})},\
  \Eprint {http://arxiv.org/abs/1502.01700} {arXiv:1502.01700 [hep-ph]}
  \BibitemShut {NoStop}%
\bibitem [{\citenamefont {Kaczmarek}\ \emph {et~al.}(2004)\citenamefont
  {Kaczmarek}, \citenamefont {Karsch}, \citenamefont {Zantow},\ and\
  \citenamefont {Petreczky}}]{Kaczmarek:2004gv}%
  \BibitemOpen
  \bibfield  {author} {\bibinfo {author} {\bibfnamefont {O.}~\bibnamefont
  {Kaczmarek}}, \bibinfo {author} {\bibfnamefont {F.}~\bibnamefont {Karsch}},
  \bibinfo {author} {\bibfnamefont {F.}~\bibnamefont {Zantow}}, \ and\ \bibinfo
  {author} {\bibfnamefont {P.}~\bibnamefont {Petreczky}},\ }\href {\doibase
  10.1103/PhysRevD.70.074505, 10.1103/PhysRevD.72.059903} {\bibfield  {journal}
  {\bibinfo  {journal} {Phys. Rev.}\ }\textbf {\bibinfo {volume} {D70}},\
  \bibinfo {pages} {074505} (\bibinfo {year} {2004})},\ \bibinfo {note}
  {[Erratum: Phys. Rev.D72,059903(2005)]},\ \Eprint
  {http://arxiv.org/abs/hep-lat/0406036} {arXiv:hep-lat/0406036 [hep-lat]}
  \BibitemShut {NoStop}%
\bibitem [{\citenamefont {Borsanyi}\ \emph {et~al.}(2012)\citenamefont
  {Borsanyi}, \citenamefont {Endrodi}, \citenamefont {Fodor}, \citenamefont
  {Katz}, \citenamefont {Krieg}, \citenamefont {Ratti},\ and\ \citenamefont
  {Szabo}}]{Borsanyi:2012cr}%
  \BibitemOpen
  \bibfield  {author} {\bibinfo {author} {\bibfnamefont {S.}~\bibnamefont
  {Borsanyi}}, \bibinfo {author} {\bibfnamefont {G.}~\bibnamefont {Endrodi}},
  \bibinfo {author} {\bibfnamefont {Z.}~\bibnamefont {Fodor}}, \bibinfo
  {author} {\bibfnamefont {S.~D.}\ \bibnamefont {Katz}}, \bibinfo {author}
  {\bibfnamefont {S.}~\bibnamefont {Krieg}}, \bibinfo {author} {\bibfnamefont
  {C.}~\bibnamefont {Ratti}}, \ and\ \bibinfo {author} {\bibfnamefont {K.~K.}\
  \bibnamefont {Szabo}},\ }\href {\doibase 10.1007/JHEP08(2012)053} {\bibfield
  {journal} {\bibinfo  {journal} {JHEP}\ }\textbf {\bibinfo {volume} {08}},\
  \bibinfo {pages} {053} (\bibinfo {year} {2012})},\ \Eprint
  {http://arxiv.org/abs/1204.6710} {arXiv:1204.6710 [hep-lat]} \BibitemShut
  {NoStop}%
\bibitem [{\citenamefont {Borsanyi}\ \emph {et~al.}(2014)\citenamefont
  {Borsanyi}, \citenamefont {Fodor}, \citenamefont {Hoelbling}, \citenamefont
  {Katz}, \citenamefont {Krieg},\ and\ \citenamefont
  {Szabo}}]{Borsanyi:2013bia}%
  \BibitemOpen
  \bibfield  {author} {\bibinfo {author} {\bibfnamefont {S.}~\bibnamefont
  {Borsanyi}}, \bibinfo {author} {\bibfnamefont {Z.}~\bibnamefont {Fodor}},
  \bibinfo {author} {\bibfnamefont {C.}~\bibnamefont {Hoelbling}}, \bibinfo
  {author} {\bibfnamefont {S.~D.}\ \bibnamefont {Katz}}, \bibinfo {author}
  {\bibfnamefont {S.}~\bibnamefont {Krieg}}, \ and\ \bibinfo {author}
  {\bibfnamefont {K.~K.}\ \bibnamefont {Szabo}},\ }\href {\doibase
  10.1016/j.physletb.2014.01.007} {\bibfield  {journal} {\bibinfo  {journal}
  {Phys. Lett.}\ }\textbf {\bibinfo {volume} {B730}},\ \bibinfo {pages} {99}
  (\bibinfo {year} {2014})},\ \Eprint {http://arxiv.org/abs/1309.5258}
  {arXiv:1309.5258 [hep-lat]} \BibitemShut {NoStop}%
\bibitem [{\citenamefont {Karsch}(2013)}]{Karsch:2013fga}%
  \BibitemOpen
  \bibfield  {author} {\bibinfo {author} {\bibfnamefont {F.}~\bibnamefont
  {Karsch}},\ }\bibfield  {booktitle} {\emph {\bibinfo {booktitle}
  {{Proceedings, 8th International Workshop on Critical Point and Onset of
  Deconfinement (CPOD 2013): Napa, CA, USA, March 11-15, 2013}}},\ }\href
  {\doibase 10.22323/1.185.0046} {\bibfield  {journal} {\bibinfo  {journal}
  {PoS}\ }\textbf {\bibinfo {volume} {CPOD2013}},\ \bibinfo {pages} {046}
  (\bibinfo {year} {2013})},\ \Eprint {http://arxiv.org/abs/1307.3978}
  {arXiv:1307.3978 [hep-ph]} \BibitemShut {NoStop}%
\bibitem [{\citenamefont {Bazavov}\ \emph {et~al.}(2017)\citenamefont
  {Bazavov}, \citenamefont {Ding}, \citenamefont {Hegde}, \citenamefont
  {Kaczmarek}, \citenamefont {Karsch}, \citenamefont {Laermann}, \citenamefont
  {Maezawa}, \citenamefont {Mukherjee}, \citenamefont {Ohno}, \citenamefont
  {Petreczky} \emph {et~al.}}]{Bazavov:2017dus}%
  \BibitemOpen
  \bibfield  {author} {\bibinfo {author} {\bibfnamefont {A.}~\bibnamefont
  {Bazavov}}, \bibinfo {author} {\bibfnamefont {H.-T.}\ \bibnamefont {Ding}},
  \bibinfo {author} {\bibfnamefont {P.}~\bibnamefont {Hegde}}, \bibinfo
  {author} {\bibfnamefont {O.}~\bibnamefont {Kaczmarek}}, \bibinfo {author}
  {\bibfnamefont {F.}~\bibnamefont {Karsch}}, \bibinfo {author} {\bibfnamefont
  {E.}~\bibnamefont {Laermann}}, \bibinfo {author} {\bibfnamefont
  {Y.}~\bibnamefont {Maezawa}}, \bibinfo {author} {\bibfnamefont
  {S.}~\bibnamefont {Mukherjee}}, \bibinfo {author} {\bibfnamefont
  {H.}~\bibnamefont {Ohno}}, \bibinfo {author} {\bibfnamefont {P.}~\bibnamefont
  {Petreczky}},  \emph {et~al.},\ }\href {\doibase 10.1103/PhysRevD.95.054504}
  {\bibfield  {journal} {\bibinfo  {journal} {Phys. Rev. D}\ }\textbf {\bibinfo
  {volume} {95}},\ \bibinfo {pages} {054504} (\bibinfo {year}
  {2017})}\BibitemShut {NoStop}%
\bibitem [{\citenamefont {Steinert}\ and\ \citenamefont
  {Cassing}(2018)}]{Steinert:2018bma}%
  \BibitemOpen
  \bibfield  {author} {\bibinfo {author} {\bibfnamefont {T.}~\bibnamefont
  {Steinert}}\ and\ \bibinfo {author} {\bibfnamefont {W.}~\bibnamefont
  {Cassing}},\ }\bibfield  {booktitle} {\emph {\bibinfo {booktitle}
  {{Proceedings, 5th FAIR NExt generation ScientistS (FAIRNESS 2017): Sitges,
  Barcelona, Spain, May 28-June 3, 2018}}},\ }\href {\doibase
  10.1088/1742-6596/1024/1/012029} {\bibfield  {journal} {\bibinfo  {journal}
  {J. Phys. Conf. Ser.}\ }\textbf {\bibinfo {volume} {1024}},\ \bibinfo {pages}
  {012029} (\bibinfo {year} {2018})}\BibitemShut {NoStop}%
\bibitem [{\citenamefont {Adamczyk}\ \emph {et~al.}(2017)\citenamefont
  {Adamczyk} \emph {et~al.}}]{Adamczyk:2017iwn}%
  \BibitemOpen
  \bibfield  {author} {\bibinfo {author} {\bibfnamefont {L.}~\bibnamefont
  {Adamczyk}} \emph {et~al.} (\bibinfo {collaboration} {STAR}),\ }\href
  {\doibase 10.1103/PhysRevC.96.044904} {\bibfield  {journal} {\bibinfo
  {journal} {Phys. Rev.}\ }\textbf {\bibinfo {volume} {C96}},\ \bibinfo {pages}
  {044904} (\bibinfo {year} {2017})},\ \Eprint
  {http://arxiv.org/abs/1701.07065} {arXiv:1701.07065 [nucl-ex]} \BibitemShut
  {NoStop}%
\bibitem [{\citenamefont {Cassing}(2007)}]{Cassing:2007nb}%
  \BibitemOpen
  \bibfield  {author} {\bibinfo {author} {\bibfnamefont {W.}~\bibnamefont
  {Cassing}},\ }\href {\doibase 10.1016/j.nuclphysa.2007.08.010} {\bibfield
  {journal} {\bibinfo  {journal} {Nucl. Phys.}\ }\textbf {\bibinfo {volume}
  {A795}},\ \bibinfo {pages} {70} (\bibinfo {year} {2007})},\ \Eprint
  {http://arxiv.org/abs/0707.3033} {arXiv:0707.3033 [nucl-th]} \BibitemShut
  {NoStop}%
\bibitem [{\citenamefont {Bellac}(2011)}]{Bellac:2011kqa}%
  \BibitemOpen
  \bibfield  {author} {\bibinfo {author} {\bibfnamefont {M.~L.}\ \bibnamefont
  {Bellac}},\ }\href {\doibase 10.1017/CBO9780511721700} {\emph {\bibinfo
  {title} {{Thermal Field Theory}}}},\ Cambridge Monographs on Mathematical
  Physics\ (\bibinfo  {publisher} {Cambridge University Press},\ \bibinfo
  {year} {2011})\BibitemShut {NoStop}%
\bibitem [{\citenamefont {Pisarski}(1989)}]{Pisarski:1989cs}%
  \BibitemOpen
  \bibfield  {author} {\bibinfo {author} {\bibfnamefont {R.~D.}\ \bibnamefont
  {Pisarski}},\ }\bibfield  {booktitle} {\emph {\bibinfo {booktitle} {{Physica
  A158 (1989) 146-157}}},\ }\href@noop {} {\bibfield  {journal} {\bibinfo
  {journal} {Physica}\ }\textbf {\bibinfo {volume} {A158}},\ \bibinfo {pages}
  {146} (\bibinfo {year} {1989})}\BibitemShut {NoStop}%
\bibitem [{\citenamefont {Blaizot}, \citenamefont {Iancu},\ and\ \citenamefont
  {Rebhan}(2001)}]{Blaizot:2000fc}%
  \BibitemOpen
  \bibfield  {author} {\bibinfo {author} {\bibfnamefont {J.~P.}\ \bibnamefont
  {Blaizot}}, \bibinfo {author} {\bibfnamefont {E.}~\bibnamefont {Iancu}}, \
  and\ \bibinfo {author} {\bibfnamefont {A.}~\bibnamefont {Rebhan}},\ }\href
  {\doibase 10.1103/PhysRevD.63.065003} {\bibfield  {journal} {\bibinfo
  {journal} {Phys. Rev.}\ }\textbf {\bibinfo {volume} {D63}},\ \bibinfo {pages}
  {065003} (\bibinfo {year} {2001})},\ \Eprint
  {http://arxiv.org/abs/hep-ph/0005003} {arXiv:hep-ph/0005003 [hep-ph]}
  \BibitemShut {NoStop}%
\bibitem [{\citenamefont {Rauber}\ and\ \citenamefont
  {Cassing}(2014)}]{Rauber:2014mca}%
  \BibitemOpen
  \bibfield  {author} {\bibinfo {author} {\bibfnamefont {L.}~\bibnamefont
  {Rauber}}\ and\ \bibinfo {author} {\bibfnamefont {W.}~\bibnamefont
  {Cassing}},\ }\href {\doibase 10.1103/PhysReVD.89.065008} {\bibfield
  {journal} {\bibinfo  {journal} {Phys. Rev.}\ }\textbf {\bibinfo {volume}
  {D89}},\ \bibinfo {pages} {065008} (\bibinfo {year} {2014})},\ \Eprint
  {http://arxiv.org/abs/1401.5381} {arXiv:1401.5381 [nucl-th]} \BibitemShut
  {NoStop}%
\bibitem [{\citenamefont {Braaten}\ and\ \citenamefont
  {Thoma}(1991)}]{Braaten:1991jj}%
  \BibitemOpen
  \bibfield  {author} {\bibinfo {author} {\bibfnamefont {E.}~\bibnamefont
  {Braaten}}\ and\ \bibinfo {author} {\bibfnamefont {M.~H.}\ \bibnamefont
  {Thoma}},\ }\href {\doibase 10.1103/PhysRevD.44.1298} {\bibfield  {journal}
  {\bibinfo  {journal} {Phys. Rev.}\ }\textbf {\bibinfo {volume} {D44}},\
  \bibinfo {pages} {1298} (\bibinfo {year} {1991})}\BibitemShut {NoStop}%
\bibitem [{\citenamefont {Chakraborty}\ and\ \citenamefont
  {Kapusta}(2011)}]{Chakraborty:2010fr}%
  \BibitemOpen
  \bibfield  {author} {\bibinfo {author} {\bibfnamefont {P.}~\bibnamefont
  {Chakraborty}}\ and\ \bibinfo {author} {\bibfnamefont {J.~I.}\ \bibnamefont
  {Kapusta}},\ }\href {\doibase 10.1103/PhysRevC.83.014906} {\bibfield
  {journal} {\bibinfo  {journal} {Phys. Rev.}\ }\textbf {\bibinfo {volume}
  {C83}},\ \bibinfo {pages} {014906} (\bibinfo {year} {2011})},\ \Eprint
  {http://arxiv.org/abs/1006.0257} {arXiv:1006.0257 [nucl-th]} \BibitemShut
  {NoStop}%
\bibitem [{\citenamefont {Kubo}(1957)}]{Kubo:1957mj}%
  \BibitemOpen
  \bibfield  {author} {\bibinfo {author} {\bibfnamefont {R.}~\bibnamefont
  {Kubo}},\ }\href {\doibase 10.1143/JPSJ.12.570} {\bibfield  {journal}
  {\bibinfo  {journal} {J. Phys. Soc. Jap.}\ }\textbf {\bibinfo {volume}
  {12}},\ \bibinfo {pages} {570} (\bibinfo {year} {1957})}\BibitemShut
  {NoStop}%
\bibitem [{\citenamefont {Zubarev}, \citenamefont {Morozov},\ and\
  \citenamefont {R{\"o}pke}(1996)}]{zubarev1996statistical}%
  \BibitemOpen
  \bibfield  {author} {\bibinfo {author} {\bibfnamefont {D.}~\bibnamefont
  {Zubarev}}, \bibinfo {author} {\bibfnamefont {V.}~\bibnamefont {Morozov}}, \
  and\ \bibinfo {author} {\bibfnamefont {G.}~\bibnamefont {R{\"o}pke}},\ }\href
  {https://books.google.de/books?id=vxTMQwAACAAJ} {\emph {\bibinfo {title}
  {Statistical Mechanics of Nonequilibrium Processes: Relaxation and
  hydrodynamic processes}}},\ Statistical Mechanics of Nonequilibrium
  Processes\ (\bibinfo  {publisher} {Akademie Verlag},\ \bibinfo {year}
  {1996})\BibitemShut {NoStop}%
\bibitem [{\citenamefont {Aarts}\ and\ \citenamefont
  {Martinez~Resco}(2002)}]{Aarts:2002cc}%
  \BibitemOpen
  \bibfield  {author} {\bibinfo {author} {\bibfnamefont {G.}~\bibnamefont
  {Aarts}}\ and\ \bibinfo {author} {\bibfnamefont {J.~M.}\ \bibnamefont
  {Martinez~Resco}},\ }\href {\doibase 10.1088/1126-6708/2002/04/053}
  {\bibfield  {journal} {\bibinfo  {journal} {JHEP}\ }\textbf {\bibinfo
  {volume} {04}},\ \bibinfo {pages} {053} (\bibinfo {year} {2002})},\ \Eprint
  {http://arxiv.org/abs/hep-ph/0203177} {arXiv:hep-ph/0203177 [hep-ph]}
  \BibitemShut {NoStop}%
\bibitem [{\citenamefont {Iwasaki}, \citenamefont {Ohnishi},\ and\
  \citenamefont {Fukutome}(2008)}]{Iwasaki:2007iv}%
  \BibitemOpen
  \bibfield  {author} {\bibinfo {author} {\bibfnamefont {M.}~\bibnamefont
  {Iwasaki}}, \bibinfo {author} {\bibfnamefont {H.}~\bibnamefont {Ohnishi}}, \
  and\ \bibinfo {author} {\bibfnamefont {T.}~\bibnamefont {Fukutome}},\ }\href
  {\doibase 10.1088/0954-3899/35/3/035003} {\bibfield  {journal} {\bibinfo
  {journal} {J. Phys.}\ }\textbf {\bibinfo {volume} {G35}},\ \bibinfo {pages}
  {035003} (\bibinfo {year} {2008})},\ \Eprint
  {http://arxiv.org/abs/hep-ph/0703271} {arXiv:hep-ph/0703271 [hep-ph]}
  \BibitemShut {NoStop}%
\bibitem [{\citenamefont {Lang}, \citenamefont {Kaiser},\ and\ \citenamefont
  {Weise}(2012)}]{Lang:2012tt}%
  \BibitemOpen
  \bibfield  {author} {\bibinfo {author} {\bibfnamefont {R.}~\bibnamefont
  {Lang}}, \bibinfo {author} {\bibfnamefont {N.}~\bibnamefont {Kaiser}}, \ and\
  \bibinfo {author} {\bibfnamefont {W.}~\bibnamefont {Weise}},\ }\href
  {\doibase 10.1140/epja/i2012-12109-3} {\bibfield  {journal} {\bibinfo
  {journal} {Eur. Phys. J.}\ }\textbf {\bibinfo {volume} {A48}},\ \bibinfo
  {pages} {109} (\bibinfo {year} {2012})},\ \Eprint
  {http://arxiv.org/abs/1205.6648} {arXiv:1205.6648 [hep-ph]} \BibitemShut
  {NoStop}%
\bibitem [{\citenamefont {Lang}\ and\ \citenamefont
  {Weise}(2014)}]{Lang:2013lla}%
  \BibitemOpen
  \bibfield  {author} {\bibinfo {author} {\bibfnamefont {R.}~\bibnamefont
  {Lang}}\ and\ \bibinfo {author} {\bibfnamefont {W.}~\bibnamefont {Weise}},\
  }\href {\doibase 10.1140/epja/i2014-14063-4} {\bibfield  {journal} {\bibinfo
  {journal} {Eur. Phys. J.}\ }\textbf {\bibinfo {volume} {A50}},\ \bibinfo
  {pages} {63} (\bibinfo {year} {2014})},\ \Eprint
  {http://arxiv.org/abs/1311.4628} {arXiv:1311.4628 [hep-ph]} \BibitemShut
  {NoStop}%
\bibitem [{\citenamefont {Haas}, \citenamefont {Fister},\ and\ \citenamefont
  {Pawlowski}(2014)}]{Haas:2013hpa}%
  \BibitemOpen
  \bibfield  {author} {\bibinfo {author} {\bibfnamefont {M.}~\bibnamefont
  {Haas}}, \bibinfo {author} {\bibfnamefont {L.}~\bibnamefont {Fister}}, \ and\
  \bibinfo {author} {\bibfnamefont {J.~M.}\ \bibnamefont {Pawlowski}},\ }\href
  {\doibase 10.1103/PhysRevD.90.091501} {\bibfield  {journal} {\bibinfo
  {journal} {Phys. Rev.}\ }\textbf {\bibinfo {volume} {D90}},\ \bibinfo {pages}
  {091501} (\bibinfo {year} {2014})},\ \Eprint {http://arxiv.org/abs/1308.4960}
  {arXiv:1308.4960 [hep-ph]} \BibitemShut {NoStop}%
\bibitem [{\citenamefont {Christiansen}\ \emph {et~al.}(2015)\citenamefont
  {Christiansen}, \citenamefont {Haas}, \citenamefont {Pawlowski},\ and\
  \citenamefont {Strodthoff}}]{Christiansen:2014ypa}%
  \BibitemOpen
  \bibfield  {author} {\bibinfo {author} {\bibfnamefont {N.}~\bibnamefont
  {Christiansen}}, \bibinfo {author} {\bibfnamefont {M.}~\bibnamefont {Haas}},
  \bibinfo {author} {\bibfnamefont {J.~M.}\ \bibnamefont {Pawlowski}}, \ and\
  \bibinfo {author} {\bibfnamefont {N.}~\bibnamefont {Strodthoff}},\ }\href
  {\doibase 10.1103/PhysRevLett.115.112002} {\bibfield  {journal} {\bibinfo
  {journal} {Phys. Rev. Lett.}\ }\textbf {\bibinfo {volume} {115}},\ \bibinfo
  {pages} {112002} (\bibinfo {year} {2015})},\ \Eprint
  {http://arxiv.org/abs/1411.7986} {arXiv:1411.7986 [hep-ph]} \BibitemShut
  {NoStop}%
\bibitem [{\citenamefont {Ozvenchuk}\ \emph
  {et~al.}(2013{\natexlab{b}})\citenamefont {Ozvenchuk}, \citenamefont
  {Linnyk}, \citenamefont {Gorenstein}, \citenamefont {Bratkovskaya},\ and\
  \citenamefont {Cassing}}]{Ozvenchuk:2012kh}%
  \BibitemOpen
  \bibfield  {author} {\bibinfo {author} {\bibfnamefont {V.}~\bibnamefont
  {Ozvenchuk}}, \bibinfo {author} {\bibfnamefont {O.}~\bibnamefont {Linnyk}},
  \bibinfo {author} {\bibfnamefont {M.~I.}\ \bibnamefont {Gorenstein}},
  \bibinfo {author} {\bibfnamefont {E.~L.}\ \bibnamefont {Bratkovskaya}}, \
  and\ \bibinfo {author} {\bibfnamefont {W.}~\bibnamefont {Cassing}},\ }\href
  {\doibase 10.1103/PhysRevC.87.064903} {\bibfield  {journal} {\bibinfo
  {journal} {Phys. Rev.}\ }\textbf {\bibinfo {volume} {C87}},\ \bibinfo {pages}
  {064903} (\bibinfo {year} {2013}{\natexlab{b}})},\ \Eprint
  {http://arxiv.org/abs/1212.5393} {arXiv:1212.5393 [hep-ph]} \BibitemShut
  {NoStop}%
\bibitem [{\citenamefont {Sasaki}\ and\ \citenamefont
  {Redlich}(2009)}]{Sasaki:2008fg}%
  \BibitemOpen
  \bibfield  {author} {\bibinfo {author} {\bibfnamefont {C.}~\bibnamefont
  {Sasaki}}\ and\ \bibinfo {author} {\bibfnamefont {K.}~\bibnamefont
  {Redlich}},\ }\href {\doibase 10.1103/PhysRevC.79.055207} {\bibfield
  {journal} {\bibinfo  {journal} {Phys. Rev.}\ }\textbf {\bibinfo {volume}
  {C79}},\ \bibinfo {pages} {055207} (\bibinfo {year} {2009})},\ \Eprint
  {http://arxiv.org/abs/0806.4745} {arXiv:0806.4745 [hep-ph]} \BibitemShut
  {NoStop}%
\bibitem [{\citenamefont {Sasaki}\ and\ \citenamefont
  {Redlich}(2010)}]{Sasaki:2008um}%
  \BibitemOpen
  \bibfield  {author} {\bibinfo {author} {\bibfnamefont {C.}~\bibnamefont
  {Sasaki}}\ and\ \bibinfo {author} {\bibfnamefont {K.}~\bibnamefont
  {Redlich}},\ }\href {\doibase 10.1016/j.nuclphysa.2009.11.005} {\bibfield
  {journal} {\bibinfo  {journal} {Nucl. Phys.}\ }\textbf {\bibinfo {volume}
  {A832}},\ \bibinfo {pages} {62} (\bibinfo {year} {2010})},\ \Eprint
  {http://arxiv.org/abs/0811.4708} {arXiv:0811.4708 [hep-ph]} \BibitemShut
  {NoStop}%
\bibitem [{\citenamefont {Bluhm}, \citenamefont {K{\"a}mpfer},\ and\
  \citenamefont {Redlich}(2009)}]{Bluhm:2009ef}%
  \BibitemOpen
  \bibfield  {author} {\bibinfo {author} {\bibfnamefont {M.}~\bibnamefont
  {Bluhm}}, \bibinfo {author} {\bibfnamefont {B.}~\bibnamefont {K{\"a}mpfer}},
  \ and\ \bibinfo {author} {\bibfnamefont {K.}~\bibnamefont {Redlich}},\
  }\bibfield  {booktitle} {\emph {\bibinfo {booktitle} {{Proceedings, 21st
  International Conference on Ultra-Relativistic nucleus nucleus collisions
  (Quark matter 2009): Knoxville, USA, March 30-April 4, 2009}}},\ }\href
  {\doibase 10.1016/j.nuclphysa.2009.10.065} {\bibfield  {journal} {\bibinfo
  {journal} {Nucl. Phys.}\ }\textbf {\bibinfo {volume} {A830}},\ \bibinfo
  {pages} {737C} (\bibinfo {year} {2009})},\ \Eprint
  {http://arxiv.org/abs/0907.3841} {arXiv:0907.3841 [hep-ph]} \BibitemShut
  {NoStop}%
\bibitem [{\citenamefont {Bluhm}, \citenamefont {K{\"a}mpfer},\ and\
  \citenamefont {Redlich}(2011)}]{Bluhm:2010qf}%
  \BibitemOpen
  \bibfield  {author} {\bibinfo {author} {\bibfnamefont {M.}~\bibnamefont
  {Bluhm}}, \bibinfo {author} {\bibfnamefont {B.}~\bibnamefont {K{\"a}mpfer}},
  \ and\ \bibinfo {author} {\bibfnamefont {K.}~\bibnamefont {Redlich}},\ }\href
  {\doibase 10.1103/PhysRevC.84.025201} {\bibfield  {journal} {\bibinfo
  {journal} {Phys. Rev.}\ }\textbf {\bibinfo {volume} {C84}},\ \bibinfo {pages}
  {025201} (\bibinfo {year} {2011})},\ \Eprint {http://arxiv.org/abs/1011.5634}
  {arXiv:1011.5634 [hep-ph]} \BibitemShut {NoStop}%
\bibitem [{\citenamefont {Albright}\ and\ \citenamefont
  {Kapusta}(2016)}]{Albright:2015fpa}%
  \BibitemOpen
  \bibfield  {author} {\bibinfo {author} {\bibfnamefont {M.}~\bibnamefont
  {Albright}}\ and\ \bibinfo {author} {\bibfnamefont {J.~I.}\ \bibnamefont
  {Kapusta}},\ }\href {\doibase 10.1103/PhysRevC.93.014903} {\bibfield
  {journal} {\bibinfo  {journal} {Phys. Rev.}\ }\textbf {\bibinfo {volume}
  {C93}},\ \bibinfo {pages} {014903} (\bibinfo {year} {2016})},\ \Eprint
  {http://arxiv.org/abs/1508.02696} {arXiv:1508.02696 [nucl-th]} \BibitemShut
  {NoStop}%
\bibitem [{\citenamefont {Policastro}, \citenamefont {Son},\ and\ \citenamefont
  {Starinets}(2001)}]{Policastro:2001yc}%
  \BibitemOpen
  \bibfield  {author} {\bibinfo {author} {\bibfnamefont {G.}~\bibnamefont
  {Policastro}}, \bibinfo {author} {\bibfnamefont {D.~T.}\ \bibnamefont {Son}},
  \ and\ \bibinfo {author} {\bibfnamefont {A.~O.}\ \bibnamefont {Starinets}},\
  }\href {\doibase 10.1103/PhysRevLett.87.081601} {\bibfield  {journal}
  {\bibinfo  {journal} {Phys. Rev. Lett.}\ }\textbf {\bibinfo {volume} {87}},\
  \bibinfo {pages} {081601} (\bibinfo {year} {2001})},\ \Eprint
  {http://arxiv.org/abs/hep-th/0104066} {arXiv:hep-th/0104066 [hep-th]}
  \BibitemShut {NoStop}%
\bibitem [{\citenamefont {Kovtun}, \citenamefont {Son},\ and\ \citenamefont
  {Starinets}(2005)}]{Kovtun:2004de}%
  \BibitemOpen
  \bibfield  {author} {\bibinfo {author} {\bibfnamefont {P.}~\bibnamefont
  {Kovtun}}, \bibinfo {author} {\bibfnamefont {D.~T.}\ \bibnamefont {Son}}, \
  and\ \bibinfo {author} {\bibfnamefont {A.~O.}\ \bibnamefont {Starinets}},\
  }\href {\doibase 10.1103/PhysRevLett.94.111601} {\bibfield  {journal}
  {\bibinfo  {journal} {Phys. Rev. Lett.}\ }\textbf {\bibinfo {volume} {94}},\
  \bibinfo {pages} {111601} (\bibinfo {year} {2005})},\ \Eprint
  {http://arxiv.org/abs/hep-th/0405231} {arXiv:hep-th/0405231 [hep-th]}
  \BibitemShut {NoStop}%
\bibitem [{\citenamefont {Astrakhantsev}, \citenamefont {Braguta},\ and\
  \citenamefont {Kotov}(2017)}]{Astrakhantsev:2017nrs}%
  \BibitemOpen
  \bibfield  {author} {\bibinfo {author} {\bibfnamefont {N.}~\bibnamefont
  {Astrakhantsev}}, \bibinfo {author} {\bibfnamefont {V.}~\bibnamefont
  {Braguta}}, \ and\ \bibinfo {author} {\bibfnamefont {A.}~\bibnamefont
  {Kotov}},\ }\href {\doibase 10.1007/JHEP04(2017)101} {\bibfield  {journal}
  {\bibinfo  {journal} {JHEP}\ }\textbf {\bibinfo {volume} {04}},\ \bibinfo
  {pages} {101} (\bibinfo {year} {2017})},\ \Eprint
  {http://arxiv.org/abs/1701.02266} {arXiv:1701.02266 [hep-lat]} \BibitemShut
  {NoStop}%
\bibitem [{\citenamefont {G{\"u}nther}\ \emph {et~al.}(2017)\citenamefont
  {G{\"u}nther}, \citenamefont {Bellwied}, \citenamefont {Borsanyi},
  \citenamefont {Fodor}, \citenamefont {Katz}, \citenamefont {Pasztor},\ and\
  \citenamefont {Ratti}}]{Gunther:2017sxn}%
  \BibitemOpen
  \bibfield  {author} {\bibinfo {author} {\bibfnamefont {J.}~\bibnamefont
  {G{\"u}nther}}, \bibinfo {author} {\bibfnamefont {R.}~\bibnamefont
  {Bellwied}}, \bibinfo {author} {\bibfnamefont {S.}~\bibnamefont {Borsanyi}},
  \bibinfo {author} {\bibfnamefont {Z.}~\bibnamefont {Fodor}}, \bibinfo
  {author} {\bibfnamefont {S.~D.}\ \bibnamefont {Katz}}, \bibinfo {author}
  {\bibfnamefont {A.}~\bibnamefont {Pasztor}}, \ and\ \bibinfo {author}
  {\bibfnamefont {C.}~\bibnamefont {Ratti}},\ }\bibfield  {booktitle} {\emph
  {\bibinfo {booktitle} {{Proceedings, 12th Conference on Quark Confinement and
  the Hadron Spectrum (Confinement XII): Thessaloniki, Greece}}},\ }\href
  {\doibase 10.1051/epjconf/201713707008} {\bibfield  {journal} {\bibinfo
  {journal} {EPJ Web Conf.}\ }\textbf {\bibinfo {volume} {137}},\ \bibinfo
  {pages} {07008} (\bibinfo {year} {2017})}\BibitemShut {NoStop}%
\bibitem [{\citenamefont {Xu}\ \emph {et~al.}(2017)\citenamefont {Xu},
  \citenamefont {Moreau}, \citenamefont {Song}, \citenamefont {Nahrgang},
  \citenamefont {Bass},\ and\ \citenamefont {Bratkovskaya}}]{Xu:2017pna}%
  \BibitemOpen
  \bibfield  {author} {\bibinfo {author} {\bibfnamefont {Y.}~\bibnamefont
  {Xu}}, \bibinfo {author} {\bibfnamefont {P.}~\bibnamefont {Moreau}}, \bibinfo
  {author} {\bibfnamefont {T.}~\bibnamefont {Song}}, \bibinfo {author}
  {\bibfnamefont {M.}~\bibnamefont {Nahrgang}}, \bibinfo {author}
  {\bibfnamefont {S.~A.}\ \bibnamefont {Bass}}, \ and\ \bibinfo {author}
  {\bibfnamefont {E.}~\bibnamefont {Bratkovskaya}},\ }\href {\doibase
  10.1103/PhysRevC.96.024902} {\bibfield  {journal} {\bibinfo  {journal} {Phys.
  Rev.}\ }\textbf {\bibinfo {volume} {C96}},\ \bibinfo {pages} {024902}
  (\bibinfo {year} {2017})},\ \Eprint {http://arxiv.org/abs/1703.09178}
  {arXiv:1703.09178 [nucl-th]} \BibitemShut {NoStop}%
\bibitem [{\citenamefont {Ryblewski}\ and\ \citenamefont
  {Florkowski}(2012)}]{Ryblewski:2012rr}%
  \BibitemOpen
  \bibfield  {author} {\bibinfo {author} {\bibfnamefont {R.}~\bibnamefont
  {Ryblewski}}\ and\ \bibinfo {author} {\bibfnamefont {W.}~\bibnamefont
  {Florkowski}},\ }\href {\doibase 10.1103/PhysRevC.85.064901} {\bibfield
  {journal} {\bibinfo  {journal} {Phys. Rev.}\ }\textbf {\bibinfo {volume}
  {C85}},\ \bibinfo {pages} {064901} (\bibinfo {year} {2012})},\ \Eprint
  {http://arxiv.org/abs/1204.2624} {arXiv:1204.2624 [nucl-th]} \BibitemShut
  {NoStop}%
\bibitem [{\citenamefont {Newton}\ and\ \citenamefont {Colson}(1736)}]{newton}%
  \BibitemOpen
  \bibfield  {author} {\bibinfo {author} {\bibfnamefont {I.}~\bibnamefont
  {Newton}}\ and\ \bibinfo {author} {\bibfnamefont {J.}~\bibnamefont
  {Colson}},\ }\href {https://www.loc.gov/item/42048007/} {\emph {\bibinfo
  {title} {The Method of Fluxions and Infinite Series;: With Its Application to
  the Geometry of Curve-Lines}}}\ (\bibinfo  {publisher} {London:: Printed by
  Henry Woodfall; and sold by John Nourse},\ \bibinfo {year}
  {1736})\BibitemShut {NoStop}%
\bibitem [{\citenamefont {Raphson}(1697)}]{raphson}%
  \BibitemOpen
  \bibfield  {author} {\bibinfo {author} {\bibfnamefont {J.}~\bibnamefont
  {Raphson}},\ }\href {\doibase 10.3931/e-rara-13516} {\emph {\bibinfo {title}
  {{Analysis aequationum universalis, seu, Ad aequationes algebraicas
  resolvendas methodus generalis, \& expedita, ex nova infinitarum serierum
  methodo, deducta ac demonstrata}}}},\ \bibinfo {edition} {editio secunda cum
  appendice ; cui annexum est, de spatio reali, seu entre infinito conamen
  mathematico-metaphysicum.}\ ed.\ (\bibinfo  {publisher} {Typis T. Braddyll,
  prostant venales apud Johannem Taylor Londini},\ \bibinfo {year}
  {1697})\BibitemShut {NoStop}%
\bibitem [{\citenamefont {Cleymans}\ \emph {et~al.}(2006)\citenamefont
  {Cleymans}, \citenamefont {Oeschler}, \citenamefont {Redlich},\ and\
  \citenamefont {Wheaton}}]{Cleymans:2005xv}%
  \BibitemOpen
  \bibfield  {author} {\bibinfo {author} {\bibfnamefont {J.}~\bibnamefont
  {Cleymans}}, \bibinfo {author} {\bibfnamefont {H.}~\bibnamefont {Oeschler}},
  \bibinfo {author} {\bibfnamefont {K.}~\bibnamefont {Redlich}}, \ and\
  \bibinfo {author} {\bibfnamefont {S.}~\bibnamefont {Wheaton}},\ }\href
  {\doibase 10.1103/PhysRevC.73.034905} {\bibfield  {journal} {\bibinfo
  {journal} {Phys. Rev.}\ }\textbf {\bibinfo {volume} {C73}},\ \bibinfo {pages}
  {034905} (\bibinfo {year} {2006})},\ \Eprint
  {http://arxiv.org/abs/hep-ph/0511094} {arXiv:hep-ph/0511094 [hep-ph]}
  \BibitemShut {NoStop}%
\bibitem [{\citenamefont {Reiter}\ \emph {et~al.}(1998)\citenamefont {Reiter},
  \citenamefont {Dumitru}, \citenamefont {Brachmann}, \citenamefont {Maruhn},
  \citenamefont {Stoecker},\ and\ \citenamefont {Greiner}}]{Reiter:1998uq}%
  \BibitemOpen
  \bibfield  {author} {\bibinfo {author} {\bibfnamefont {M.}~\bibnamefont
  {Reiter}}, \bibinfo {author} {\bibfnamefont {A.}~\bibnamefont {Dumitru}},
  \bibinfo {author} {\bibfnamefont {J.}~\bibnamefont {Brachmann}}, \bibinfo
  {author} {\bibfnamefont {J.~A.}\ \bibnamefont {Maruhn}}, \bibinfo {author}
  {\bibfnamefont {H.}~\bibnamefont {Stoecker}}, \ and\ \bibinfo {author}
  {\bibfnamefont {W.}~\bibnamefont {Greiner}},\ }\href {\doibase
  10.1016/S0375-9474(98)00556-9} {\bibfield  {journal} {\bibinfo  {journal}
  {Nucl. Phys.}\ }\textbf {\bibinfo {volume} {A643}},\ \bibinfo {pages} {99}
  (\bibinfo {year} {1998})},\ \Eprint {http://arxiv.org/abs/nucl-th/9806010}
  {arXiv:nucl-th/9806010 [nucl-th]} \BibitemShut {NoStop}%
\bibitem [{\citenamefont {Cassing}\ and\ \citenamefont
  {Bratkovskaya}(1999)}]{Cassing:1999es}%
  \BibitemOpen
  \bibfield  {author} {\bibinfo {author} {\bibfnamefont {W.}~\bibnamefont
  {Cassing}}\ and\ \bibinfo {author} {\bibfnamefont {E.~L.}\ \bibnamefont
  {Bratkovskaya}},\ }\href {\doibase 10.1016/S0370-1573(98)00028-3} {\bibfield
  {journal} {\bibinfo  {journal} {Phys. Rept.}\ }\textbf {\bibinfo {volume}
  {308}},\ \bibinfo {pages} {65} (\bibinfo {year} {1999})}\BibitemShut
  {NoStop}%
\bibitem [{\citenamefont {Akiba}\ \emph {et~al.}(1996)\citenamefont {Akiba}
  \emph {et~al.}}]{Akiba:1996xf}%
  \BibitemOpen
  \bibfield  {author} {\bibinfo {author} {\bibfnamefont {Y.}~\bibnamefont
  {Akiba}} \emph {et~al.} (\bibinfo {collaboration} {E802}),\ }\bibfield
  {booktitle} {\emph {\bibinfo {booktitle} {{Quark matter '96. Proceedings,
  12th International Conference on Ultrarelativistic Nucleus Nucleus
  Collisions, Heidelberg, Germany, May 20-24, 1996}}},\ }\href {\doibase
  10.1016/S0375-9474(96)00350-8} {\bibfield  {journal} {\bibinfo  {journal}
  {Nucl. Phys.}\ }\textbf {\bibinfo {volume} {A610}},\ \bibinfo {pages} {139C}
  (\bibinfo {year} {1996})}\BibitemShut {NoStop}%
\bibitem [{\citenamefont {Lacasse}\ \emph {et~al.}(1996)\citenamefont {Lacasse}
  \emph {et~al.}}]{Lacasse:1996gb}%
  \BibitemOpen
  \bibfield  {author} {\bibinfo {author} {\bibfnamefont {R.}~\bibnamefont
  {Lacasse}} \emph {et~al.} (\bibinfo {collaboration} {E877}),\ }\bibfield
  {booktitle} {\emph {\bibinfo {booktitle} {{Quark matter '96. Proceedings,
  12th International Conference on Ultrarelativistic Nucleus Nucleus
  Collisions, Heidelberg, Germany, May 20-24, 1996}}},\ }\href {\doibase
  10.1016/S0375-9474(96)00351-X} {\bibfield  {journal} {\bibinfo  {journal}
  {Nucl. Phys.}\ }\textbf {\bibinfo {volume} {A610}},\ \bibinfo {pages} {153C}
  (\bibinfo {year} {1996})},\ \Eprint {http://arxiv.org/abs/nucl-ex/9609001}
  {arXiv:nucl-ex/9609001 [nucl-ex]} \BibitemShut {NoStop}%
\bibitem [{\citenamefont {Ahmad}\ \emph {et~al.}(1996)\citenamefont {Ahmad},
  \citenamefont {Bonner}, \citenamefont {Chan}, \citenamefont {Clement},
  \citenamefont {Efremov}, \citenamefont {Efstathiadis}, \citenamefont
  {Eiseman}, \citenamefont {Etkin}, \citenamefont {Foley}, \citenamefont
  {Hackenburg} \emph {et~al.}}]{Ahmad:1991nv}%
  \BibitemOpen
  \bibfield  {author} {\bibinfo {author} {\bibfnamefont {S.}~\bibnamefont
  {Ahmad}}, \bibinfo {author} {\bibfnamefont {B.}~\bibnamefont {Bonner}},
  \bibinfo {author} {\bibfnamefont {C.}~\bibnamefont {Chan}}, \bibinfo {author}
  {\bibfnamefont {J.}~\bibnamefont {Clement}}, \bibinfo {author} {\bibfnamefont
  {S.}~\bibnamefont {Efremov}}, \bibinfo {author} {\bibfnamefont
  {E.}~\bibnamefont {Efstathiadis}}, \bibinfo {author} {\bibfnamefont
  {S.}~\bibnamefont {Eiseman}}, \bibinfo {author} {\bibfnamefont
  {A.}~\bibnamefont {Etkin}}, \bibinfo {author} {\bibfnamefont
  {K.}~\bibnamefont {Foley}}, \bibinfo {author} {\bibfnamefont
  {R.}~\bibnamefont {Hackenburg}},  \emph {et~al.},\ }\href {\doibase
  https://doi.org/10.1016/0370-2693(96)00642-9} {\bibfield  {journal} {\bibinfo
   {journal} {Physics Letters B}\ }\textbf {\bibinfo {volume} {382}},\ \bibinfo
  {pages} {35 } (\bibinfo {year} {1996})}\BibitemShut {NoStop}%
\bibitem [{\citenamefont {Barrette}\ \emph {et~al.}(2001)\citenamefont
  {Barrette} \emph {et~al.}}]{Barrette:2000cb}%
  \BibitemOpen
  \bibfield  {author} {\bibinfo {author} {\bibfnamefont {J.}~\bibnamefont
  {Barrette}} \emph {et~al.} (\bibinfo {collaboration} {E877}),\ }\href
  {\doibase 10.1103/PhysRevC.63.014902} {\bibfield  {journal} {\bibinfo
  {journal} {Phys. Rev.}\ }\textbf {\bibinfo {volume} {C63}},\ \bibinfo {pages}
  {014902} (\bibinfo {year} {2001})},\ \Eprint
  {http://arxiv.org/abs/nucl-ex/0007007} {arXiv:nucl-ex/0007007 [nucl-ex]}
  \BibitemShut {NoStop}%
\bibitem [{\citenamefont {Albergo}\ \emph {et~al.}(2002)\citenamefont
  {Albergo}, \citenamefont {Bellwied}, \citenamefont {Bennett}, \citenamefont
  {Boemi}, \citenamefont {Bonner}, \citenamefont {Caines}, \citenamefont
  {Christie}, \citenamefont {Costa}, \citenamefont {Crawford}, \citenamefont
  {Cronqvist} \emph {et~al.}}]{Albergo:2002tn}%
  \BibitemOpen
  \bibfield  {author} {\bibinfo {author} {\bibfnamefont {S.}~\bibnamefont
  {Albergo}}, \bibinfo {author} {\bibfnamefont {R.}~\bibnamefont {Bellwied}},
  \bibinfo {author} {\bibfnamefont {M.}~\bibnamefont {Bennett}}, \bibinfo
  {author} {\bibfnamefont {D.}~\bibnamefont {Boemi}}, \bibinfo {author}
  {\bibfnamefont {B.}~\bibnamefont {Bonner}}, \bibinfo {author} {\bibfnamefont
  {H.}~\bibnamefont {Caines}}, \bibinfo {author} {\bibfnamefont
  {W.}~\bibnamefont {Christie}}, \bibinfo {author} {\bibfnamefont
  {S.}~\bibnamefont {Costa}}, \bibinfo {author} {\bibfnamefont {H.~J.}\
  \bibnamefont {Crawford}}, \bibinfo {author} {\bibfnamefont {M.}~\bibnamefont
  {Cronqvist}},  \emph {et~al.},\ }\href {\doibase
  10.1103/PhysRevLett.88.062301} {\bibfield  {journal} {\bibinfo  {journal}
  {Phys. Rev. Lett.}\ }\textbf {\bibinfo {volume} {88}},\ \bibinfo {pages}
  {062301} (\bibinfo {year} {2002})}\BibitemShut {NoStop}%
\bibitem [{\citenamefont {Alt}\ \emph {et~al.}(2006)\citenamefont {Alt} \emph
  {et~al.}}]{Alt:2006dk}%
  \BibitemOpen
  \bibfield  {author} {\bibinfo {author} {\bibfnamefont {C.}~\bibnamefont
  {Alt}} \emph {et~al.} (\bibinfo {collaboration} {NA49}),\ }\href {\doibase
  10.1103/PhysRevC.73.044910} {\bibfield  {journal} {\bibinfo  {journal} {Phys.
  Rev.}\ }\textbf {\bibinfo {volume} {C73}},\ \bibinfo {pages} {044910}
  (\bibinfo {year} {2006})}\BibitemShut {NoStop}%
\bibitem [{\citenamefont {Alt}\ \emph {et~al.}(2008{\natexlab{a}})\citenamefont
  {Alt} \emph {et~al.}}]{Alt:2007aa}%
  \BibitemOpen
  \bibfield  {author} {\bibinfo {author} {\bibfnamefont {C.}~\bibnamefont
  {Alt}} \emph {et~al.} (\bibinfo {collaboration} {NA49}),\ }\href {\doibase
  10.1103/PhysRevC.77.024903} {\bibfield  {journal} {\bibinfo  {journal} {Phys.
  Rev.}\ }\textbf {\bibinfo {volume} {C77}},\ \bibinfo {pages} {024903}
  (\bibinfo {year} {2008}{\natexlab{a}})},\ \Eprint
  {http://arxiv.org/abs/0710.0118} {arXiv:0710.0118 [nucl-ex]} \BibitemShut
  {NoStop}%
\bibitem [{\citenamefont {Alt}\ \emph {et~al.}(2008{\natexlab{b}})\citenamefont
  {Alt} \emph {et~al.}}]{Alt:2008qm}%
  \BibitemOpen
  \bibfield  {author} {\bibinfo {author} {\bibfnamefont {C.}~\bibnamefont
  {Alt}} \emph {et~al.} (\bibinfo {collaboration} {NA49}),\ }\href {\doibase
  10.1103/PhysRevC.78.034918} {\bibfield  {journal} {\bibinfo  {journal} {Phys.
  Rev.}\ }\textbf {\bibinfo {volume} {C78}},\ \bibinfo {pages} {034918}
  (\bibinfo {year} {2008}{\natexlab{b}})},\ \Eprint
  {http://arxiv.org/abs/0804.3770} {arXiv:0804.3770 [nucl-ex]} \BibitemShut
  {NoStop}%
\bibitem [{\citenamefont {Afanasiev}\ \emph {et~al.}(2002)\citenamefont
  {Afanasiev} \emph {et~al.}}]{Afanasiev:2002mx}%
  \BibitemOpen
  \bibfield  {author} {\bibinfo {author} {\bibfnamefont {S.~V.}\ \bibnamefont
  {Afanasiev}} \emph {et~al.} (\bibinfo {collaboration} {NA49}),\ }\href
  {\doibase 10.1103/PhysRevC.66.054902} {\bibfield  {journal} {\bibinfo
  {journal} {Phys. Rev.}\ }\textbf {\bibinfo {volume} {C66}},\ \bibinfo {pages}
  {054902} (\bibinfo {year} {2002})},\ \Eprint
  {http://arxiv.org/abs/nucl-ex/0205002} {arXiv:nucl-ex/0205002 [nucl-ex]}
  \BibitemShut {NoStop}%
\bibitem [{\citenamefont {Anticic}\ \emph {et~al.}(2004)\citenamefont {Anticic}
  \emph {et~al.}}]{Anticic:2003ux}%
  \BibitemOpen
  \bibfield  {author} {\bibinfo {author} {\bibfnamefont {T.}~\bibnamefont
  {Anticic}} \emph {et~al.} (\bibinfo {collaboration} {NA49}),\ }\href
  {\doibase 10.1103/PhysRevLett.93.022302} {\bibfield  {journal} {\bibinfo
  {journal} {Phys. Rev. Lett.}\ }\textbf {\bibinfo {volume} {93}},\ \bibinfo
  {pages} {022302} (\bibinfo {year} {2004})},\ \Eprint
  {http://arxiv.org/abs/nucl-ex/0311024} {arXiv:nucl-ex/0311024 [nucl-ex]}
  \BibitemShut {NoStop}%
\bibitem [{\citenamefont {Anticic}\ \emph {et~al.}(2011)\citenamefont {Anticic}
  \emph {et~al.}}]{Anticic:2010mp}%
  \BibitemOpen
  \bibfield  {author} {\bibinfo {author} {\bibfnamefont {T.}~\bibnamefont
  {Anticic}} \emph {et~al.} (\bibinfo {collaboration} {NA49}),\ }\href
  {\doibase 10.1103/PhysRevC.83.014901} {\bibfield  {journal} {\bibinfo
  {journal} {Phys. Rev.}\ }\textbf {\bibinfo {volume} {C83}},\ \bibinfo {pages}
  {014901} (\bibinfo {year} {2011})},\ \Eprint {http://arxiv.org/abs/1009.1747}
  {arXiv:1009.1747 [nucl-ex]} \BibitemShut {NoStop}%
\bibitem [{\citenamefont {Anticic}\ \emph {et~al.}(2012)\citenamefont {Anticic}
  \emph {et~al.}}]{Anticic:2012ay}%
  \BibitemOpen
  \bibfield  {author} {\bibinfo {author} {\bibfnamefont {T.}~\bibnamefont
  {Anticic}} \emph {et~al.} (\bibinfo {collaboration} {NA49}),\ }\href
  {\doibase 10.1103/PhysRevC.86.054903} {\bibfield  {journal} {\bibinfo
  {journal} {Phys. Rev.}\ }\textbf {\bibinfo {volume} {C86}},\ \bibinfo {pages}
  {054903} (\bibinfo {year} {2012})},\ \Eprint {http://arxiv.org/abs/1207.0348}
  {arXiv:1207.0348 [nucl-ex]} \BibitemShut {NoStop}%
\bibitem [{\citenamefont {Ahle}\ \emph
  {et~al.}(2000{\natexlab{a}})\citenamefont {Ahle} \emph
  {et~al.}}]{Ahle:1999uy}%
  \BibitemOpen
  \bibfield  {author} {\bibinfo {author} {\bibfnamefont {L.}~\bibnamefont
  {Ahle}} \emph {et~al.} (\bibinfo {collaboration} {E917, E866}),\ }\href
  {\doibase 10.1016/S0370-2693(00)00037-X} {\bibfield  {journal} {\bibinfo
  {journal} {Phys. Lett.}\ }\textbf {\bibinfo {volume} {B476}},\ \bibinfo
  {pages} {1} (\bibinfo {year} {2000}{\natexlab{a}})},\ \Eprint
  {http://arxiv.org/abs/nucl-ex/9910008} {arXiv:nucl-ex/9910008 [nucl-ex]}
  \BibitemShut {NoStop}%
\bibitem [{\citenamefont {Ahle}\ \emph
  {et~al.}(2000{\natexlab{b}})\citenamefont {Ahle} \emph
  {et~al.}}]{Ahle:2000wq}%
  \BibitemOpen
  \bibfield  {author} {\bibinfo {author} {\bibfnamefont {L.}~\bibnamefont
  {Ahle}} \emph {et~al.} (\bibinfo {collaboration} {E917, E866}),\ }\href
  {\doibase 10.1016/S0370-2693(00)00916-3} {\bibfield  {journal} {\bibinfo
  {journal} {Phys. Lett.}\ }\textbf {\bibinfo {volume} {B490}},\ \bibinfo
  {pages} {53} (\bibinfo {year} {2000}{\natexlab{b}})},\ \Eprint
  {http://arxiv.org/abs/nucl-ex/0008010} {arXiv:nucl-ex/0008010 [nucl-ex]}
  \BibitemShut {NoStop}%
\bibitem [{\citenamefont {Bearden}\ \emph {et~al.}(2005)\citenamefont {Bearden}
  \emph {et~al.}}]{Bearden:2004yx}%
  \BibitemOpen
  \bibfield  {author} {\bibinfo {author} {\bibfnamefont {I.~G.}\ \bibnamefont
  {Bearden}} \emph {et~al.} (\bibinfo {collaboration} {BRAHMS}),\ }\href
  {\doibase 10.1103/PhysRevLett.94.162301} {\bibfield  {journal} {\bibinfo
  {journal} {Phys. Rev. Lett.}\ }\textbf {\bibinfo {volume} {94}},\ \bibinfo
  {pages} {162301} (\bibinfo {year} {2005})},\ \Eprint
  {http://arxiv.org/abs/nucl-ex/0403050} {arXiv:nucl-ex/0403050 [nucl-ex]}
  \BibitemShut {NoStop}%
\bibitem [{\citenamefont {Arsene}\ \emph {et~al.}(2005)\citenamefont {Arsene}
  \emph {et~al.}}]{Arsene:2005mr}%
  \BibitemOpen
  \bibfield  {author} {\bibinfo {author} {\bibfnamefont {I.}~\bibnamefont
  {Arsene}} \emph {et~al.} (\bibinfo {collaboration} {BRAHMS}),\ }\href
  {\doibase 10.1103/PhysRevC.72.014908} {\bibfield  {journal} {\bibinfo
  {journal} {Phys. Rev.}\ }\textbf {\bibinfo {volume} {C72}},\ \bibinfo {pages}
  {014908} (\bibinfo {year} {2005})},\ \Eprint
  {http://arxiv.org/abs/nucl-ex/0503010} {arXiv:nucl-ex/0503010 [nucl-ex]}
  \BibitemShut {NoStop}%
\bibitem [{\citenamefont {Adler}\ \emph {et~al.}(2004)\citenamefont {Adler}
  \emph {et~al.}}]{Adler:2003cb}%
  \BibitemOpen
  \bibfield  {author} {\bibinfo {author} {\bibfnamefont {S.~S.}\ \bibnamefont
  {Adler}} \emph {et~al.} (\bibinfo {collaboration} {PHENIX}),\ }\href
  {\doibase 10.1103/PhysRevC.69.034909} {\bibfield  {journal} {\bibinfo
  {journal} {Phys. Rev.}\ }\textbf {\bibinfo {volume} {C69}},\ \bibinfo {pages}
  {034909} (\bibinfo {year} {2004})},\ \Eprint
  {http://arxiv.org/abs/nucl-ex/0307022} {arXiv:nucl-ex/0307022 [nucl-ex]}
  \BibitemShut {NoStop}%
\bibitem [{\citenamefont {Agakishiev}\ \emph {et~al.}(2012)\citenamefont
  {Agakishiev} \emph {et~al.}}]{Agakishiev:2011ar}%
  \BibitemOpen
  \bibfield  {author} {\bibinfo {author} {\bibfnamefont {G.}~\bibnamefont
  {Agakishiev}} \emph {et~al.} (\bibinfo {collaboration} {STAR}),\ }\href
  {\doibase 10.1103/PhysRevLett.108.072301} {\bibfield  {journal} {\bibinfo
  {journal} {Phys. Rev. Lett.}\ }\textbf {\bibinfo {volume} {108}},\ \bibinfo
  {pages} {072301} (\bibinfo {year} {2012})},\ \Eprint
  {http://arxiv.org/abs/1107.2955} {arXiv:1107.2955 [nucl-ex]} \BibitemShut
  {NoStop}%
\bibitem [{\citenamefont {Cutler}\ and\ \citenamefont
  {Sivers}(1978)}]{Cutler:1977qm}%
  \BibitemOpen
  \bibfield  {author} {\bibinfo {author} {\bibfnamefont {R.}~\bibnamefont
  {Cutler}}\ and\ \bibinfo {author} {\bibfnamefont {D.~W.}\ \bibnamefont
  {Sivers}},\ }\href {\doibase 10.1103/PhysRevD.17.196} {\bibfield  {journal}
  {\bibinfo  {journal} {Phys. Rev.}\ }\textbf {\bibinfo {volume} {D17}},\
  \bibinfo {pages} {196} (\bibinfo {year} {1978})}\BibitemShut {NoStop}%
\bibitem [{\citenamefont {Bengtsson}\ and\ \citenamefont
  {Ingelman}(1985)}]{Bengtsson:1984yx}%
  \BibitemOpen
  \bibfield  {author} {\bibinfo {author} {\bibfnamefont {H.~U.}\ \bibnamefont
  {Bengtsson}}\ and\ \bibinfo {author} {\bibfnamefont {G.}~\bibnamefont
  {Ingelman}},\ }\href {\doibase 10.1016/0010-4655(85)90003-7} {\bibfield
  {journal} {\bibinfo  {journal} {Comput. Phys. Commun.}\ }\textbf {\bibinfo
  {volume} {34}},\ \bibinfo {pages} {251} (\bibinfo {year} {1985})}\BibitemShut
  {NoStop}%
\bibitem [{\citenamefont {Gell-Mann}(1962)}]{GellMann:1962xb}%
  \BibitemOpen
  \bibfield  {author} {\bibinfo {author} {\bibfnamefont {M.}~\bibnamefont
  {Gell-Mann}},\ }\href {\doibase 10.1103/PhysRev.125.1067} {\bibfield
  {journal} {\bibinfo  {journal} {Phys. Rev.}\ }\textbf {\bibinfo {volume}
  {125}},\ \bibinfo {pages} {1067} (\bibinfo {year} {1962})}\BibitemShut
  {NoStop}%
\bibitem [{\citenamefont {Schwartz}(2014)}]{Schwartz:2013pla}%
  \BibitemOpen
  \bibfield  {author} {\bibinfo {author} {\bibfnamefont {M.~D.}\ \bibnamefont
  {Schwartz}},\ }\href
  {http://www.cambridge.org/us/academic/subjects/physics/theoretical-physics-and-mathematical-physics/quantum-field-theory-and-standard-model}
  {\emph {\bibinfo {title} {{Quantum Field Theory and the Standard Model}}}}\
  (\bibinfo  {publisher} {Cambridge University Press},\ \bibinfo {year}
  {2014})\BibitemShut {NoStop}%
\bibitem [{\citenamefont {Mertig}, \citenamefont {B{\"o}hm},\ and\
  \citenamefont {Denner}(1991)}]{MERTIG1991345}%
  \BibitemOpen
  \bibfield  {author} {\bibinfo {author} {\bibfnamefont {R.}~\bibnamefont
  {Mertig}}, \bibinfo {author} {\bibfnamefont {M.}~\bibnamefont {B{\"o}hm}}, \
  and\ \bibinfo {author} {\bibfnamefont {A.}~\bibnamefont {Denner}},\ }\href
  {\doibase https://doi.org/10.1016/0010-4655(91)90130-D} {\bibfield  {journal}
  {\bibinfo  {journal} {Computer Physics Communications}\ }\textbf {\bibinfo
  {volume} {64}},\ \bibinfo {pages} {345 } (\bibinfo {year}
  {1991})}\BibitemShut {NoStop}%
\bibitem [{\citenamefont {Shtabovenko}, \citenamefont {Mertig},\ and\
  \citenamefont {Orellana}(2016)}]{SHTABOVENKO2016432}%
  \BibitemOpen
  \bibfield  {author} {\bibinfo {author} {\bibfnamefont {V.}~\bibnamefont
  {Shtabovenko}}, \bibinfo {author} {\bibfnamefont {R.}~\bibnamefont {Mertig}},
  \ and\ \bibinfo {author} {\bibfnamefont {F.}~\bibnamefont {Orellana}},\
  }\href {\doibase https://doi.org/10.1016/j.cpc.2016.06.008} {\bibfield
  {journal} {\bibinfo  {journal} {Computer Physics Communications}\ }\textbf
  {\bibinfo {volume} {207}},\ \bibinfo {pages} {432 } (\bibinfo {year}
  {2016})}\BibitemShut {NoStop}%
\end{thebibliography}%

\end{document}